\documentclass[12pt,preprint]{aastex}

\slugcomment{in prep.}

\shorttitle{The dMe Flare Star EV~Lac in Quiescence }
\shortauthors{Osten et al.}

\begin{document}

\title{ From Radio to X-ray: The Quiescent Atmosphere of the dMe Flare Star EV~Lacertae}
 
\author{Rachel A. Osten\altaffilmark{1}}
\affil{Astronomy Department, University of Maryland, College Park, MD 20742-2421; rosten@astro.umd.edu}
\altaffiltext{1}{Hubble Fellow}

\author{Suzanne L. Hawley }
\affil{Astronomy Department, Box 351580, University of Washington,
 Seattle, WA 98195; slh@astro.washington.edu}
\author{Joel Allred}
\affil{Physics Department, Box 351560, University of Washington, Seattle, WA 98195;
jallred@u.washington.edu}

\author{Christopher M. Johns-Krull}
\affil{Physics \& Astronomy Department, Rice University, 6100 Main Street, Houston, TX 77005; cmj@rice.edu}

\author{Alexander Brown, Graham M. Harper}
\affil{Center for Astrophysics and Space Astronomy, University of Colorado, 593 UCB, Boulder, CO 80309-0593; ab@casa.colorado.edu,gmh@casa.colorado.edu}

\begin{abstract}
We report on multi-wavelength observations spanning radio to X-ray wavelengths of the M dwarf flare star,
EV~Lacertae,
probing the
characteristics of the outer atmospheric plasma from the upper chromosphere to the
corona.
We detect the star at a wavelength of 2 cm (15 GHz) for the first time.
UV and FUV line profiles show evidence of nonthermal broadening, and the velocity width
appear to peak at lower temperatures than in the Sun; this trend is confirmed
in
another active
M dwarf flare star.
Electron density measurements indicate
nearly constant electron pressures between $\log T=$5.2 and 6.4.  At higher
coronal temperatures, there is a sharp increase of two orders of magnitude
in density (n$_{e}\sim$10$^{13}$ cm$^{-3}$ at $\log T=$6.9).
X-ray, EUV, FUV and NUV spectra
constrain the DEM from the upper chromosphere through the corona.
The coronal pressures are inconsistent with
the assumption of hydrostatic equilibrium, either through EM modeling
or application of scaling laws, and imply large conductive loss
rates and a large energy input at the highest temperatures.
The timescales for radiative and conductive losses in EV~Lac's
upper atmosphere
imply that significant continued
heating must occur for the corona to maintain its quiescent properties.
The high frequency radio detection
requires the high temperature X-ray-emitting coronal plasma
to be spatially distinct
from the radio emission source. Length scales in the low-temperature corona
are markedly larger than those in the high-temperature corona, further suggestions of
an inhomogeneous mixture of thermal and nonthermal coronal plasma.
\end{abstract}


\section{Introduction}
By studying the properties of 
extreme coronae, such as on M dwarf flare stars, clues to common
physical processes arising on these stars and our low-activity Sun 
can be investigated.
dMe stars are small, 
and have intense magnetic fields covering
a large majority of the stellar disk \citep{jkv1996,saar1994}. 
The transition to fully convective stars occurs at spectral type close to M4V, so dMe stars
with spectral types earlier than this are close to being fully convective.
Despite their small size, dMe flare stars produce significant coronal
emission, in many cases near the maximum (L$_{x}$/L$_{bol}\sim$10$^{-3}$) that
any star is able to maintain in the quiescent state \citep{vilhuwalter1987,hempelmann1995}.
How different are coronal structures on dMe stars compared to the Sun, and other active stars?
%At first blush, active late-type stars are more similar to each other in X-ray (and radio) characteristics
%than they are to the Sun, something which has proved puzzling for quite awhile.  
The large magnetic filling
factors inferred for the photospheres of dMe stars
%, and presumed expansion with height to occupy all the magnetic volume
%accessible to it, 
suggest a stark contrast with the Sun's small ($<$1\%) sunspot area coverage.
%The preponderance of radio gyrosynchrotron emission,
%observed in the Sun only in active regions undergoing particle acceleration due to flare-like events, along with 
%enhanced X-ray temperatures typically only found in the Sun during flares, suggests that flare-like events might be
%occurring continuously on a small scale in active stellar coronae.  Statistical studies of stellar coronae
%(refs) indeed suggest that flaring is an important contributor to the overall observed X-ray emission.
Recent results from X-ray spectra have indicated that the elemental fractionation
pattern seen in most active stars, including active M dwarfs, is opposite that seen on 
the Sun.  In the solar corona lines with First Ionization Potential (FIP) $<$10 eV
are enhanced relative to those with greater FIP; this
is referred to as the 
First Ionization Potential (FIP) effect. 
At temperatures less
than about 1 MK, the FIP effect disappears from full-disk observations of the 
Sun \citep{ldw1995}, and lower-temperature structures do not
show the FIP effect.  
 In active stellar coronae, elements with FIP $>$10 eV appear
enhanced relative to those with lower FIP.  
The properties of time-averaged active stellar coronae appear to resemble solar flares,
with elevated temperatures and densities.

The n$_{e}^{2}$
dependence of the emissivity in transitions appearing in the UV through X-ray 
skews analysis to the densest structures, but still allows an investigation of gross coronal properties.
As \citet{rtv1978} remarked, 
``spatially unresolved observations of stellar coronae represent a statistical smoothing of
widely disparate plasma structures.''  Reconstructions of the 
emission measure distribution of plasma, and
the dependence of electron density, with temperature, n$_{e}$(T$_{e}$), from high spectral resolution
grating data give constraints on stellar coronal structure and heating.  
Another motivation for a detailed investigation of the outer atmosphere of an M dwarf flare star
lies in recent studies which suggest that the occurrence of flares may be frequent enough to supply
the heating requirements of some stellar coronae \citep[][but first put forth in \citet{parker1988}]{gudel2003}.  
Recent studies on the Sun are still inconclusive about the role of micro- and nano-flares
in providing the steady X-ray luminosity \citep{kruckerhudson}.
%This is in marked contrast 
%with the Sun, where a there is a disconnect of $\approx$ XX between flare energy resources and 
%coronal heating requirements.
If flares are the dominant source providing
coronal heating in dMe flare stars, then 
the atmospheric properties should reflect that.

EV~Lac (Gliese 873) is a young disk population dM3.5e star at a distance of 5 pc.  It is the
second brightest M dwarf X-ray source seen in the ROSAT All-Sky Survey, with a 
quiescent 0.1--2.4 keV flux of $\sim$ 4$\times$ 10$^{-11}$ erg cm$^{-2}$ s$^{-1}$ \citep{hunsch1999}.
The quiescent X-ray luminosity is a significant fraction of the total
bolometric luminosity ($\sim$0.2\%), near the maximum of any star in its quiescent (non-flaring)
state.
Its low mass \citep[0.3 M$_{\odot}$;][]{delfosse1998} puts it
near the dividing line for fully convective stars.  EV~Lac
has very strong, 4 kG magnetic fields covering $>$ 50\% of the stellar surface 
\citep{jkv1996,saar1994} 
so coronal phenomena are dominated by the interactions of these close-packed magnetic field lines.
Recent investigations of the H Lyman-$\alpha$ profile have determined the existence of
a stellar wind through detection of an astrosphere \citep{wood1}, with an estimated mass loss rate very similar
to the solar mass loss rate \citep{wood2}.

Based on its previous record for large and dramatic flare variability, we studied
the multi-wavelength behavior of EV~Lac with observations utilizing the {\it Chandra X-ray
Observatory}, the {\it Hubble Space Telescope}, ground-based optical photometry and spectroscopy,
and ground-based radio interferometry, over a period of two days in September 2001.
Many flares were detected; 
a separate paper \citep{osten2005}
discusses the flare observations.
The focus of this paper is an investigation of EV~Lac's
quiescent atmospheric properties from X-ray to radio wavelengths. 

\section{Observations and Data Reduction }
%The subsequent sections
%describe the individual programs.

\subsection{VLA }
EV~Lac was observed with the NRAO\footnote{The National Radio Astronomy Observatory
is a facility of the National Science Foundation operated under cooperative agreement by 
Associated Universities, Inc.} Very Large Array on 19 and 20 September 2001.  
The array was moving from C to D configuration.
The phase calibrator was 2255+420, and the flux density calibrator was 0137+331 (3C 48).
The observations on 19 September 2001 were performed using the entire array, and
frequency switched between 6, 3.6, and 2 cm (4.9, 8.4, and 15.0 GHz), for a total of 10 hours
of observations.  On 20 September, the observing strategy changed:  we observed
simultaneously in two subarrays at 6 and 3.6 cm, to characterize the dual frequency
behavior of any flares that might occur during the Chandra observations.  
Processing of the data was done in AIPS (version 31 Dec 03).
The field
around EV~Lac is crowded; there is a large radio galaxy $\sim$ 2.7 arcminutes away, and numerous
radio sources.  We performed multi-field cleaning, taking advantage of the NVSS radio
survey to locate possible bright radio sources in the primary beam and sidelobes.  
After imaging the field, visibilities of sources not identified as EV~Lac were removed,
and the field was re-imaged.  
The quantities measured are the total
flux, I, the amount of circularly polarized flux, V, 
and the percentage of circular polarization, $\pi_c$ is defined as \\
\begin{equation}
\pi_c (\%)= \frac{V}{I} \times 100 \;\; .
\end{equation}
%where I is the Stokes parameter for total flux, V is the Stokes paramter for
%circularly polarized flux.  

\subsection{HST}
EV~Lac was observed with the {\it HST-STIS} 
for 4 orbits on 20 September 2001. 
The data were obtained in the Cycle 9 program 8880 with the E140M grating, centered at 1425 \AA\ 
using the FUV-MAMA detector and the 0.2x0.2 arcsec aperture. 
The wavelength range
covered was 1140--1735 \AA\ in 44 orders, with an approximate resolving power of $R=45,000$.
The data were acquired in TIME-TAG mode to enable investigation
of short time-scale variability at high spectral resolution.  

Spectra were extracted in 60 second intervals to search for significant
variability; the order location
solutions from the pipeline-processed data were used to extract spectra 
and assign wavelengths.
The background and spectral extraction widths were 7 pixels, and the background was fit
by a second order polynomial; the background was then subtracted to obtain a net spectrum. 
We constructed a light curve by summing the background-subtracted counts 
in each 60 second spectrum,
using Poisson statistics to generate error bars.  
A quiescent spectrum was constructed by excluding times of flaring activity.  The net quiescent
exposure time was 10,320 seconds (total exposure 10,920 s).

\subsection{FUSE}
EV~Lac was observed with the {\it Far Ultraviolet Spectroscopic Explorer (FUSE)}\footnote{Based on observations made with the NASA-CNES-CSA Far Ultraviolet Spectroscopic Explorer.  
{\it FUSE} is operated for NASA by the Johns-Hopkins University under NASA contract NAS5-32985.} in the Cycle 3
program C1140201 (PI A. Brown) on 1 July 2002.  An overview of the {\it FUSE} mission and the on-orbit performance
are described in \citet{moosetal2000} and \citet{sahnow2000}, respectively.

The data for EV~Lac were collected in time-tagged mode through the 30$\times$30 arcsec
LWRS apertures chosen to maximize the data collected in the SiC channels ($920-1100$\AA) which
can be lost from smaller apertures owing to thermal misalignment of the telescope.
%The LWRS aperture is large enough that the target remains within the aperture despite the
%normal level of pointing jitter and target drift experienced during {\it FUSE} observing.
Our subsequent examination of the stellar signal suggests that the target were within
the apertures throughout the observation.
The spectral resolution using the LWRS is typically $R\sim 15,000$ ($20\>{\rm km\>s}^{-1}$).
%The data from C1140201, which were collected as 18 sub-exposures (1--16, 23 and 24),
%were retrieved from MAST.  
The raw spectra were calibrated with CalFUSE v2.4.0 \citep{calfuseref};
the location of the extraction windows were manually selected and 
optimally aligned, with the orbital day
and night spectra being extracted separately.  The 38 ks observation contained
approximately 16 ks and 20 ks of orbital day and night, respectively.  We adopted the default
burst screening parameters and background subtraction.

\subsection{EUVE}
We make use of an earlier observation of EV~Lac with the Extreme Ultraviolet Explorer (EUVE), 
in order to fill in critical
temperature regions.  
EV~Lac was observed by EUVE in 1993, from 9--17 September.  
For more information on the EUVE mission and instruments see \citet{euveref}.
%Preliminary results from the EUVE observation were made by \citet{ambruster} {\it do I need to
%mention this?}
Processing of the data
used specialized reduction procedures in IRAF and IDL\footnote{IRAF is distributed
by the National Optical Astronomy Observatories, which is operated by the Association of 
Universities for Research in Astronomy, Inc., under contract to the National Science Foundation (USA).} 
\citep[see][]{osten2003};
briefly, times of large background
were excluded, and events falling within a circular region around the source were
tagged for source spectral extraction; 
an annular region around the source region was tagged as background.
A light curve was generated by binning source and background photons and subtracting 
the background.
Times of obvious
flaring were excluded.  A net spectrum was extracted for the short wavelength (SW) and
medium wavelength (MW) spectrometers, respectively.  The net exposure time was 105ks
for each spectrometer.

Inspection of the MW spectrum revealed the presence of two transitions of \ion{Fe}{16}, at 
$\lambda$ 335.41
and $\lambda$ 360.76, whose ratio can be used to constrain the column density of intervening
interstellar hydrogen.  Previous high energy missions 
\citep[e.g.,][]{sciortino1999}
have used
N$_{H}$ of 10$^{19}$ cm$^{-2}$; the ratio of the two \ion{Fe}{16} lines gives N$_{H} \sim$ 4$\times$10$^{18}$ cm$^{-2}$ based on interstellar cross sections in \citet{mm1983} and the
contribution functions tabulated in \citet{brs1995}.
%a 3 $\sigma$ upper limit
%of N$_{H} \leq$ 6 $\times$ 10$^{18}$, 
We adopt this value for correction of observed EUV and X-ray 
emission lines.

\subsection{ Chandra Observations}
The primary focus of the investigation was a unique 100 ks Chandra 
observation, from 2001 September 19.8--21.0.
The observations were made using the High Energy Transmission
Grating Spectrometer (HETGS) in conjunction with the ACIS-S detector array\footnote{
More information about Chandra and its instruments can be found at http://cxc.harvard.edu/proposer/POG/html/MPOG.html}.
The HETGS provides an undispersed image of the object from 0.3--10 keV
and spectra through two 
gratings:  the Medium-Energy Grating (MEG) covering the wavelength range 1.7--31 \AA\
in first order and the High Energy Grating (HEG) covering the wavelength range 1.2--18.5 \AA\
in first order, with almost twice the spectral resolution.  
%The observation
%was taken in timed exposure mode, for which CCD events are accumulated every 3.24 s 
%before being read out.  This sets the minimum time resolution.
Processing the Chandra data utilized CIAO, Version 2.2, ``threads'', i.e.,
processing recipes, for different aspects of the data reduction. 
We eliminated bad aspect times and confirmed that the observation was not affected
by ACIS ``background flares''.  The new ACIS pixel size (0.0239870 mm) and 
focal length (10070.0 mm) were updated in processing the ACIS events.  Events with
energies less than 300 eV and greater than 10,000 eV were excluded, as these are not
well-calibrated; also excluded were events whose pulse-height invariant (PI)
values were 0,1, or 1024 (overflow/underflow values).  Only ASCA grades 0,2,3,4,6
were kept.  The events were resolved into spectral events using region filtering
(rectangles around the MEG and HEG stripes), and a level 2 event file was generated.
One of the CCD chips in the ACIS-S array (S4) suffered increased scatter (streaking)
in the horizontal (CHIPY) direction, apparently caused by a flaw in the serial
readout; a destreak filter was applied to remove these data.  Updated sensitivity
files were used to correct for contamination on the ACIS chip which affects sensitivity
at low energies.

We used custom IDL procedures to determine light curve variations and subsequent
spectral analysis.  The level 2 events file was filtered for MEG events falling within
the spectral extraction window ($\pm$0.1 mm in the cross-diffraction
coordinate).  
The first $\sim$40 ks of the observation were unaffected by flares; the remaining
60 ks was characterized by numerous small enhancement events and a few high contrast,
but short duration, events.  A separate paper details the X-ray flare variations in 
context of 
the multi-wavelength nature of the monitoring campaign \citep{osten2005}.  Little
difference is apparent in the spectra of time intervals affected by small flares and 
those characterized by quiescence, so we extracted spectra corresponding 
to the sum of the initial quiescent
period and a later interval marked by small enhancements 
\citep[flares 3--6 and 9 in the terminology of][]{osten2005} for this study.  
The total exposure time of the extracted interval is 65 ks.

\section{Analysis \label{analsec}}
\subsection{Radio}
\begin{figure}
\begin{center}
\includegraphics[scale=0.5,angle=90]{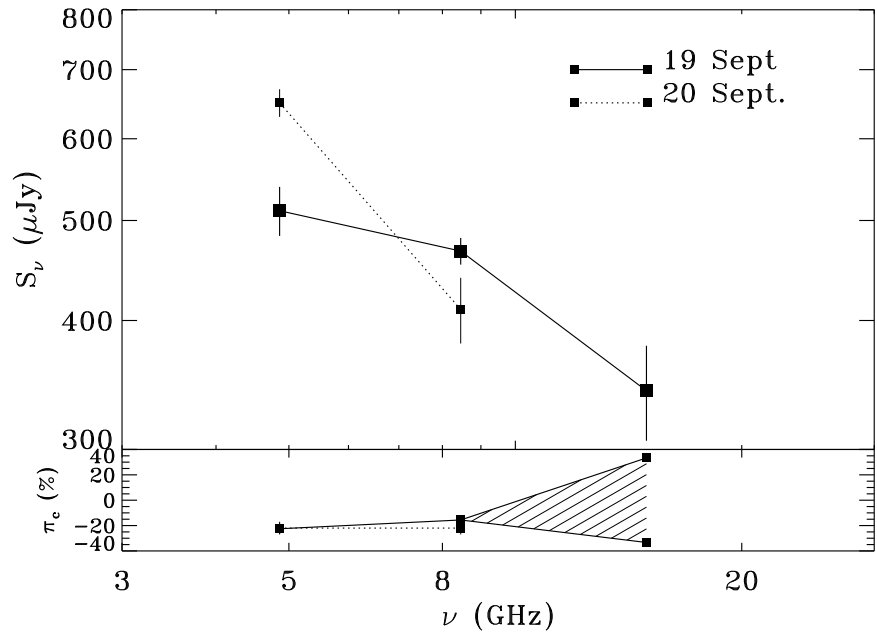}
\caption{ Observed radio flux densities of EV~Lac on two consecutive days in 2001 September.
Bottom panel shows the circular polarization spectrum.
%Overplotted are two model scenarios which can reproduce the general flux
%characteristics.  
\label{fig:radiospec}}
\end{center}
\end{figure}

The flare campaign described by \citet{osten2005} focused on the large flare observed at
3.6 and 6 cm on 20 September 2001.  
Here
we concentrate on the 19 September and quiescent part of the 20 September radio observations.  
Flux and 
polarization parameters for the data on 19 September are listed in Table~\ref{tbl:vla};
values for 20 September are listed in \citet{osten2005}.  
Figure~\ref{fig:radiospec} plots the flux and polarization spectra.
In addition to determining the
average flux density of the source at  three frequencies on 19 September, we investigated
the temporal variations to ensure that no large flare events occurred.
We detect the source at 2 cm
for the first time.  The spectral index between 6 and 3.6 cm (assuming S$_{\nu} \propto \nu^{\alpha}$)
 is $-0.16\pm$0.11 (1$\sigma$), while between 3.6 and 2 cm it is $-0.55\pm$0.18 (1 $\sigma$).  
The 6 and 3.6 cm circular polarization values 
are almost the same as the ``preflare'' values recorded on 20 September, and reported in
\citet{osten2005}, and the 3.6 cm fluxes are very similar as well.  The largest variation
is in the 6 cm flux on 19 and 20 September, but it is less than 30\%.
Thus we have confidence that
the quiescent conditions in the radio-emitting corona were at a characteristic level during 
all the 
observations described here.

\subsection{Line Fluxes}
\subsubsection{HST}
\begin{figure}
\begin{center}
\includegraphics[scale=0.7]{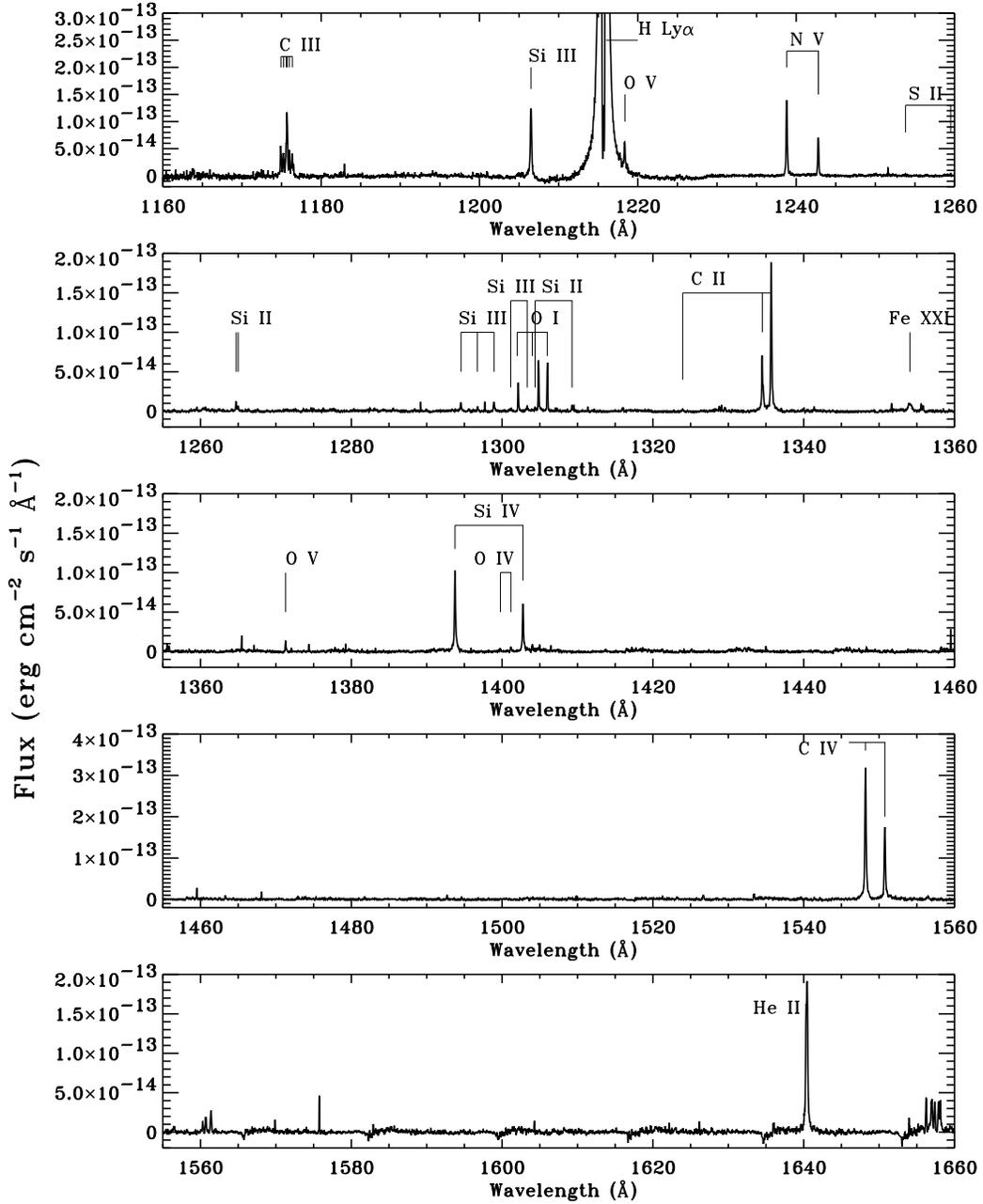}
\caption{HST/STIS spectrum of EV~Lac.  Spectrum has been smoothed over 4 wavelength bins for clarity.  Bright emission
lines and those used in the DEM/density analyses are identified.  \label{fig:stisspect}}
\end{center}
\end{figure}

%A light curve of the STIS time-tagged data was made by extracting spectra in 60 second
%intervals, and calculating the total net source counts in each 60 s spectrum, to 
%determine time intervals affected by flares.  After separating these, we co-added the
%quiescent intervals to form a good S/N estimate of the quiescent UV spectrum, with an
%exposure time of 10,320 seconds.  
The quiescent STIS spectrum, characterized by bright emission lines, is shown
in Figure~\ref{fig:stisspect}.
We fit moderate and strong emission lines using a Gaussian line profile analysis, convolving
the line profile fit with the empirical line spread function to account
for instrumental broadening appropriate for the aperture and grating.
\footnote{Tabulated line spread function data is available from 
http://www.stsci.edu/hst/stis/performance/spectral\_resolution}  
We use the laboratory wavelengths in the stellar rest frame, corrected for
EV~Lac's radial velocity of $-1.5$ km s$^{-1}$ \citep{rvref}.
Table~\ref{tbl:stislines} lists the derived velocities, widths, and fluxes.

In the case of \ion{O}{5} $\lambda$1218, we 
fitted the wing of H~Ly$\alpha$
as a second order polynomial to subtract and fit the \ion{O}{5} line.  
The transition of \ion{Fe}{24} $\lambda$1354 is blended with a \ion{C}{1} transition, 
and requires careful modeling of the expected flux from the \ion{C}{1} line to remove
its influence on the \ion{Fe}{24} profile.  We used the line flux listed in 
\citet{ayres2003}, who included EV~Lac in their survey. 
The contaminating influence of the
\ion{C}{1} transition is weak in this case, as estimates from 
simple Gaussian fitting to the line profile
are similar to those obtained from the more sophisticated approach.
We fit the \ion{C}{3} multiplet from 1174.9--1176.4 \AA\ as the sum of six Gaussians,
with central wavelengths fixed to the laboratory values, and FWHM and amplitudes allowed to vary.

%The width of the broad component of $\lambda$1238
%is the only broad component line profile to display accompanying nonthermal velocity which 
%is subsonic.

We also checked that line ratios of transitions selected for use in emission measure analysis 
were consistent with the effectively thin flux ratios.  Table~\ref{tbl:linerat} lists the
transitions, the expected effectively thin ratios, as well as the observed values.  
The observed ratio of \ion{Si}{2} $\lambda$1533.430/$\lambda$1526.706 is slightly lower
than the effectively thin ratio, as is \ion{C}{3} $\lambda$1175.713/$\lambda$1174.935.
This is consistent with trends reported in the Sun and other stars \citep[][and references therein]{delzanna2002}.
The \ion{C}{3} transitions share a common upper level so their ratio will depart from the optically
thin values when there is opacity in the line --- the deviation is that expected from line scattering
in these systems.  They can still be considered effectively thin if no photons are destroyed.
The magnitude of the departures is small, and we use the total line fluxes to estimate the
DEM at these temperatures (see \S 5.2).
For line profiles
which were fit better by the sum of two Gaussian line profiles, we compared the flux ratios of
the narrow and broad components separately.  For the \ion{Si}{4} and \ion{C}{4} doublets,
the flux ratio of each narrow component, as well as the flux ratio of each broad component,
is consistent within the errors with the effectively thin flux ratio.  
%{\it Linsky et al. (1995) found that the widths of Capella's intersystem lines were
%closer to the widths of the narrow components detected in the
%resonance lines than to the widths of the broad components.. . .}

\subsubsection{FUSE}
\begin{figure}
\begin{center}
\includegraphics[scale=0.7,angle=90]{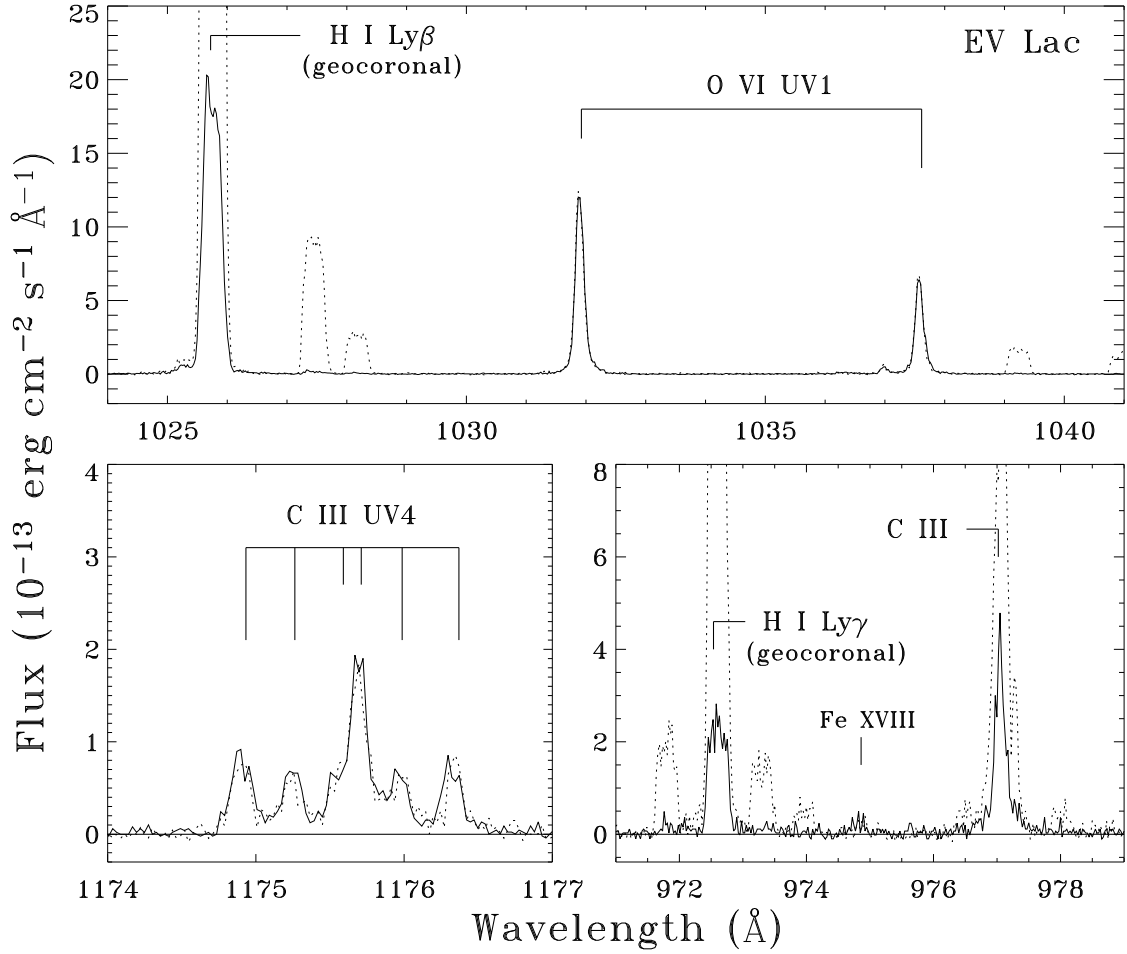}
\caption{FUSE spectrum of EV~Lac.\label{fig:fusespect}}
\end{center}
\end{figure}

The data were split into night-time and day-time data so that the effects of
air-glow emission are recognizable.  No obvious flaring was detected during the
observation.  The spectra containing the detected transition region emission lines
are shown in Figure~\ref{fig:fusespect}.  The only stellar lines detected are \ion{O}{6}
$\lambda\lambda$1031.9,1037.6, the \ion{C}{3} $\lambda$1175 multiplet and the $\lambda$
977 resonance line, and the forbidden transition \ion{Fe}{18} $\lambda$974.  A comparison of the day-time and
night-time spectra show that the \ion{O}{6} lines and the \ion{C}{3} intersystem multiplet
are unaffected by airglow features.  The day-time data for the \ion{C}{3} 977 \AA\ profile show
the presence of significant scattered solar \ion{C}{3} photons in the SiC channels, which is not 
present during orbital night.  Since EV~Lac is so near ($d=5$pc), FUV reddening is likely to be
small.
Emission line fluxes for these lines were measured by direct summation over the line profiles.
No continuum signal was detected in the region of these lines.  
Line fluxes of detected lines are listed in Table~\ref{tbl:fuselines}.

\subsubsection{EUVE and Chandra}
\begin{figure}
\begin{center}
\includegraphics[scale=0.7,angle=90]{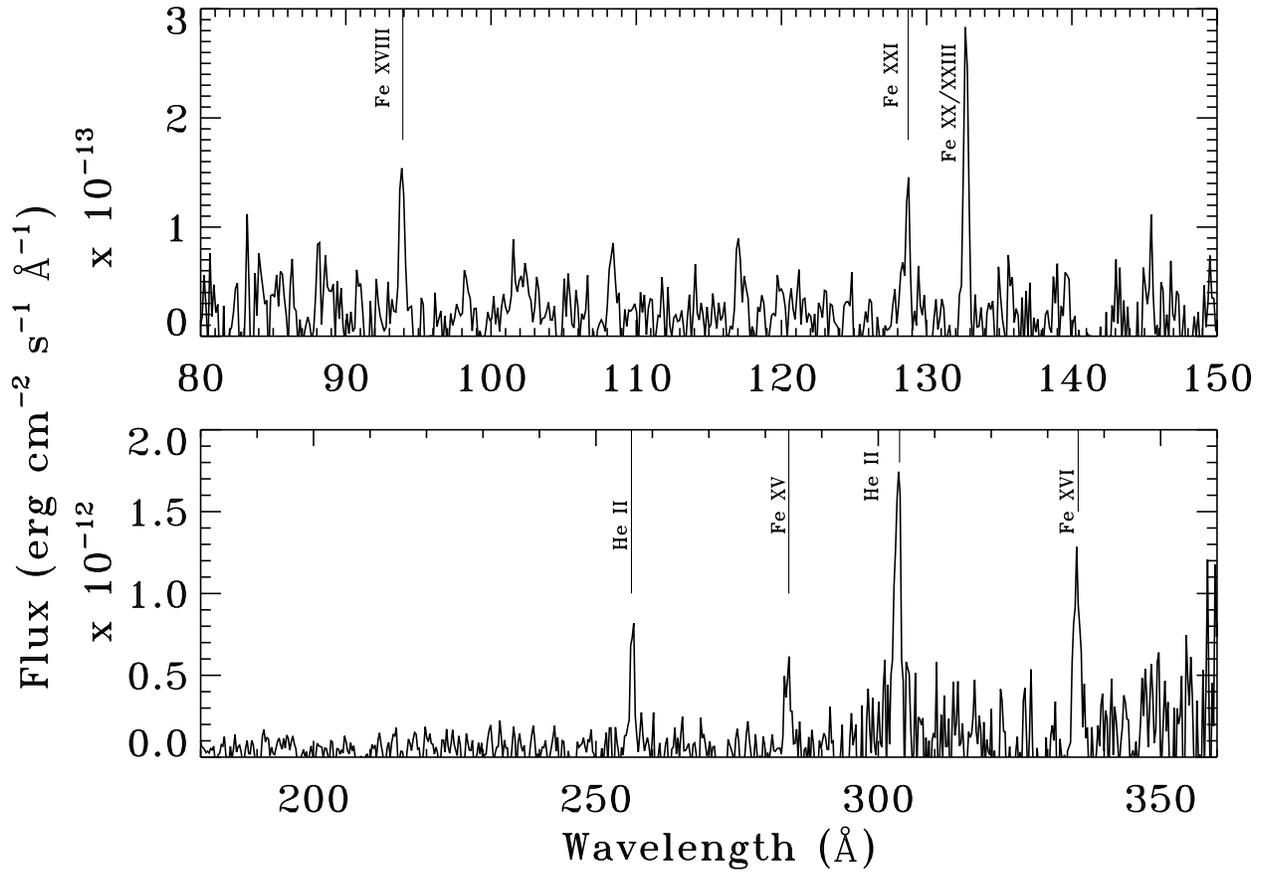}
\caption{EUVE spectrum of EV~Lac.\label{fig:euvespect}}
\end{center}
\end{figure}

\begin{figure}
\begin{center}
\includegraphics[scale=0.7]{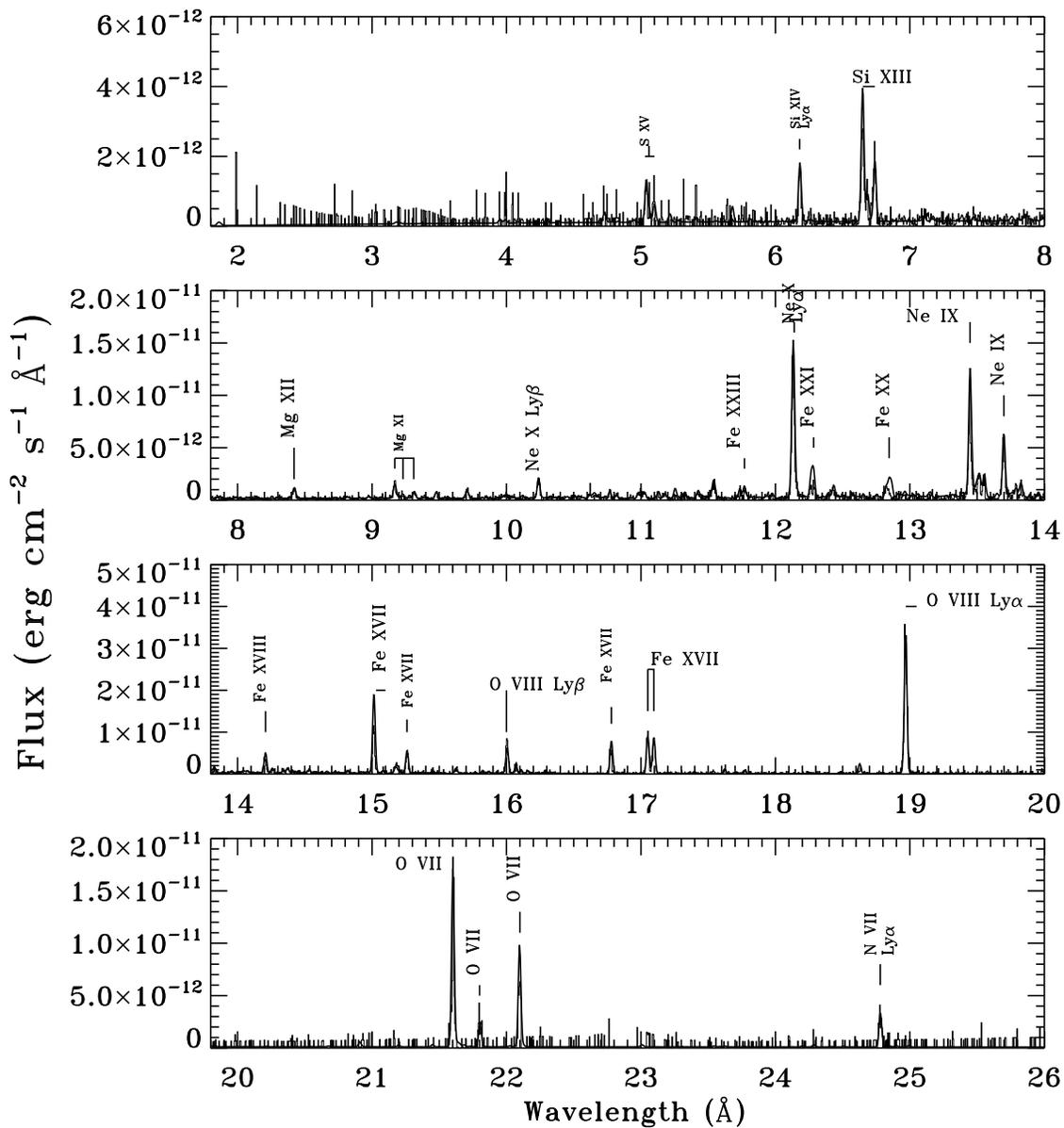}
\caption{Chandra MEG spectrum of EV~Lac (black histogram), with synthesized spectrum from
DEM and abundance analysis in \S 5.
Bright emission lines
are identified. 
\label{fig:chandraspec}}
\end{center}
\end{figure}

The EUV and X-ray spectra are dominated by lines of highly ionized iron, as well as bright lines arising
from transitions of H- and He-like ions in the X-ray region
(Figures~\ref{fig:euvespect} and \ref{fig:chandraspec}).  These lines are well described as effectively
thin transitions occurring in collisional ionization equilibrium. 
%optical depth effects have been
%investigated, and while there is disagreement between theory and observations, these seem to point to
%inaccuracies in atomic data more than optical depth effects (need references).  
At the spectral
resolution of the Chandra HETGS, the lines are unresolved; in our analysis of the line profiles 
we assume a Gaussian line profile shape with width fixed to the instrumental resolution (23 m\AA\
for MEG, 11 m\AA\ for HEG),
and we solve for line flux and central wavelength.  We generated a list of moderate to strong
transitions in the HETG bandpass from the atomic linelists of the APEC \citep{apecref} project
to use for line fitting.   
Widths of EUVE spectral line fits were kept fixed to the instrumental resolution: 0.38 \AA\ 
for SW,
1.14 \AA\ for MW.  

Continuum processes, consisting of free-free, free-bound, and two-photon 
processes from several elements (but dominated by H and He), can be non-negligible at
high energies. 
As the continuum
emission can be important in the EUV spectral region, but difficult to disentangle from the numerous weak lines,
we determined the expected amount of continuum emission in the EUV spectral range using the scaling at X-ray wavelengths 
$\lambda \ge$ 10 \AA\ and subtracted this from the observed EUV spectra to estimate continuum-free emission line fluxes.
As discussed in \S 4 in more detail, we utilized a self-consistent treatment of line and continuum emission
in order to make a first estimate of the DEM and abundances, and determine EUV/X-ray line fluxes
without contribution from continuum emission for further analysis.  
Line fluxes of detected EUV lines are listed in Table~\ref{tbl:euvelines}.
Table~\ref{tbl:xraylines} lists the detected X-ray transitions used in the DEM, abundance, and
electron density analyses described in the next sections, along with the continuum-free estimate of
line fluxes.  %Figure~\ref{fig:chandradem} shows the coronal DEM, Fe line agreement, 

%This procedure is described in
%detail in \citet{osten2003}.  Briefly, transitions of iron in the EUV and X-ray spectral range are used
%to constrain the shape of the DEM, relative to the (as yet) unknown iron to hydrogen ratio.  The DEM distribution
%is used to predict the shape of continuum emission, and also to constrain the presence of temperatures hotter
%than those indicated by observed emission lines.  Scaling between the observed continuum shape and level, and minimizing the
%disagreement between observed and predicted iron line fluxes, produces a first guess at the DEM and continuum-free emission
%line fluxes. The shape of the DEM is used to constrain the elemental abundances with respect to iron, for all non-iron 
%emission lines detected in the Chandra spectrum.  The shape and level of the continuum spectrum is then used to determine the
%iron abundance on an absolute scale, with respect to hydrogen, and the other elemental abundances bootstrapped to hydrogen.
%The process repeats with the initial estimate of the DEM and elemental abundances, until a final solution converges.
%Line fluxes of detected EUV lines are listed in Table~\ref{tbl:euvelines}.
%Table~\ref{tbl:xraylines} lists the detected X-ray transitions used in the DEM, abundance, and
%electron density analyses described in the next sections, along with the continuum-free estimate of
%line fluxes.  %Figure~\ref{fig:chandradem} shows the coronal DEM, Fe line agreement, 
%derived abundances, and shape
%and level of continuum emission.

\subsubsection{Variability \label{section:var}}
Only the STIS and Chandra spectra are strictly simultaneous; the EUVE observation was 
made $\sim$8 years earlier,
the FUSE observation $\sim$10 months later.  In light of the dramatic variability of dMe stars, 
it is worthwhile to
check whether the state of the star is roughly the same in these three different pointings.  We 
have made some initial cuts to exclude
large flares, but smaller levels of variability may exist, and we 
address that here.  

The \ion{C}{3} multiplet at $\lambda$1175 is present in both STIS and FUSE spectra, taken $\sim$10 months apart.
The line fluxes in the FUSE spectrum are $\approx$1.6 times larger than the STIS line fluxes.
The ionization fraction of the ion \ion{C}{3} is a maximum at a temperature $\approx$ 9$\times$10$^{4}$ K.
There is evidence for a small amount of variability in the upper chromospheric and transition region
plasma. If the DEM has changed slightly between these times, we could expect a difference of this magnitude.

%Even though
%the spectral ranges differ greatly, 
There are coronal transitions in the UV, FUV, EUV, and 
X-ray whose observed fluxes can be
compared against the emissivity ratios for the transitions, to quantify the level of 
variability at these different times.
In the STIS spectral range there is \ion{Fe}{21} $\lambda$1354, and \ion{Fe}{21} $\lambda$128.73 
is detected by
EUVE.  
In the FUSE spectral range there is \ion{Fe}{18} $\lambda$974, and \ion{Fe}{18}$\lambda$93.92 
is detected by
EUVE.  The observed fluxes of these lines are compared against the brightest transitions of 
\ion{Fe}{18} and \ion{Fe}{21}
detected in the Chandra bandpass, to determine whether and how much EV~Lac varied between 
the observations.  We use 
$\lambda\lambda$9.48,12.284 as the brightest transitions of \ion{Fe}{21}, and 
$\lambda\lambda$14.208,16.07 of \ion{Fe}{18}.
The emissivity ratios of these transitions have a range, due to their
temperature dependence.  Since
the observed flux is proportional to the emissivity times the DEM, and it is possible 
that the DEM has changed if the system
is variable, we examine a range of emissivity ratios to gauge the extent of variability. 
By assuming that lines are formed over $\Delta \log T=$0.3, we calculate the range of
emissivities expected.
Table~\ref{tbl:var} lists the observed flux (energy)
ratios, along with the emissivity (energy) ratios, over a temperature range of 
$\pm$0.15 dex from T$_{\rm max}$, the temperature at which the ionization fraction peaks.
The observed flux ratios are consistent within the observational errors to the range
expected from temperature variations within $\Delta \log T=$0.3.

Based on the chromospheric and coronal line flux comparisons, we hereby assume that 
by analyzing these spectra as a whole
we are diagnosing a characteristic state of EV~Lac.
%the atmosphere 
%of EV~Lac experiences variations that are 
%temperature independent; i.e., the entire atmosphere
%varies.  
%This assumption is probably not true, as evidence from other stars 
%indicates \citep[e.g. Capella,][]{ayres2003},
%but we 
%have no means of constraining variations at less than coronal 
%temperatures (T$\leq$ 4$\times$10$^{6}$K).  

\subsection{Line Profiles}
Only the STIS and FUSE spectra have sufficient spectral resolution to perform line profile
analysis.
Earlier work on active stars by \citet{linskywood1994} showed that UV emission lines are often
better described as the sum of two Gaussians, and we follow this approach in our
line profile analysis, comparing the statistical fit between a single Gaussian and double Gaussian.
%For each line profile, we used the empirical line spread function to characterize the
%instrumental broadening, and determined whether one or two Gaussian features best
%reproduced the observed line profile. 
Line profile parameters are listed in Table~\ref{tbl:stislines} for STIS lines,
and Table~\ref{tbl:fuselines} for FUSE lines. 
In the case where two Gaussians were the better fit,
we record the line profile parameters for each Gaussian.
%The cleanest
%emission line profiles are for the \ion{O}{6} lines and we performed Gaussian
%fitting for these profiles.  
%This leads to an excess above the thermal
%width of at most 13 km s$^{-1}$ (for T$_{\rm form}=$3$\times$10$^{5}$K), which
%we attribute to a nonthermal velocity.  The \ion{O}{6} data point is the open circle at 
%$\log T_{\rm form}=5.5$ in Figure~\ref{fig:vturb}.
%A cross-correlation for the FUSE spectra show in both channels (SiC? LiF?)
%a possible change in the \ion{O}{6} $\lambda$1032
%red wing between orbital day and night.

\begin{center}
\begin{figure}
\includegraphics[angle=90,scale=0.5]{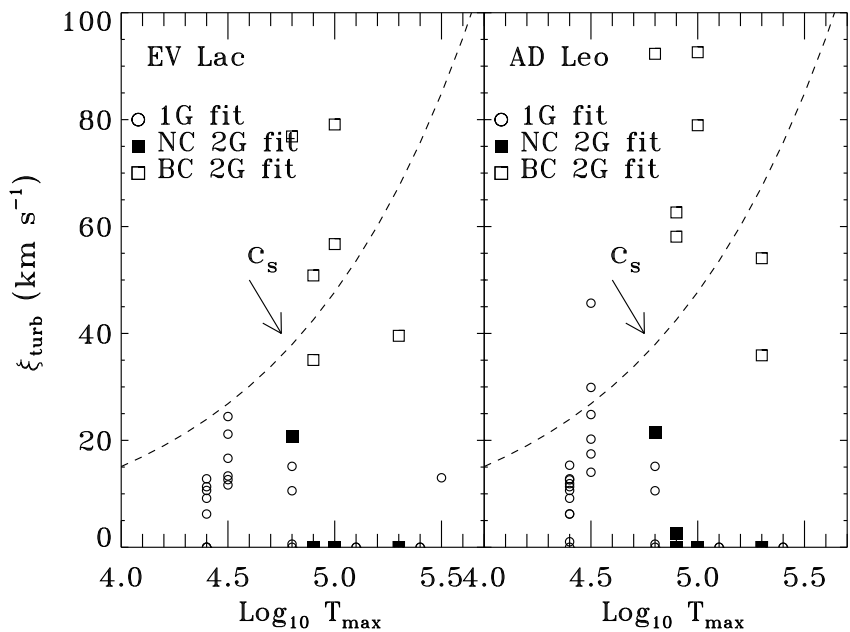}
\caption{Nonthermal velocities deduced from excess broadening in line profiles 
found in STIS spectrum.  ``1G fit'' refers to line profiles fit by a single Gaussian,
``NC 2G fit'' refers to the narrow component of a two Gaussian line profile fit,
and ``BC 2G fit'' refers to the broad component of a two Gaussian line profile fit.
The dashed line represents the sound speed of a fully ionized plasma assuming solar
abundances. {\it (Left panel)} Trends for EV~Lac; {\it (right panel)} trends for AD Leo.
\label{fig:vturb}}
\end{figure}
\end{center}

There is a problem with the STIS pipeline for TIMETAG mode observations,
in which the Doppler shift is not correctly applied.\footnote{The problem has been corrected
in CALSTIS version 2.15c; see http://www.stsci.edu/hst/stis/calibration/pipe\_soft\_hist/update215c.html
and \citet{herczeg2005} for more details.}  The magnitude of the error is 3.8 pixels, or about 12 km s$^{-1}$,
and degrades the intrinsic resolution of the spectra.  Since our spectrum is accumulated over 4 orbits,
we corrected for this effect by adding a Gaussian
of width 12 km s$^{-1}$ to the observed velocity width in quadrature to account for the decreased resolution.
The observed line widths can then be expressed as the sum in quadrature of three
velocities, \\
\begin{equation}
\left(\frac{\Delta \lambda}{\lambda}\right)^{2}=3.07\times10^{-11} \left( \frac{2k_{B}T_{\rm max}}{m_{i}}+\xi^{2}
+v_{\rm inst}^2 \right)
\end{equation}
 adapted from \citet{wood1997},
where $\Delta \lambda$ is the line profile FWHM in \AA, $\lambda$ is the line center in \AA, $k_{B}$ is
Boltzmann's constant in erg K$^{-1}$, $T_{\rm max}$ is the temperature at which the ion's ionization
fraction peaks in $K$, 
$m_{i}$ is 
the ion mass in $g$, $\xi$ is the most probable nonthermal velocity in km s$^{-1}$, and $v_{\rm inst}$
is the instrumental broadening of 12 km s$^{-1}$. 
%with those expected for a Maxwellian distribution 
%with temperature T$_{\rm form}$ corresponding to the temperature of the maximum ionization fraction, and 
%computed nonthermal velocities where excess broadening was present using the formula:\\
Figure~\ref{fig:vturb}
displays the trend of nonthermal velocity with formation temperature for the line profiles
analyzed here. 
If the line profile can be divided into a narrow and a broad component, and the
interpretation of the broad component is microflaring, then the ``quiet'' atmospheric
structures should be considered apart from the microflaring effects.
On the other hand, if the entire line profile can instead be described by a normal
distribution of line widths \citep[as implied by][]{wood1996} then the entire line flux should be used,
since the emission is integrated over the disk of the star.  
We note that apart from \ion{Si}{3} $\lambda$1206, the narrow components of the
two Gaussian fits are consistent with the thermal line widths.

\begin{figure}
\begin{center}
\includegraphics[scale=0.4,angle=90]{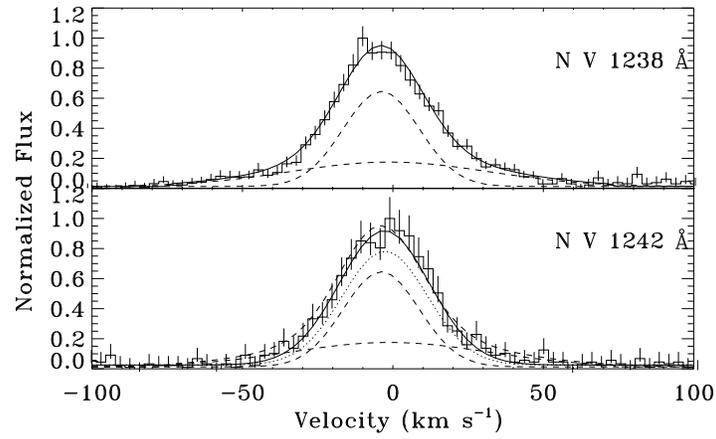}
\caption{Comparison of $\lambda$1238 and $\lambda$1242 \AA\ transitions of \ion{N}{5}.
The Gaussian line profiles are shown (dotted lines), as well as the result after convolving with
the line spread function (solid line).
The $\lambda$1238 line has higher S/N, being a factor of 2 brighter, and shows evidence for a 
broad second Gaussian.  For comparison we overplot the single Gaussian fit from the $\lambda$1242
line to illustrate the excess flux in the wings of this profile. \label{fig:n5compare}}
\end{center}
\end{figure}

Both \ion{O}{6} lines in the FUSE spectra are clearly skewed with the emission peak shifted
shortward of the line centroid and with emission extending further in the longward wing than
for the shortward wing.  A single Gaussian fits the bulk of the 1032 \AA\ line profile
reasonably and gives a line flux similar to that from direct integration with a line width
(FWHM) of 0.16 \AA, or 46 km s$^{-1}$ width.  
One of the highest temperature transitions in the STIS spectra, \ion{N}{5}, 
has one line profile ($\lambda$1238),
statistically better fit by two Gaussians, while the other line profile ($\lambda$1242)
is statistically better fit by one Gaussian.  Lines from a resonance doublet are formed in
a 2:1 ratio with same intrinsic line profile, so it is odd that only one \ion{N}{5} transition
shows evidence for two Gaussians.
We compare the line profiles in Figure~\ref{fig:n5compare}
to examine the reality of the second Gaussian in the $\lambda$1238 line.  Since this line is
brighter than its sister transition by a factor of two, the absence of evidence for a second Gaussian
in the wings of the $\lambda$1242 line may simply
reflect the poorer S/N; the two Gaussians which fit the $\lambda$1238 line are compatible with
the flux-scaled $\lambda$ 1242 line profile.  

\citet{wood1996} noted that
the discrepancy between the widths of the \ion{O}{4}$]$ intersystem line, a
low opacity transition, and \ion{N}{5}, which is a strong scattering line,
in the active star HR~1099
suggests that these lines are not originating entirely from the same regions,
despite the fact that the temperatures at which the lines are formed are nearly identical.
In contrast, the widths of narrow component features in EV~Lac are similar to those deduced from single-Gaussian
fits to other transitions sharing the same formation temperature, indicating that
the two are likely formed in the same regions and not subject to any strong scattering effects.

In the solar transition region, \citet{jordan1991} 
showed that
the trend of nonthermal velocities with electron temperature increased as 
$\xi_{\rm turb} \propto T_{e}^{1/4}$.
\citet{deremason1993} determined that average nonthermal velocities in quiet regions
of the Sun's atmosphere peak at $\sim$27 km s$^{-1}$ at temperatures of 10$^{5}$K,
and increase up to $\log T\sim$5.5.
\citet{linskywood1994} found broad and narrow components for the \ion{Si}{4} and \ion{C}{4}
transitions of the dMe flare star, AU~Mic; the narrow components were characterized by 
nonthermal velocities of 15 km s$^{-1}$ while the broad components 
averaged 97 km s$^{-1}$.  
In contrast, it appears that the trend of turbulent velocities in EV Lac's quiescent chromosphere--lower 
transition region, for lines with single Gaussian fits or the narrow component
of lines fit by two Gaussians,
{\em decreases} with increasing temperature above $\log T\sim$4.8.  
These values appear to be
subsonic, compared with typical nonthermal values inferred from the broad components of
two Gaussians, which are mostly supersonic.

%We also considered that transitions are not formed at a single temperature, but 
%have a range (usually T$_{\rm max} \pm$0.15), and the shape of the underlying distribution
%of plasma with temperature can skew the temperature which most of the plasma experiences.
%Based on the results in \S 5.2.3, we synthesized line profiles by using the DEM,
%calculating the thermal width at each temperature, and summing up these contributions.
%This is not a huge effect.

We explored the run of nonthermal velocities with temperature in the high S$/$N spectrum
of another active M dwarf, AD~Leo, during quiescent periods, to determine
if the pattern seen on EV~Lac is consistent with other active M dwarfs.  
This data is described
in more detail in \citet{hawleyetal2003}.  
The quiescent STIS spectrum of AD~Leo was obtained using the same setup as for EV~Lac,
and also suffers from the 12 km s$^{-1}$ Doppler correction error.  The exposure time
for AD~Leo's quiescent spectrum is 52963 s.  Line profile parameters were measured in the same
way; i.e. one or two Gaussian line profiles were fit, and the line widths were interpreted
as the sum in quadrature of a thermal width, nonthermal width, and instrumental 12 km s$^{-1}$
width.  The right hand panel of Figure~\ref{fig:vturb} displays the trend of nonthermal
velocities for AD~Leo.  The same lines, \ion{Si}{3} $\lambda$1206, \ion{Si}{4} $\lambda\lambda$
1393,1402, \ion{C}{4} $\lambda\lambda$1548,1550, and \ion{N}{5} $\lambda$1238,1242 display
evidence for two Gaussian profiles and show the same trend of broad component
and narrow component velocities with temperature as EV~Lac.  

\section{Electron Densities}
We investigated the electron densities in the outer atmosphere of EV~Lac by making use
of density-sensitive line ratios in the X-ray and ultraviolet.  The X-ray portion of the
spectrum has numerous density-sensitive ratios in the forbidden to intercombination
transitions of helium-like ions; the ions \ion{Si}{13}, \ion{Mg}{11},
\ion{Ne}{9}, and \ion{O}{8} are detected in either the MEG or HEG grating and can place 
constraints on n$_{e}$ between 10$^{9}$ --- 10$^{14}$ cm$^{-3}$ at temperatures of
$\sim$2$\times$ 10$^{6}$--3$\times$ 10$^{7}$ K.
The ultraviolet portion of the spectrum also has density sensitive line ratios; for ease of
interpretation, we choose transitions originating from the same ion, to separate density
issues from temperature and abundance issues.  We concentrate on transitions of \ion{O}{4}
and \ion{O}{5}; the ratios with the best signal to noise are \ion{O}{4} $\lambda$ 1401.171/
$\lambda$1399.779, and \ion{O}{5} $\lambda$1218.390/$\lambda$1371.292.  These ratios place constraints
for electron densities in the region n$_{e}$ of 10$^{10}$- 10$^{14}$ cm$^{-3}$ at temperatures of $\sim$ 
10$^{5}$--2.5 $\times$ 10$^{5}$ K.  %Thus the combination of the X-ray and ultraviolet density
%diagnostics can place great constraints on the atmospheric pressures encountered in the 
%outer atmosphere of EV~Lac.

\begin{figure}
\begin{center}
\includegraphics[scale=0.7]{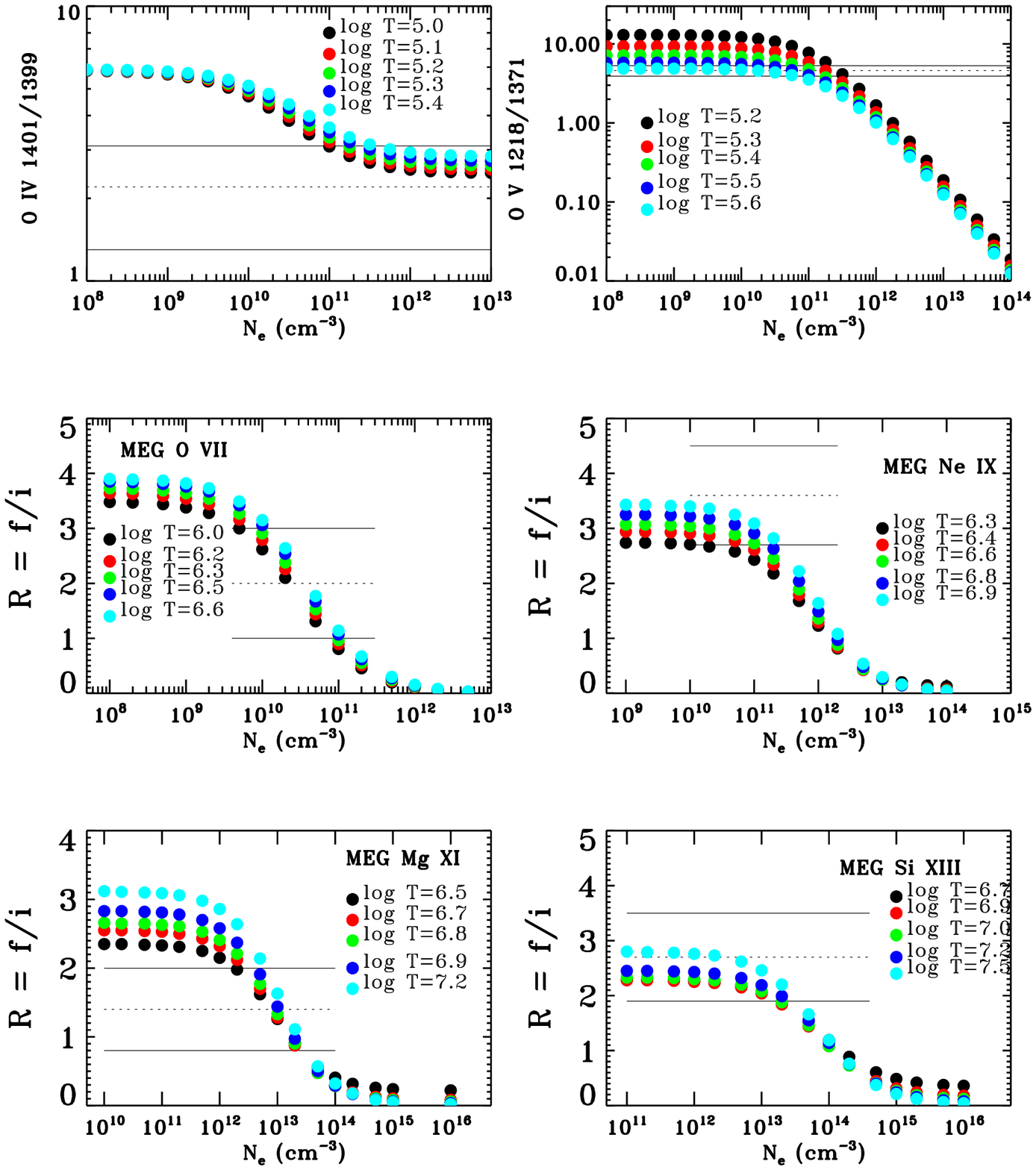}
\caption{Electron densities in the transition region and corona of EV~Lac,
determined from line ratios of Chandra HEG/MEG and HST/STIS emission lines.
Theoretical curves for \ion{Si}{13}, \ion{Mg}{11}, \ion{Ne}{9}, and \ion{O}{7} are taken from 
\citet{porquet2001};
theoretical curves for \ion{O}{5} and \ion{O}{4} are from CHIANTI v4.2 \citep{chiantiv1,chiantiv4}. \label{fig:densities}}
\end{center}
\end{figure}

\begin{figure}
\begin{center}
\includegraphics[scale=0.6]{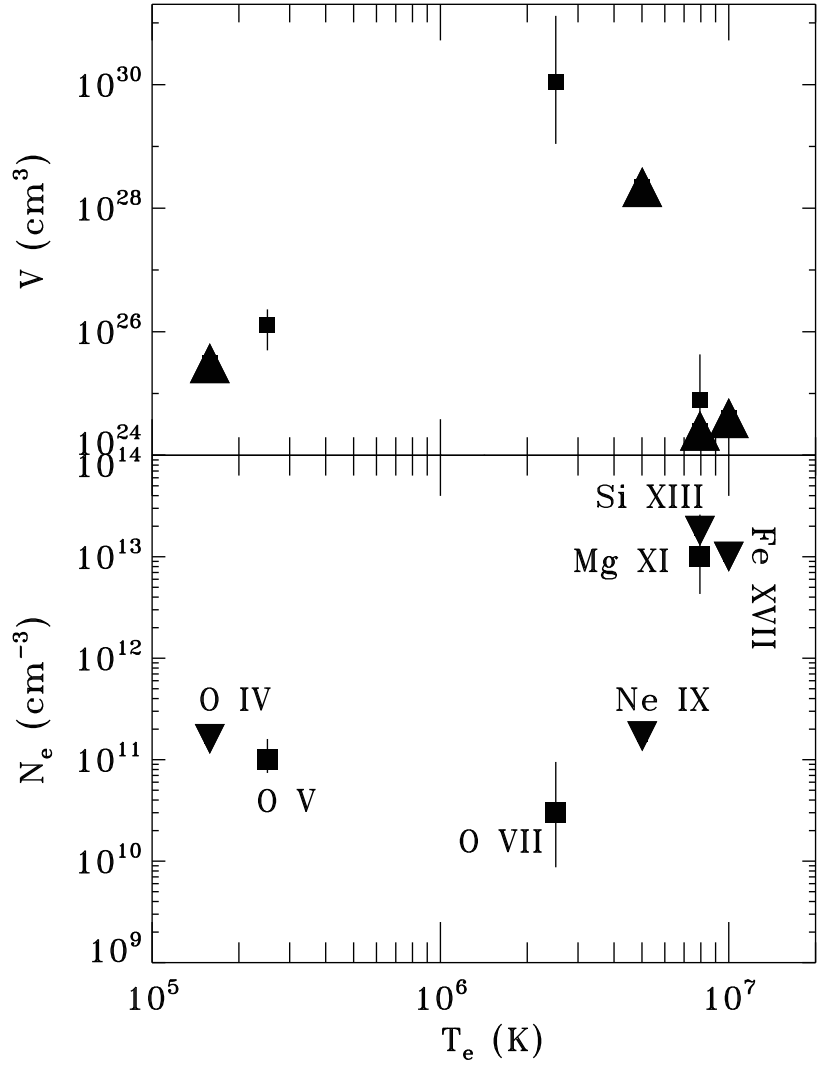}
\caption{{\it (lower)} Dependence of electron density on electron temperature,
using density diagnostics shown in Figure~\ref{fig:densities}.  
{\it (upper)} Volume of plasma contributing to emission in the transition region
and corona, using electron density estimates and emission measures. \label{fig:nete}}
\end{center}
\end{figure}

The theoretical line ratios are shown in Figure~\ref{fig:densities}, along with the observed values
and uncertainties.   For the UV transitions we used theoretical ratios from CHIANTI v4.2
\citep{chiantiv1,chiantiv4}; ratios
for the X-ray transitions come from \citet{porquet2001}.
Table~\ref{tbl:dens} lists the transitions used in the density determinations,
energy flux ratios, temperature at which the ratio was evaluated, and the corresponding
density measurement and 1$\sigma$ uncertainties (or upper limits).
The electron density determinations are plotted in Figure~\ref{fig:nete}
in the bottom panel. 
%This may illustrate a discontinuity between
%the structures producing the bulk of the transition region emission, and those contributing to
%the coronal emission.

The intercombination and forbidden transitions of \ion{Ne}{9} and \ion{Mg}{11} are blended,
with \ion{Fe}{21} and \ion{Fe}{19} in the case of Ne~IX and with the
Lyman series limit of \ion{Ne}{10} for \ion{Mg}{11}.
\citet{ness2004} and \citet{testa2004} surveyed electron density diagnostics in X-ray
spectra of a sample of active stars, including EV~Lac.  They used the entire accumulated
observation for spectral analysis, whereas we have excluded
large flaring events.  The \ion{O}{7} and \ion{Ne}{9} estimates
in Table~\ref{tbl:dens} are slightly lower than those of \citet{ness2004}, 
while the \ion{Mg}{11} density estimate in Table~\ref{tbl:dens} is slightly higher
than the values given in \citet{testa2004}.  We have not made explicit correction for the blending
in the Ne or Mg lines, and our use of a subset of the observation makes our resulting
error bars larger, but nonetheless there is good agreement within the error margins.  

\section{Differential Emission Measure Distribution and Abundances}
\begin{figure}
\begin{center}
\includegraphics[scale=0.5,angle=90]{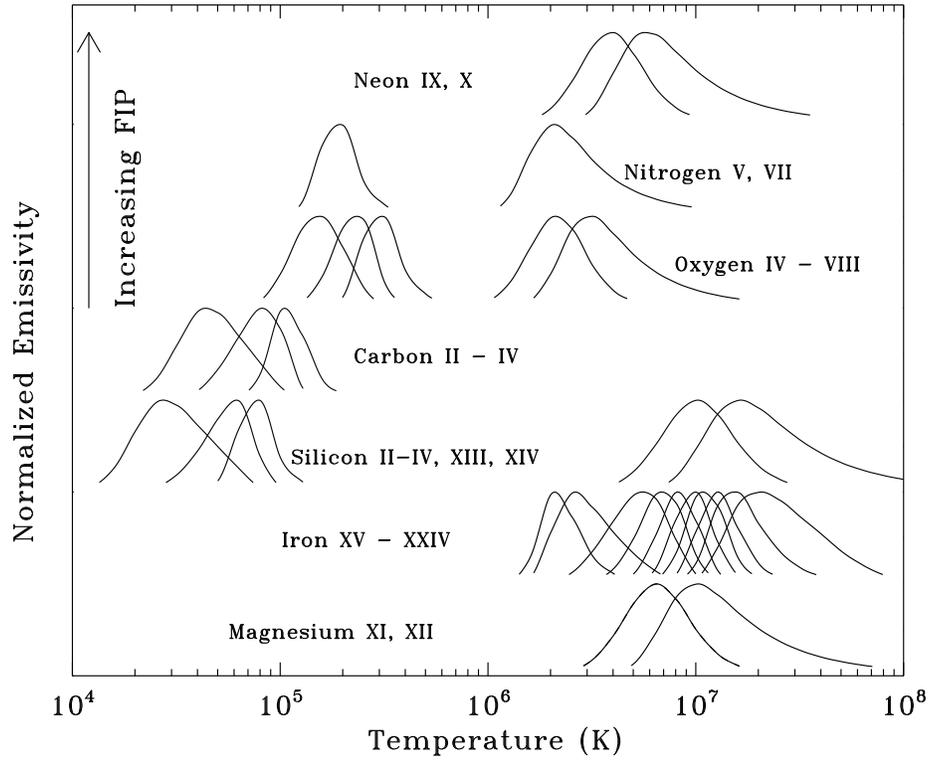}
\caption{Temperature coverage of emission lines detected in STIS, FUSE,
EUVE, and HETGS spectra.  Elements are ordered in increasing First Ionization
Potential (FIP).  Individual emissivities have been normalized to illustrate
their temperature coverage.\label{fig:tempsens}}
\end{center}
\end{figure}

The UV and X-ray line fluxes, along with the X-ray continuum, were used to constrain the shape of the 
differential emission measure distribution.  This can be done using transitions for which the following 
assumptions are met: \\
\begin{enumerate}
\item The transition is formed in coronal equilibrium, i.e. a balance between collisional excitation
and radiative de-excitation.
\item The transition is optically thin.
\item The transition is insensitive to changes in electron density.
\item The transition is detected with sufficient sensitivity in the spectrum, and is unblended
with transitions from other elements or ionization stages.
\end{enumerate}
In practice, many of our observed transitions from the upper chromosphere to the corona 
meet these criteria, and span three
orders of magnitude in temperature.  Figure~\ref{fig:tempsens} displays the temperature coverage of transitions in the
STIS, FUSE, EUVE, and HETG spectra which can be used to constrain the DEM.  Note that there is a gap in temperature
coverage from 4$\times$10$^{5}K<$T$<$10$^{6}$K.
In \S 5.1 we describe how we use the EUV and X-ray transitions (and the X-ray continuum)
to compute the coronal DEM and abundances. 
Then in \S 5.2 we extend the temperature coverage to the transition region
and upper chromosphere by using lines found in the UV and FUV.
%The coronal DEM and abundances 
%derived from EUV and X-ray transitions
%(and the X-ray continuum) provided an initial guess, and the temperature coverage was extended to
%lower values to cover the transition region and upper chromospheric lines found in the UV and FUV.
There are
a few coronal transitions at these lower wavelengths (\ion{Fe}{21} $\lambda$1354 and \ion{Fe}{18} $\lambda$978),
but since the DEM at $\log T \geq$ 6.2 is already constrained, we use these as consistency checks between the flux
calibration and possible evidence of stellar variability (see \S~\ref{section:var}).  

We constrained the choice of detected iron lines in the X-ray region
to those which do not show evidence of density sensitivity; 
Smith et al. (2002) 
\footnote{Calculations can be found at 
http://cxc.harvard.edu/atomdb/features\_density.html}
compiled a list of density-sensitive X-ray lines
between 1.2 and 31 \AA, and marked those as density-sensitive which had standard deviations of 
greater than 10\% of the mean emissivity, calculated at temperature grid points of $\log T(K)=$6,6.5,7,7.5
and densities between 10$^{5}$ and 10$^{15}$ cm$^{-3}$.

\subsection{Inverting the Chandra and EUVE Spectra: Coronal DEM and Abundance}
\begin{figure}
\begin{center}
\includegraphics[scale=0.4]{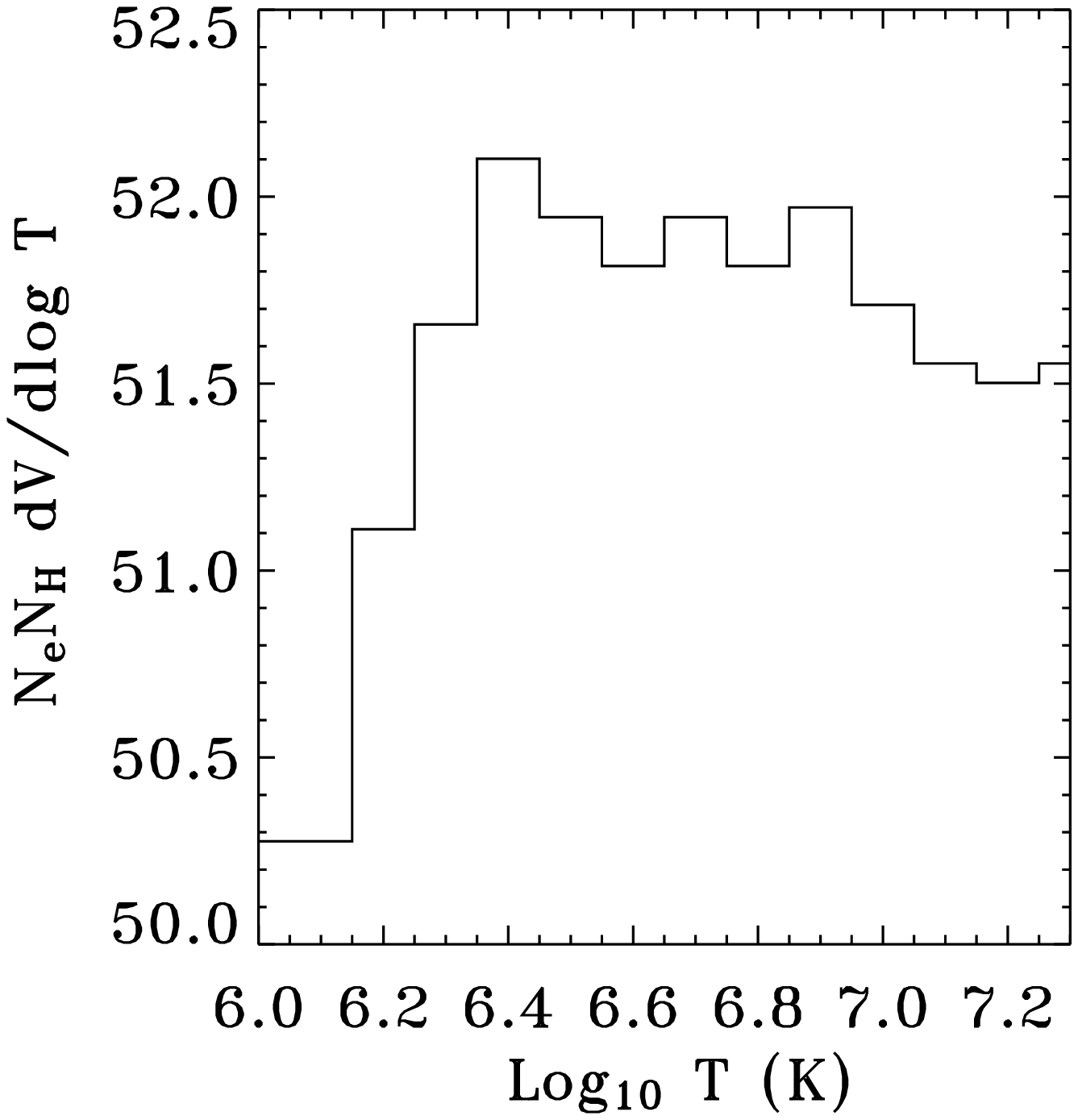}
\includegraphics[scale=0.44]{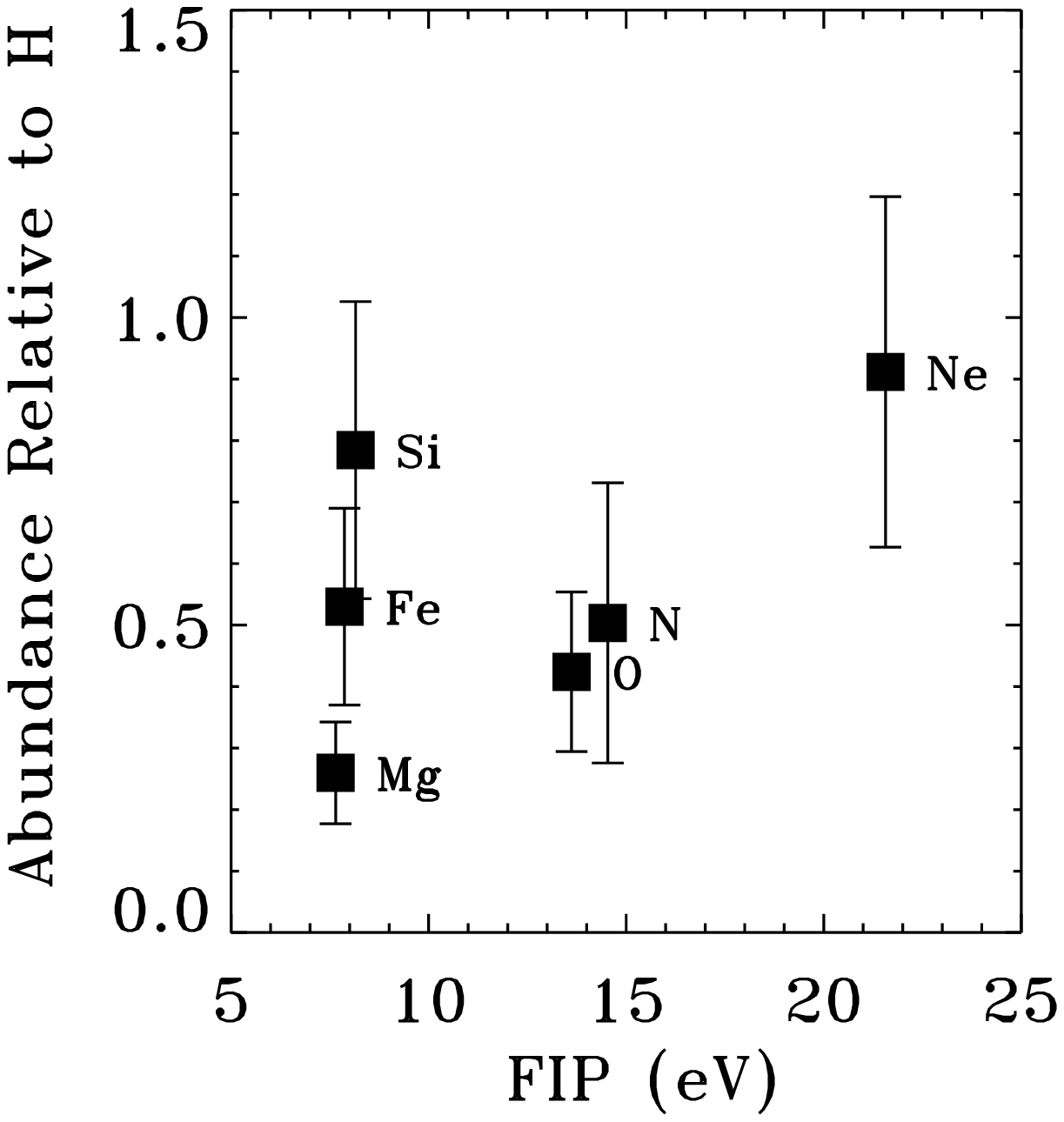}
\caption{{\it (left)} Differential emission measure (DEM) distribution derived
from the EUVE and Chandra spectra. {\it (right)} Abundances derived from Chandra spectra.
 \label{fig:chandradem}}
\end{center}
\end{figure}

%It is also important to exclude the use of density sensitive emission lines.
%in general, abundance analyses of such rapidly
%rotating active stars is difficult to accomplish, and we take the solar photospheric abundances
%of Grevesse and Sauval as a proxy for the likely photospheric abundances of EV~Lac.  
We first examined the Chandra and EUVE spectra, and constrained the DEM in the temperature
range $\sim$ 10$^{6}$--3$\times$ 10$^{7}$ 
\citep[using the self-consistent treatment of line and continuum emission described in][]{osten2003}.
We define the DEM as:\\
\begin{equation}
\phi (T)= \frac{n_{e}n_{H}dV}{d\log T}
\end{equation}
so that the observed line flux, $f$ of an element $i$'s ionic transition 
$\lambda$, can be expressed as $f_{\lambda,i}\propto A_{i} \int \phi(T) P_{\lambda,i}(T)d\log T$, 
where $A_{i}$ is the abundance of element $i$, $P_{\lambda,i}$ is the emissivity of the transition,
and $\phi(T)$ is the DEM.
Once a guess at the DEM was made, using only iron lines present in the
X-ray and EUV spectra, the abundances of N, O, Ne, Mg, and Si relative to Fe were determined,
and the abundance of iron with respect to hydrogen was determined
by using the shape and level of continuum emission in the X-ray to constrain the Fe/H ratio.
Errors on the abundances relative to iron reflect the signal to noise statistics of the 
emission lines, while the error on the iron to hydrogen ratio was calculated by 
Monte Carlo simulations, perturbing the continuum flux values a random fraction of the noise level
and recalculating the Fe/H ratio -- the uncertainty in the Fe/H ratio is the 1 $\sigma$ value
of the resulting
distribution of Fe/H values.
Abundances of other elements with respect to hydrogen were then calculated, propagating
both sources of error.
The DEM at coronal temperatures and derived coronal abundance pattern are plotted 
in Figure~\ref{fig:chandradem}, and Table~\ref{tbl:abund} lists the derived metal abundances.

\subsection{DEM and Abundances from Chromosphere through Corona}

\subsubsection{DEM}
The DEM was then solved for iteratively, using all lines detected in the EUV and X-ray
spectra, along with transitions determined from UV and FUV spectra (which include a few
coronal transitions).  The emissivities were calculated in the low density limit, using 
APEC \citep{apecref},
and any density-sensitive lines were excluded from this portion of the analysis.  We attempted
to minimize the ratio f$_{obs}$/f$_{pred}$ for all transitions used.  
%The APEC code does 
%incorporate the CHIANTI database v2.0 \citep{chiantiv1,chiantiv2} 
%to include ions formed at lower temperatures; however we discovered improvements
%between the current version of CHIANTI v4.2 \citep{chiantiv1,chiantiv4} and v2.0, which results from improvements/refinements
%in the atomic data being used for UV transitions, so we used the contribution functions from CHIANTI
%v4.2 where the atomic data was more up to date.
The APEC code incorporates atomic data from the CHIANTI project; we used a pre-release update
 which incorporates the most recent version of CHIANTI v4.2 \citep{chiantiv1,chiantiv4}.
As discussed in \citet{delzanna2002} and \S 5.3, there appear to be discrepancies for
Li- and Na- isoelectronic sequences (\ion{N}{5}, \ion{O}{6}, \ion{C}{4}, \ion{Si}{4})
so we relied more heavily on other transitions.
We computed the effective temperature of each transition, taking into account the effect 
that the shape of the DEM has in skewing the dominant plasma which the transition experiences.
The effective temperature is computed as \\
\begin{equation}
\log T_{\rm eff} = \frac{\int P_{\lambda}(T_{e})\phi(T_{e}) \log T_{e} dT_{e}}{\int P_{\lambda}(T_{e}) \phi(T_{e}) dT_{e}}
\end{equation}
where $P_{\lambda}(T)$ is the plasma emissivity and $\phi(T)$ is the DEM.
The volume emission measure (VEM) was estimated by integrating the DEM in each $\Delta \log T=$0.1 bin,
and the volume occupied by the plasma was estimated at temperatures where electron densities could be
constrained (\S 4) -- the resulting volume estimates ($V\propto VEM/n_{e}^{2}$) are plotted in Figure~\ref{fig:nete}.

\subsubsection{Abundances}
Our measurements of coronal abundances in EV~Lac agree well with those determined from XMM-Newton
observations by \citet{rs2005}.  The X-ray luminosity determined from
their RGS spectrum was 4.26$\times$10$^{28}$ erg s$^{-1}$ (5--35 \AA), which is higher than our
X-ray luminosity of 1.8$\times$10$^{28}$ erg s$^{-1}$ (1.8--26 \AA).  
The XMM-Newton observation also included several flares, and the RGS spectrum appears to 
include both flaring and quiescent time intervals, which would explain the higher luminosity.
Despite the $\sim$
factor of two change in luminosity, the overall characteristics of the spectra (dominant
temperatures, abundances) appear similar.

In general there is overlap between elements whose ionic transitions appear in the coronal spectra
and those ionic transitions appearing in the UV/FUV spectra; Figure~\ref{fig:tempsens} illustrates the
temperature coverage and element distribution of the lines in the present analysis.  
The element
carbon is the only element which we cannot constrain using our coronal spectrum, since
the helium- and hydrogen-like transitions occur at $\lambda >$30 \AA.
It is not known whether the elemental abundance behavior found in the corona continues to
lower temperatures. The photospheric abundances of EV~Lac are not severely depleted 
\citep{pmsu1,gizis1997}
but could be mildly depleted.
We examined flux ratios of two sets of oxygen and nitrogen lines to investigate the 
abundance ratios in the corona and transition region.  Tables~\ref{tbl:stislines} and ~\ref{tbl:xraylines}
indicate that \ion{O}{5} and \ion{N}{5} have nearly the same T$_{\rm eff}$, and
\ion{O}{7} and \ion{N}{7} have identical values of T$_{\rm eff}$.  By combining the observed
flux ratios with the DEM-weighted emissivity ratios, a constraint on the abundance ratio in the
corona and transition region can be made: assuming that the error in observational flux
measurement dominates, we find:\\
\begin{eqnarray}
\left( \frac{A_{O}} {A_{N}} \right)_{TR}= \frac{f_{O~V }} {f_{N~V}} 
\left( \frac{\int P_{N~V}(T) \phi(T) d\log T}{\int P_{O~V}(T) \phi(T) d\log T} \right) \\
\left( \frac{A_{O}}{A_{N}} \right)_{C} = \frac{f_{O~VII}}{f_{N~VII}}
\left( \frac{\int P_{N~VII}(T) \phi(T) d\log T}{\int P_{O~VII}(T) \phi(T) d\log T} \right) 
\end{eqnarray}
where e.g. $P_{\rm O~V}(T)$ is the emissivity of the \ion{O}{5} transition, 
and e.g. $f_{O~V}$ is the observed flux of the transition of \ion{O}{5}.
Combining the line fluxes and emissivities, we determine 
$(A_{O}/A_{N})_{TR}=$0.2$\pm$0.03 and  $(A_{O}/A_{N})_C=$0.6$\pm$0.2. 
These ratios are with respect to the solar photospheric abundances of \citet{gs1998,gs1999}.
This implies
an enhancement of the coronal abundances by about a factor of three compared to the transition
region.  
Nitrogen and oxygen 
have nearly the same FIP (14.5, 13.6 eV for N, O respectively) and do not display any 
differential behavior in the Solar corona compared to the lower atmosphere 
\citep{feldmanlaming2000}, so we would not expect to see any large difference in the ratio
in EV~Lac if it behaves like the Sun.  
There is a systematic effect which appears to affect transitions
of Li- and Na-like isosequences, however (see \S5.3 and Table~\ref{tbl:ncbc}), in which the observed flux for \ion{N}{5} 
is 2--5 times the flux predicted using the DEM and emissivities,
depending on which abundance patterns are assumed to occur in the transition region, as well as
the role of dynamics in line formation.  Since the origin of this effect is unexplained,
and is of the same magnitude as the abundance ratio (but working in the opposite direction), it is
not possible to determine for certain whether abundance gradients exist in the outer atmosphere
of EV~Lac.
We considered two alternatives regarding the elemental abundance behavior in the outer
atmosphere of EV~Lac.

The first possibility was 
that the same abundances which exist in the corona also
exist in the chromosphere (and possibly into the photosphere as well).
To model this, we took the coronal elemental abundances, and applied that to the lower
temperature transitions in computing predicted fluxes from the DEM.  
For C, we found it necessary 
to reduce the abundance to force the observed fluxes to agree with those predicted
from the DEM using other elements. 
%We determined the value after iterations with the DEM, and found 
The carbon abundance
value which minimized the $\chi^{2}$ statistic between observed fluxes and those predicted from the
DEM %, with the C abundance as a free parameter.
using all transitions of carbon in the STIS and FUSE spectra, was 0.45 Solar.
%carbon abundance relative to the photospheric carbon abundance was 0.45. 
Excluding the \ion{C}{4} transitions, which appear to have some undetermined discrepancies
(see \S 5.3),
increases the abundance depletion to approximately 0.35 times the solar photospheric value.

The second possibility is that EV~Lac is similar to the Sun. 
%As the density trend displays a rather marked discontinuity between
%transition region and coronal temperatures (see \S 4) it is entirely possible
%that different emitting structures are responsible for the bulk of the transition region
%and coronal plasma.  There is then no reason for these structures to have the same abundances
%as found higher in the atmosphere.  
Chemical fractionation in the solar atmosphere can be found at temperatures above 2.5$\times$10$^{4}$K
\citep{feldmanlaming2000}, and there is no reason to expect that the outer atmosphere of EV~Lac
has the same abundance pattern at all temperatures.
In order to accommodate this, we introduced a hybrid abundance
pattern, using our coronal abundances above 10$^{6}$K, and 
the solar photospheric abundances of \citet{gs1998,gs1999} below 10$^{6}$K. 
We note that the \citet{delzanna2002} investigation of AU Mic only used coronal iron line diagnostics,
so they were not able to determine if coronal abundance anomalies
existed, or explore the possibility of an abundance gradient between coronal and photospheric
abundances.  
%They used the solar photospheric abundances of \citet{gs1998,gs1999}, with oxygen
%corrected to the value in {\bf Grevesse (2002 Adv. Space Research)  }.
%The corona of EV~Lac is depleted of metals in comparison with the solar corona.  A
%comparison with EV~Lac's photospheric composition is not possible because of
%the lack of reliable abundance determinations.  
%Another active dMe flare star, AD~Leo,
%has been shown to have photospheric metal depletion \citep{jones1996} 
%although this result could be due to chromospheric activity effects 
%filling in the line absorption rather than metal depletion (M. Giampapa, private communication).  
%We have 
%explored the possibility of the existence of an abundance gradient in the atmosphere
%of EV~Lac through our DEM modelling. 

\begin{figure}
\begin{center}
\includegraphics[scale=0.6]{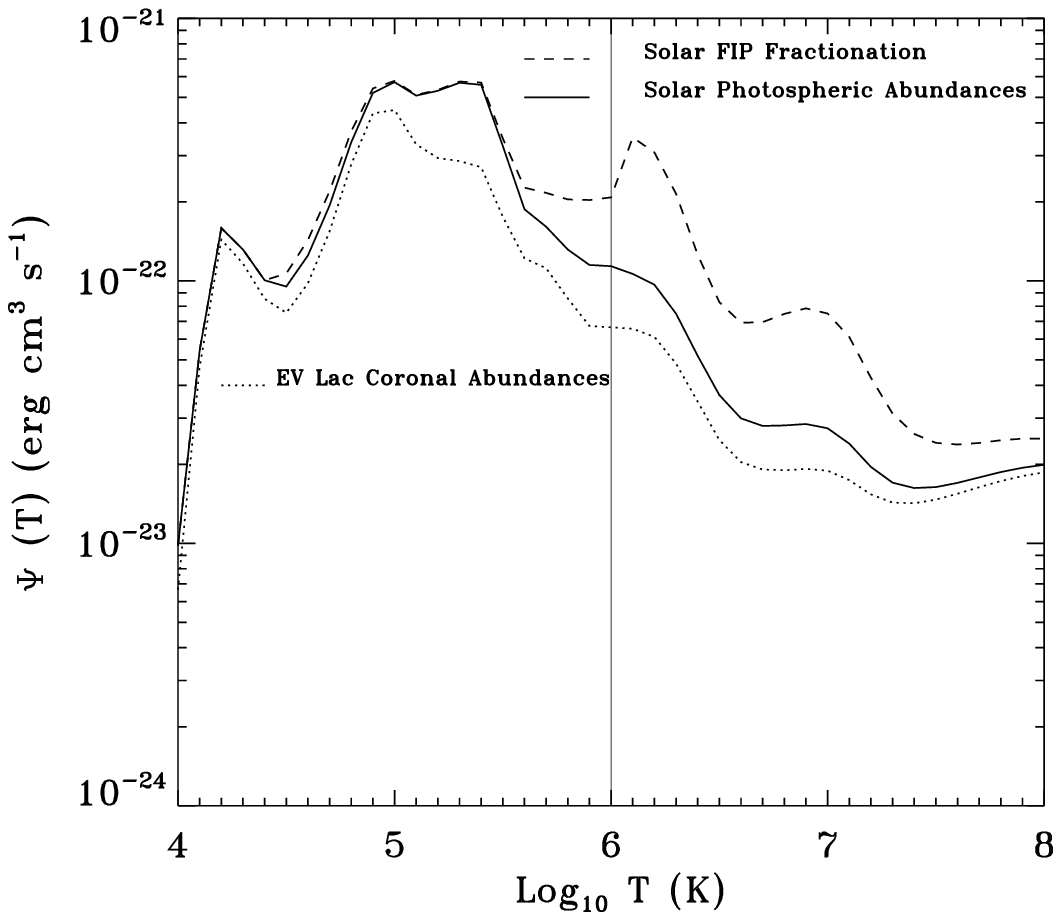}
\includegraphics[scale=0.6]{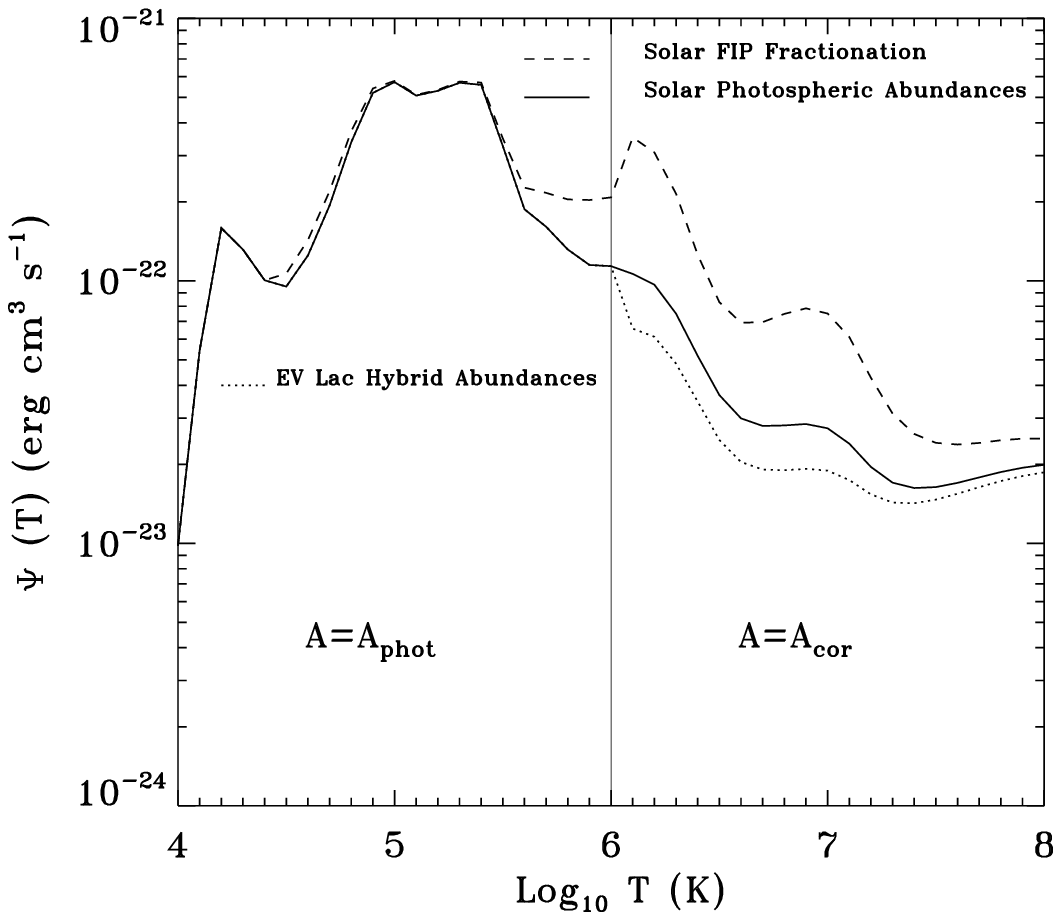}
\caption{{\it (left)} Radiative loss function computed using atomic database
from APEC v.1.3.1, with EV Lac's coronal abundance used at all temperatures.
{\it (right)} Radiative loss function computed with a hybrid abundance pattern,
consisting of EV Lac's coronal abundance at temperatures $>$1MK, and
the solar photospheric abundances at lower temperatures.  Insets to both
figures illustrate the stellar coronal abundance pattern with FIP. \label{fig:radloss}}
\end{center}
\end{figure}

The observed flux
of an emission line is proportional to both the elemental abundance and the differential emission 
measure.  It is well known in the solar outer atmosphere that an abundance fractionation mechanism
exists, which depends upon the first ionization potential (FIP) of the element: those with a 
FIP less than 10 eV are preferentially enhanced in the corona compared with those having a FIP greater
than 10 eV \citep[see][for a review]{feldmanlaming2000}.  The interplay of the abundances and 
DEM are important because both tie into
estimating the radiative losses of the upper atmospheric material.  
Figure~\ref{fig:radloss}
%computes the radiative loss function $\psi(T_{e})$ for the two abundance 
%scenarios, and 
shows the estimates of radiative losses in EV~Lac's atmosphere using the
two abundance scenarios
with the radiative losses from solar photospheric abundances, and the solar
abundance fractionation described in \citet{feldmanlaming2000}.  
The radiative loss function has been calculated here from the APEC database, and correcting the 
emissivities in those databases to the abundances of \citet{gs1998,gs1999}.  At a
given temperature, the emissivity from lines and continuum are summed, then multiplied by the appropriate
atmospheric abundance being assumed.  

\subsubsection{DEM and Abundances}
\begin{figure}
\begin{center}
\includegraphics[scale=0.5]{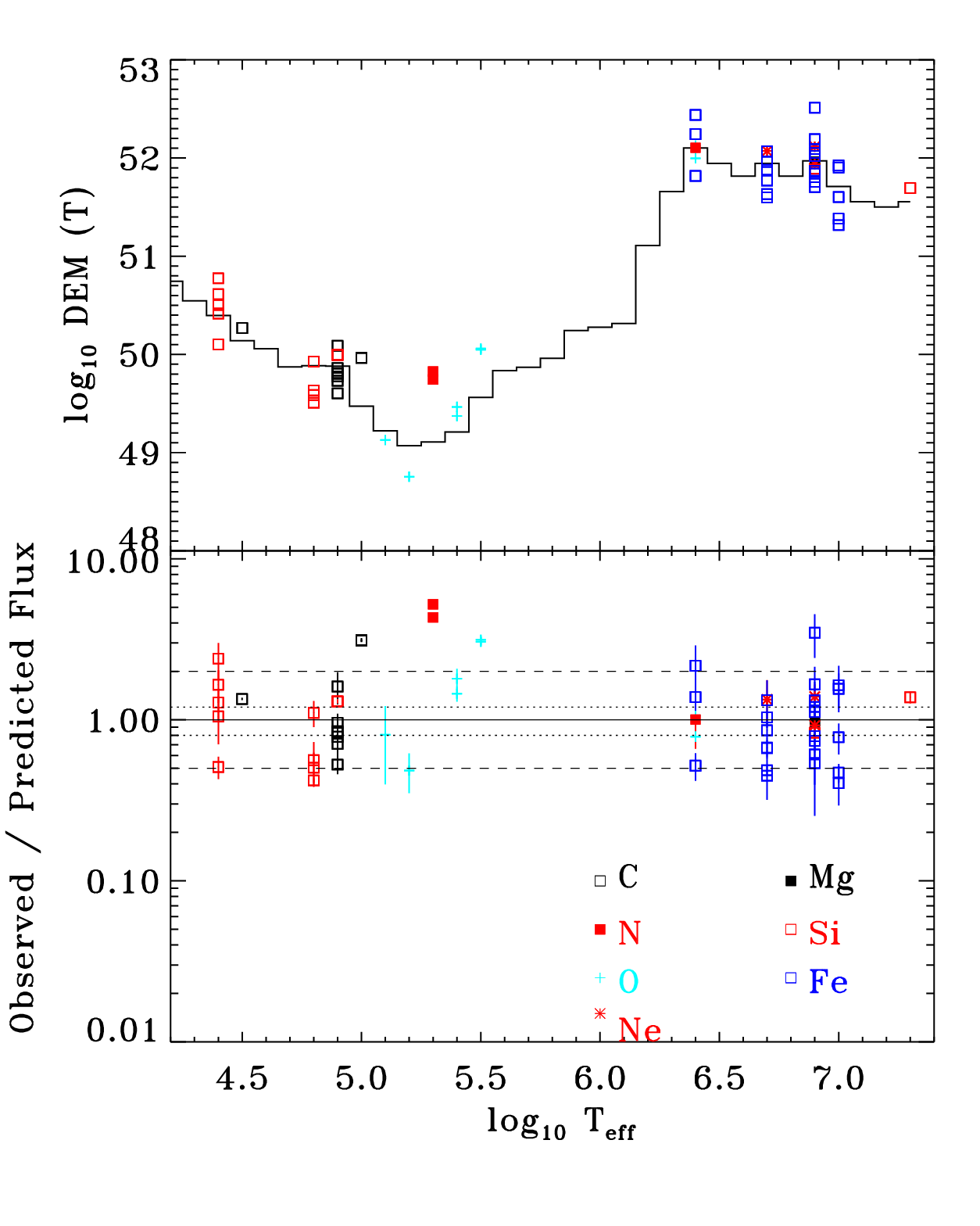}
\includegraphics[scale=0.5]{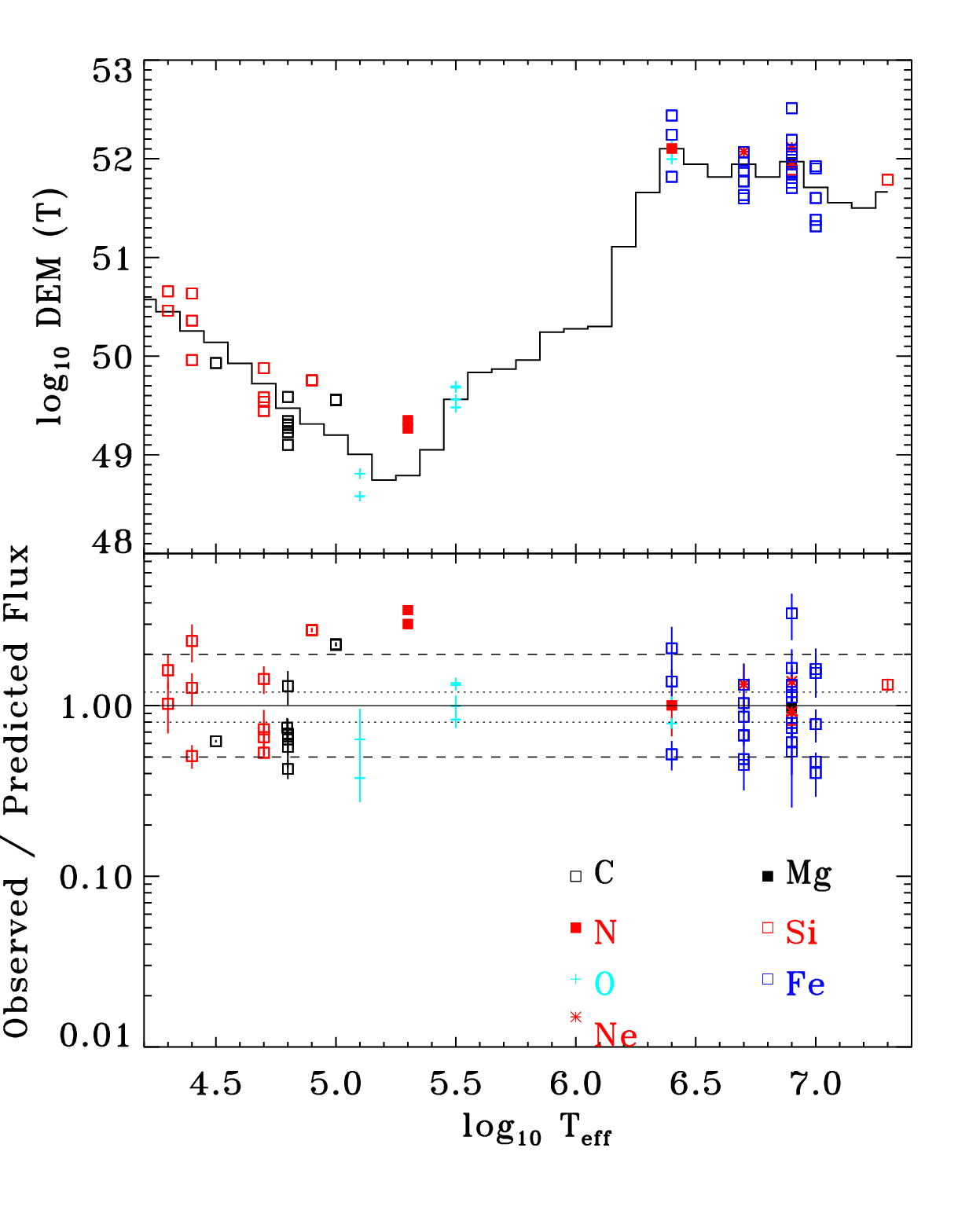}
\caption{{\it (top left)} DEM reconstructed from STIS, FUSE, EUVE, and Chandra spectra
using abundances derived from coronal spectra.  Each point represents the transitions used
in the DEM analysis: the points are plotted at the effective temperature, and the ordinate
is the DEM at that temperature, multiplied by $f_{\rm obs}/f_{\rm pred}$.  Legend is in the lower panel.
{\it (bottom left)}  Agreement between
observed and predicted fluxes, plotted against effective temperature.  
Legend indicates which elements are being plotted. \label{fig:coronaldem}
{\it (top right)} DEM reconstructed from STIS, FUSE, EUVE, and Chandra spectra
using a hybrid abundance:  coronal abundance pattern at T$>$1MK, and
solar photospheric for T$<$1MK. Symbols and legend are as in top left.
{\it (bottom right)} Same as bottom left, for the hybrid abundance case.}
\end{center}
\end{figure}

Figure~\ref{fig:coronaldem} displays the resulting DEM,
and agreement between observed and predicted line fluxes, for transitions used in this analysis.
Overall there is good agreement between the general shapes of the two DEMs: the hybrid abundance case has
a slightly deeper trough at a temperature of $\sim$1.6$\times$10$^{5}$K, which is due to the 
higher abundances.
The temperature range over which the two DEMs differ most significantly is 
5.3$\times$10$^{4}<$T(K)$<$2.7$\times$10$^{5}$, with a maximum difference of less than 
a factor of 4 in a given temperature
bin.  Table~\ref{tbl:ncbc} lists the ratio of observed to predicted flux
for lines in the Li- and Na- isoelectronic sequences 
(\ion{N}{5}, \ion{O}{6}, \ion{C}{4}, \ion{Si}{4}).
For both abundance cases, the fluxes 
%Li- and Na- isoelectronic sequences (\ion{N}{5}, \ion{O}{6}, \ion{C}{4}, \ion{Si}{4})
are underestimated, but by differing amounts.
%in the coronal case, the factors are 3.1--5.3, while in the hybrid case, they are 1.3--3.6.
%Table~\ref{tbl:ncbc} summarizes the results of using different abundances.
The magnitude of the coronal fractionation pattern is not severe, 
as the elements are depleted by factors less
than 4.  
Thus we cannot conclude whether there is an indication for EV~Lac's chromosphere
or transition region to be metal-depleted as is found in the corona.
Lacking any other constraints on the atmospheric abundance patterns, we cannot say which model is preferred, and consider
both alternatives in subsequent sections.

\subsection{Discrepancies in the Li- and Na- Isoelectronic Sequences}
In \S 5.2.1 and \S 5.2.3 we mentioned the discrepancy between
lines in the Li- and Na- isosequences and transitions
arising from other isosequences in determining the DEM:  
%Several transition region lines suffer from this
%effect: as pointed out in \S 5.2, lines of \ion{C}{4}, \ion{N}{5}, \ion{Si}{4}, and \ion{O}{6} have a
the predicted flux is systematically less than that observed for the Li- and Na- isosequences.  
In \S3.3 we described how the line profiles of several strong lines, notably
the discrepant lines of \ion{C}{4}, \ion{N}{5}, and \ion{Si}{4}, display evidence for
two Gaussian features in the line profile; the narrow component has been interpreted
as a quiescent line profile, while the broad component is due to micro-flaring.
We tested the effect that selectively using the narrow component flux or
both narrow and broad component fluxes had on producing agreement with
the fluxes predicted from the DEM.
%In combining the line fluxes of the UV transitions to determine the DEM, 
%the entire line flux has been used.  In this section we explore the effect of
%separating the broad and narrow components of the transition region line profiles.
Table~\ref{tbl:ncbc} displays the change in agreement between observed and predicted
flux when only the narrow component integrated line flux is used, for the two abundance cases.
The DEMs used are those solved for in the previous section, so the only parameter being changed 
is the observed flux.
For the lines originating from Li- and Na- isoelectronic sequences
the reduction in flux makes the
agreement better, but not perfect.  
\ion{Si}{3}$\lambda$1206 is included because it shows evidence for two Gaussian line
profiles, but has the opposite behavior in $f_{\rm obs}/f_{\rm pred}$.

%gave an approach that
%involved using both narrow and broad components of the line profile in determining the DEM
%for these transitions, as 
%these are also the transitions which display evidence for two Gaussian 
%features in the line profile.  While this does not resolve the discrepancy, it does raise
%other questions regarding the origin of the discrepancy, which we address here.

This discrepancy has been addressed for several decades in interpreting solar transition
region spectra.
%\citet{judge1995} 
%raised this issue, pointing out that the discrepancy in the solar
%transition region only applied to strong lines above 1000 \AA\ from Li and
%Na isoelectronic sequences.  
%They suggested that a possible explanation is inaccuracies in atomic data
%for these sequences. 
Allowing for density-dependence of
the ionization fraction could remove some, but not all, of the discrepancy. 
\citet{delzanna2002} suggested the origin had to 
do with the temperature corresponding to the peak emission for these ions.
We note that the current calculations being used for the stellar case are the ionization
equilibrium calculations of \citet{mm1998}, done in the low density limit.  Based on solar 
work, we might expect that some of the discrepancy may be removed by considering the density-dependence
of the ionization fractions, particularly
as EV~Lac's quiescent transition region densities are higher than typical
quiet Sun transition region densities 
\citep{solartrden}. This approach is used in a recent work by \citet{simjordan2005} who note that
the high transition region pressures reduce the dielectronic recombination rate by merging
the highly excited states into the continuum; the effect is to move the ionization
balance to lower temperatures.  

These two isosequences have long timescales due to recombination
from He-like and Ne-like (closed shell) ionization stages, and so also could
be more responsive to non-equilibrium effects.  
\citet{judge1995} suggested
%preferred a non-equilibrium explanation,
%due to dynamical effects, 
that diffusion of ions
due to steep temperature gradients in the transition region could explain
the observed behavior.  
\citet{wikstol1998} explored the properties of dynamic plasmas from their
emitted spectra, and concluded that a number of erroneous conclusions could be
drawn by making the assumption of an intrinsically static atmosphere. 
%In their
%simulations, acoustic and MHD waves propagating through a loop (simulation of the effect
%of nano-flare heating) cause large nonthermal widths in all coronal and transition region
%lines.  
Because of the steep temperature gradient in the transition region, these zones
respond dramatically to the passage of waves, with large redshifts and blueshifts due to
compression and relaxation.  
This could be observed as extra broadening in transition region lines averaged over
many wave passages.
%In contrast, the coronal lines show less dynamic response because
%of the smaller temperature gradients in the corona and the larger spatial scales
%over which coronal emission originates. 
%\citet{wikstol1998} conclude 
%that transition region lines may be a better
%indicator of the dynamics and heating of the corona than coronal lines.

%In the EUV and X-ray spectrum, most 
%detected transitions are from Hydrogenic- and helium-like stages, but Na- and Li-like Iron
%is detected (\ion{Fe}{16} and \ion{Fe}{24}).  Neither of these ions displays any discrepancy
%with neighboring ionization stages, or transitions from other elements which overlap in temperature.
%The non-equilibrium explanation in the stellar transition region may be related to the presence of
%turbulent broadening evident in the transition region of EV~Lac, as the turbulent pressure can be up to
%twice that of the gas pressure at 10$^{5}$K (Figure~\ref{fig:ptot}).  
%If the dynamics of transition region emission are responding to a multitude of heating
%events, as the simulations of \citet{wikstol1998} imply, then the lack of agreement between
%a model static atmosphere and a spatially and temporally averaged real dynamic atmosphere is
%not unusual. 

As discussed in \citet{hawleyetal2003} for UV flares in AD~Leo, these transitions
are also the brightest and show the most response to large flares.
Weaker transitions don't show as large a flare response.  
The observed broadening (and discrepancy)
may therefore be related to micro-flaring, and reflect a non-equilibrium situation.  
This could be
addressed by detailed calculations of the dynamical response of a flare-heated atmosphere.
This explanation cannot account for the \ion{Si}{3} transition at 1206 \AA, which also
shows broad components, but belongs to a different iso-sequence 
\citep[similarly for \ion{C}{3}
transitions in the FUSE bandpass, which show broad components in other active dwarfs;][]{redfield2002}.
Nor does the \ion{O}{6} doublet, where no broad
components are found, fit this scenario, although broad components
have been found in \ion{O}{6} transitions in other active dwarf stars \citep{ake2000,redfield2002}.
%and \ion{N}{5} $\lambda$1242, which was statistically fit better by a single
%Gaussian, cannot be explained using the microflaring hypothesis.
%While we have no constraints on
%the presence/absence of turbulent broadening at shorter wavelengths
%due to spectral resolution limitations,
%the apparent decrease of nonthermal velocities in the transition region 
%suggests that this will be
%a small effect at most at higher temperatures.  The spectral resolution required to
%constrain supra-thermal line widths of 
%Li-like \ion{Ne}{8}, \ion{Mg}{10}, and \ion{Si}{12} in 
%the soft X-ray transitions of  $1s^{2}3p ^{2}P_{3/2} - 1s^{2}2s ^{2}S_{1/2}$
%($\sim$88, 58, 41 \AA, respectively)
%exceed $\lambda/\Delta \lambda=$6200, which is 
%out of the reach of current (and future) instrumentation.  

%The effect of a dynamic plasma could also be investigated further by examining the impact on ionzation balance calculations
%of adding a power-law
%distribution of electrons to the Maxwellian temperature distribution, and determining if that would
%affect the temperature at peak ionization fraction.

\section{Discussion: Inferring Atmospheric Properties}
We determined several fundamental quantities --- $n_{e}$, $A_{i}$, and the DEM --- in \S 4 and \S 5
which describe the atmosphere.  Now we turn to computing the atmospheric properties based on 
these measurements.

\subsection{Electron Pressures}
\begin{figure}
\begin{center}
\includegraphics[scale=0.6]{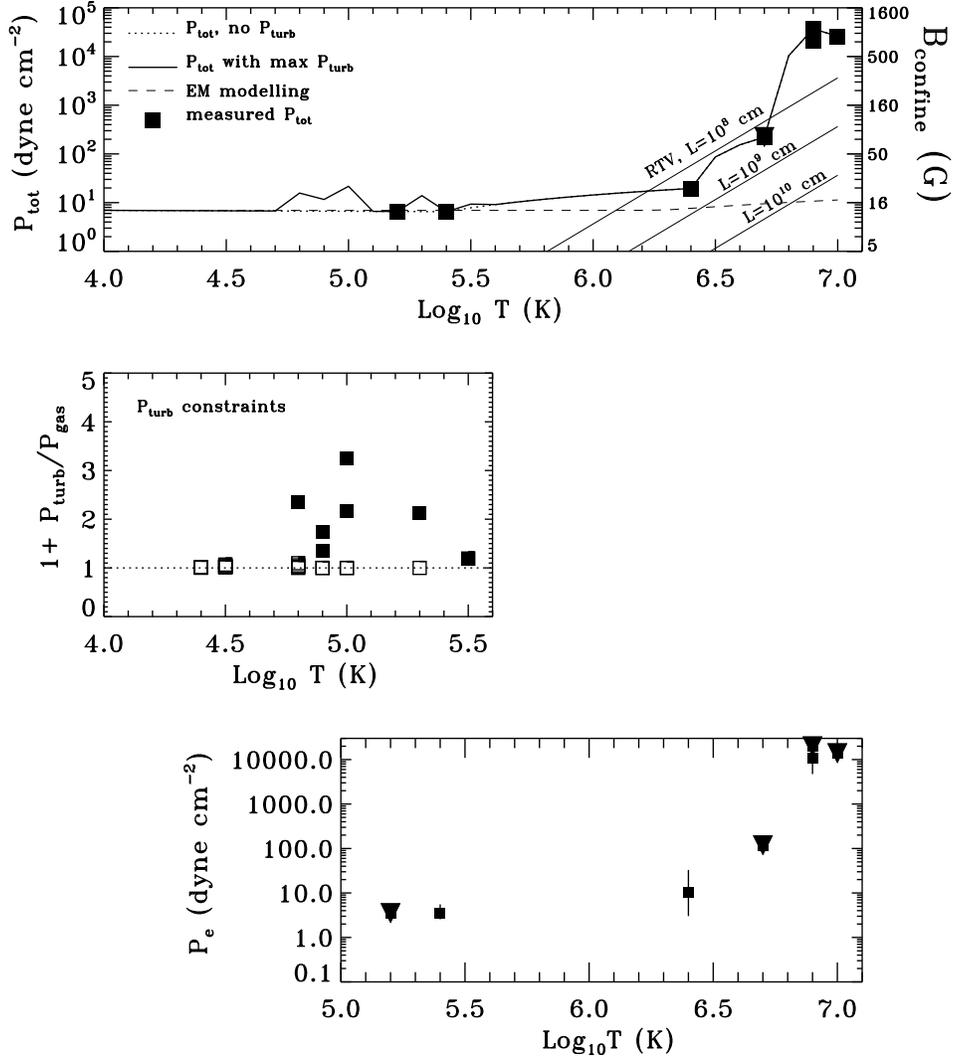}
\caption{{\it (bottom)} Electron pressure constraints versus temperature, for the
densities determined emission line ratios in \S4.
{\it (middle)}Calculation of enhancement over gas pressure afforded by turbulent
pressure, deduced from the broadening of line profiles in the STIS spectrum.
Circles refer to single Gaussian fits to line profiles; open squares refer to 
the narrow component of line profiles fit by two Gaussians; and filled squares refer
to the broad components of such line profiles.
{\it (top)} Total pressure versus temperature, using electron density constraints from 
line ratios and turbulent pressure calculations.  
\label{fig:ptot}}
\end{center}
\end{figure}

The bottom panel of Figure~\ref{fig:ptot} displays the dependence of electron pressure 
on electron temperature calculated from density-sensitive ratios in \S4.
We combine these estimates of the electron pressure with turbulent pressure in estimating
the total pressure at different points in EV~Lac's transition region and corona.
Upper chromospheric and transition region line profiles
show excess broadening, 
interpreted as nonthermal turbulent velocities ($\xi$) and discussed in \S 3.3.
We estimate the 
magnitude of the turbulent pressure
using the following equation: \\
\begin{equation}
Y=\frac{P_{\rm turb}}{P_{\rm gas}} = \frac{\mu m_{p}}{2k_{B}T} \xi^{2},
\end{equation}
where $\mu=1.4X/(1+1.1X)$ is the average particle mass, $X=n_{H}/n_{e}$, P$_{\rm gas}$
is the gas pressure (related to electron pressure P$_{e}$ via P$_{g}=P_{e}(1+1.1X)$ \citep[from][]{jordan1996}).
The middle panel of Figure~\ref{fig:ptot} displays the amount of turbulent pressure
at temperatures where line broadening indicates nonthermal velocities.
The largest $\xi$ at these temperatures is used to compute the maximum amount of
total pressure, including turbulence,
in the upper chromosphere through the corona.
%enhancement for all cases where extra broadening was deduced in the STIS {\bf (and FUSE)} line profiles
%using the widths from the narrow and broad components separately,
%and plots the run of total pressure with temperature. 
%in the atmosphere from the upper chromosphere to the corona,
%by combining the estimates of electron pressure from the STIS and HETG spectra.
The top panel of Figure~\ref{fig:ptot} displays the run of total pressure
with temperature.  This was computed two ways, with no turbulent pressure (gas pressure only)
and with the maximum amount of turbulent
pressure indicated from line widths.
Although the data are sparse, there is
an implied disconnect between the transition region pressures, roughly constant, and the coronal
pressures, which increase sharply with temperature.  

Using equations in \citet{jordan1996} and \citet{griffiths1999}, 
we investigate the run of pressure in the
atmosphere using the emission measure distribution, and a reference pressure,
assuming the transition region and corona are in hydrostatic equilibrium,
and using a spatially uniform plane-parallel approach.
For total pressure (defined as the sum of gas and turbulent pressures)
\begin{equation}
P^{2}_{\rm tot}(T)=P_{\rm tot,ref}^{2} \pm 8.5\times10^{-40} g_{\star} 
\int_{T_{\rm ref}}^{T} CEM0.3 (1+1.1X)^{2} (1+Y) dTe
\end{equation}
where $Y=P_{\rm turb}/P_{\rm gas}$. 
The DEM is converted to column emission measure (CEM; $\int n_{e} n_{H} ds$) via: \\
\begin{equation}
\int n_{e}n_{H} ds = \frac{\int \phi(T) d\log T}{4\pi R_{\star}^2}
\end{equation}
where R$_{\star}$ is the stellar radius, and the
assumption is that the emission is spherically symmetric ($ds$ is a radial path length). 
The CEM is regridded so that it is expressed as CEM0.3, which in the notation of \citet{jordan1996} is \\
\begin{equation}
CEM0.3(\log T) =\int_{\log T-0.15}^{\log T+0.15} n_{e}n_{H} ds
\end{equation}
the $\Delta \log T=$0.3 arises because emission lines are typically formed over a temperature
range of $\Delta \log T=$0.3.
The reference total pressure used is 6.5 dyne cm$^{-2}$, determined at  $\log T=5.4$, 
and consistent with the upper limit at $\log T=$5.2; 
we assume that lower in the atmosphere the pressure is constant (except where P$_{\rm turb}$ 
measurements suggest additional pressure support).
The EM modeling indicates an almost flat $P(T)$ profile which is not consistent with the observed
coronal pressures.

For temperatures above 6$\times$10$^{5}$K, the estimated trend between
electron pressure and temperature using the \citet{rtv1978} scaling laws are displayed,
for loop lengths of 10$^{8}$ -- 10$^{10}$ cm.  
This treatment assumes that the coronal loops are hydrostatic, constant
pressure with maximum temperature at the top of the loop, with the heating scale length
greater than or equal to the loop scale size. 
The high pressures in the corona can only be accommodated by small loops (10$^{8}$ cm or smaller)
and still cannot explain the large jump in pressure at $\log T=$6.9.

The steep increase in total pressure implied by the observed coronal electron densities in
Figure~\ref{fig:ptot} cannot be ascribed to the consequences of hydrostatic equilibrium.
%as the predicted run of total pressure with temperature underpredicts the 
%coronal pressures by orders of magnitude.  
Neither the emission-measure modeling
nor the loop scaling laws are able to reproduce the observed coronal pressures, except
for extremely small loop sizes (10$^{8}$cm, or $\sim$0.004R$_{\star}$) and only in the lower
corona --- the observed electron pressures near 10 MK are too high.
Figure~\ref{fig:ptot} also displays the magnitude of 
the magnetic field strengths required to magnetically confine the plasma:  
at the highest coronal temperatures ($>$ 8MK),
field strengths of up to 1kG are required. 

\subsection{Timescales}
\begin{figure}
\begin{center}
\includegraphics[scale=0.6]{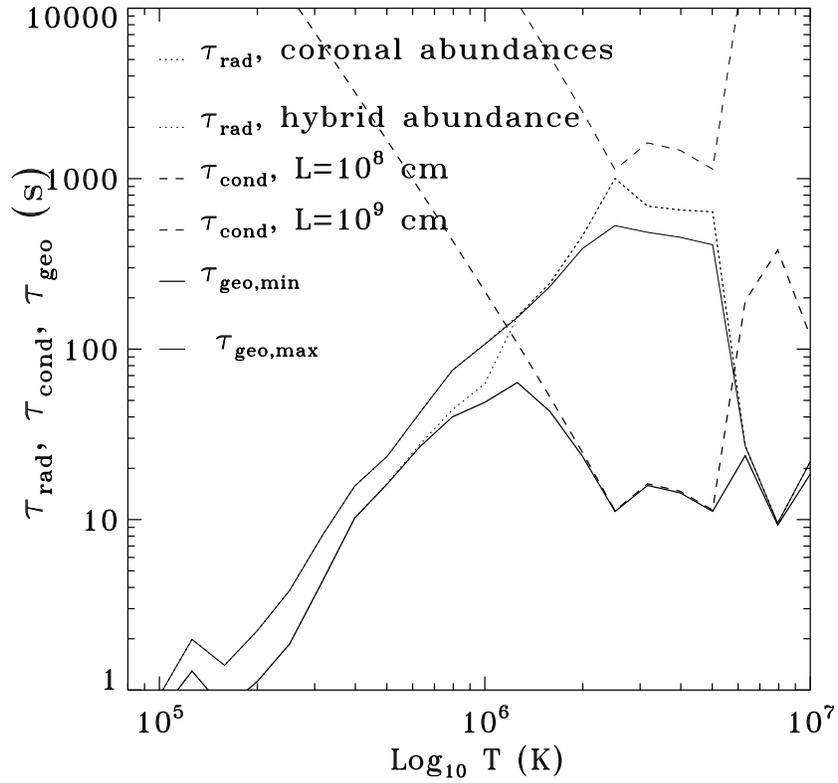}
\caption{Timescales for radiative and conductive losses, calculated using the run of electron density with temperature shown in Figure~\ref{fig:nete}, and the radiative loss functions
shown in Figure~\ref{fig:radloss}.  Conductive loss timescales are estimated using loop lengths
of 10$^{8}$ and 10$^{9}$ cm, respectively.\label{fig:times}}
\end{center}
\end{figure}

Combining the radiative loss function calculated above with the run of electron density 
versus temperature calculated in \S 4, we can calculate
the radiative loss timescale, $\tau_{\rm rad}(T_{e})$, \\
\begin{equation}
  \tau_{\rm rad} (T_{e})= \frac{3(T_{e}) k_{B}T_{e}}{n_{e}(T_{e})\psi(T_{e})}
\end{equation}
where $n_{e}(T_{e})$ is the electron density at a given electron temperature T$_{e}$, $\psi(T_{e})$
is the radiative loss function (units erg cm$^{3}$ s$^{-1}$), and $k_{B}$ is Boltzmann's constant.  We interpolated the observed
values of n$_{e}(T_{e})$ onto a finer grid with $\Delta \log T=$0.1 spacing between $\log T=$5 and 7.  
We assumed the atmosphere at temperatures below 10$^{5}$K was at constant pressure to extrapolate
the radiative loss function to lower temperatures; the density estimates from \ion{O}{4} and 
\ion{O}{5} appear to confirm this.
We also computed the timescale for conductive losses using different estimates for loop lengths,
based on \citet{vdo1988}\\
\begin{equation}
\tau_{\rm cond} = \frac{3n_{e}k_{B}L^{2}}{\kappa T_{e}^{5/2}}
\end{equation}
where $\kappa$ is Spitzer conductivity coefficient 
\citep[=8.8$\times$10$^{-7}$ erg cm$^{-1}$ s$^{-1}$ K$^{-7/2}$,][]{spitzer1962},
$L$ is the loop length scale.  The geometric mean of the two, $\tau_{\rm geo}$, is \\
\begin{equation}
\frac{1}{\tau_{\rm geo}} = \frac{1}{\tau_{\rm rad}} + \frac{1}{\tau_{\rm cond}}
\end{equation}
The dependence of  $\tau_{\rm rad}$ and $\tau_{\rm cond}$
on electron temperature between these temperatures is depicted in Figure~\ref{fig:times}.
The two loop length estimates are based on L of 10$^{8}$ and 10$^{9}$ cm derived from the 
pressure-temperature scaling discussed in Section 6.1.
The total timescale in the transition region and corona is between 10 and 1000 s, which is small compared to 
the timescale for solar coronal radiative
losses, $\sim$10$^{4}$s \citep{golub1989}. 
%and implies that a continuous supply of energy
%(heating) is necessary to maintain the quiescent atmosphere.
The depression in radiative losses from EV~Lac's corona compared to the
Solar corona, due to abundance depletion, is more than compensated for by the high
inferred electron densities, and the timescales for radiative loss are consequently
short. For the shorter loop length inferred here, the conductive timescale is shorter than
the radiative timescales. The dependence of conductive timescale on loop length is L$^{2}$
so conductive timescales rise rapidly with an increase in loop length.
In order for the corona to maintain its quiescence, the heating timescale must be of the same
order as the timescale on which energy is lost.
This implies that there must be significant continued heating for the corona to
maintain its quiescence.

\subsection{Energy Balance}
Because we have measured the DEM from the upper chromosphere to corona, we can use it and the
observational constraints on pressures to investigate
how much energy is being lost and therefore how much is required to heat the corona.
In the absence of flows, the atmosphere at each point is in a balance between radiative and conductive losses,
and mechanical heating.  We estimate the magnitude of each of these terms, using \\
\begin{equation}
 \nabla \cdot F_{c} + \nabla \cdot F_{r} =  \nabla \cdot F_{h} 
\end{equation}
where 
radiative losses $F_{r}$ are estimated as \\
\begin{equation}
F_{r}(T_{e}) =\int n_{e}n_{H} \psi(T_{e}) ds,
\end{equation}
and the classical expression for conductive losses is \\
\begin{equation}
F_{c,cl}(T_{e})=-\kappa T_{e}^{5/2} \frac{dT_{e}}{ds},
\end{equation}
where $ds$ is the emitting region.
%{\bf If the mean free path is greater than the density scale height
%then is the classical expression for conductivity applicable? Or does this mean that the corona
%isn't in hydrostatic equilibriium?}
The equation defining the emission measure distribution can be
re-written to describe the temperature stratification, \\
\begin{equation}
\frac{dT_{e}}{ds} = \frac{n_{H}}{n_{e}} \frac{P_{e}^{2}}{\sqrt(2) T_{e} CEM k_{B}^{2}}
\end{equation}
from \citet{jordan1996},
where all quantities except temperature are assumed to be constant over the region
of line formation, and a plane-parallel atmosphere is assumed.
The classical expression for conductive flux uses the Spitzer conductivity times the
local temperature gradient to approximate the heat lost.  However, a limiting value of
the conductive flux occurs when the electrons are free-streaming, in which case the 
saturated conductive flux can be written as \\
\begin{equation}
F_{c,sat}=sgn(F_{cl}) \frac{n_{e} (k_{B}T_{e})^{3/2}}{4\sqrt{m_{e}}}
\end{equation}
where $F_{cl}$ is the classical expression.
The classical conductive fluxes in the corona become unphysically large, due to
the high coronal pressures discussed in \S 6.1 through the
$P_{e}^{2}$ dependence of $dT_{e}/ds$.
At most temperature bins above $\log T=$5., the conductive fluxes exceed the saturated
levels, the maximum being in the corona, by factors of up to 10$^{3}$.
Following the discussion in \citet{sa1980} and \citet{fisher1985}, we compute the
following expression for evaluating the conductive fluxes:\\
\begin{equation}
F_{c}= \frac{F_{c,cl}}{1+F_{c,cl}/F_{c,sat}}
\end{equation}
which has the advantage of achieving the limit $F_{c}\rightarrow F_{c,cl}$ when $F_{c,cl}\gg F_{c,sat}$
and $F_{c}\rightarrow F_{c,sat}$ when F$_{c,cl}\ll F_{c,sat}$.
We ignore anomalous heat conduction due to ion-acoustic instability.

Our goal is to compare the run of radiative and conductive losses with temperature, so we 
compute divergences of the analytic expressions above (see Appendix).
%\begin{eqnarray}
%\nabla \cdot F_{r}(T_{e}) =  \frac{\phi(T)\psi(T_{e})\Delta \log T}{4\pi R_{\star}^2} \;\;\; \rm erg\; cm^{-3}\;s^{-1}\\
%\nabla \cdot F_{c,cl}(T_{e}) = -\kappa T^{3/2} \left( \frac{dT}{ds} \right)^{2}
%\left[ \frac{3}{2} +\frac{d\ln P}{d\ln T}+\frac{d\ln CEM}{d\ln T} \right] \;\;\; \rm erg\; cm^{-3}\; s^{-1}
%\end{eqnarray}
We have assumed $n_{H}/n_{e}$ is constant over the temperatures being considered here.
An estimate of radiative losses has been calculated using the two scenarios
for abundance patterns in the outer atmosphere, and conductive losses are computed using the 
resulting DEMs.

\begin{figure}
\begin{center}
\includegraphics[scale=0.4]{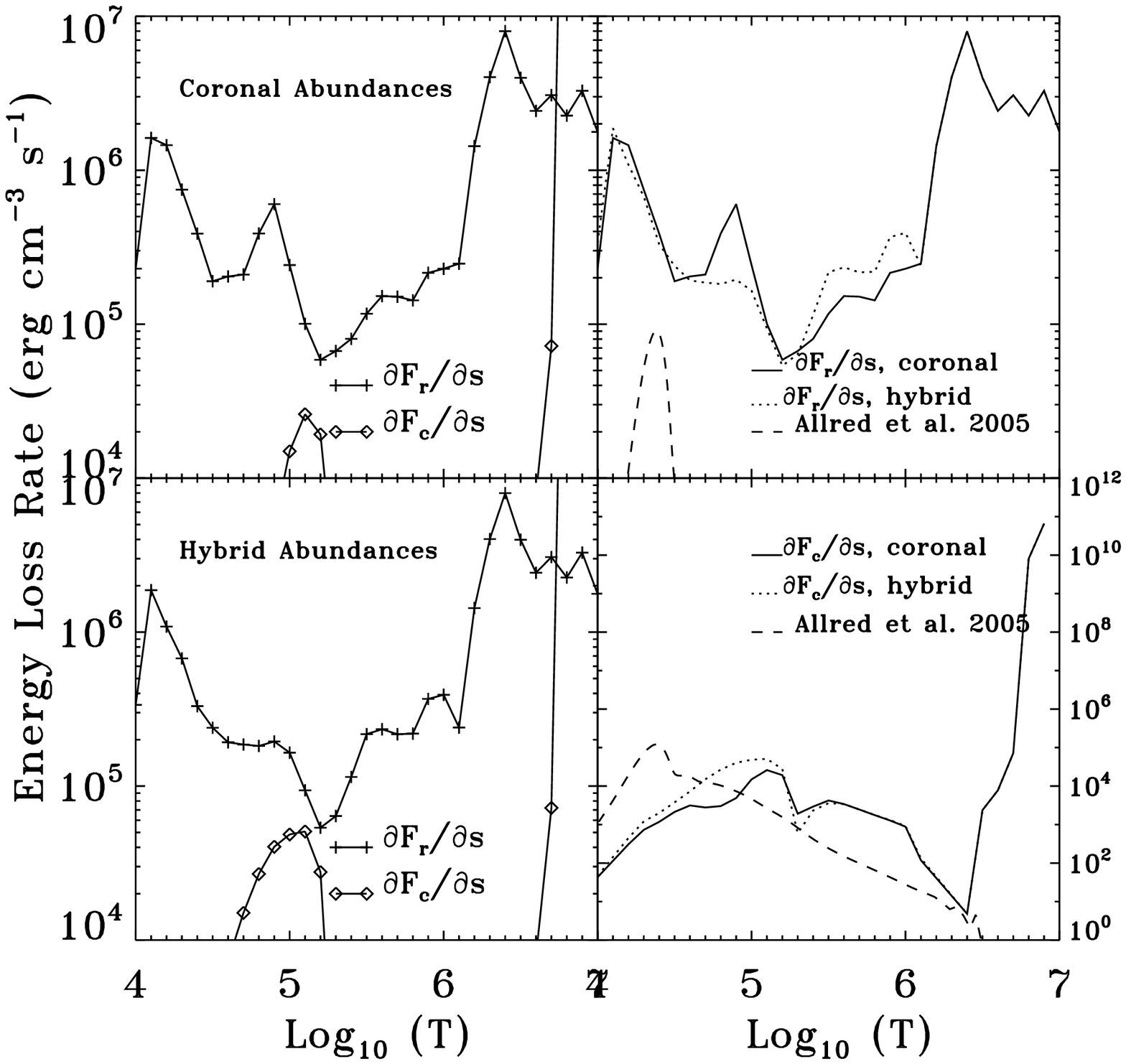}
\includegraphics[scale=0.4,height=6cm]{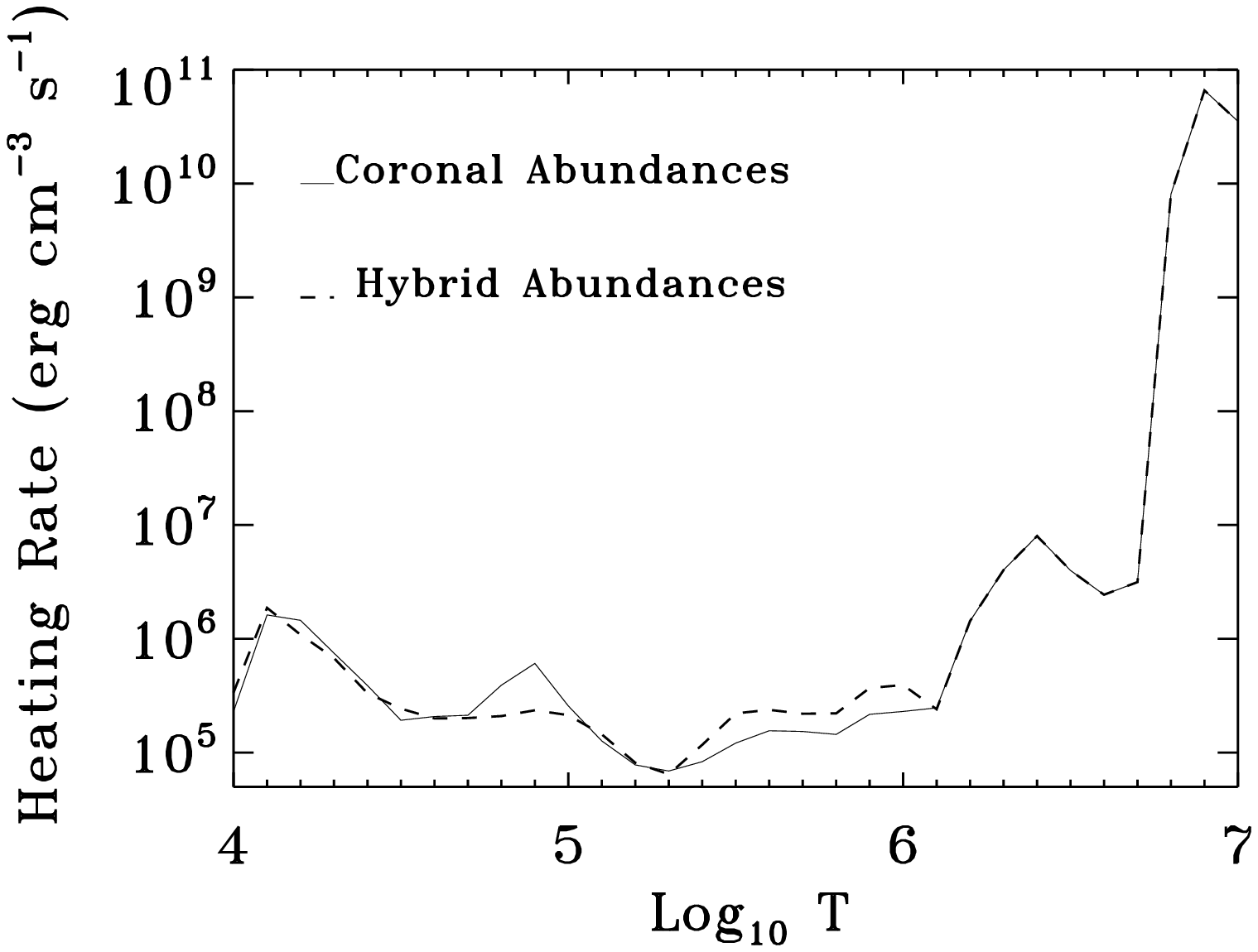}
\caption{{\it (left)} 
Radiative and conductive losses estimated using
the two DEMs, run of electron pressure with temperature, and radiative loss functions.  
The left-hand panels compare radiative and conductive losses for each abundance scenario; right-hand panels compare radiative and conductive losses with preflare model calculations of \citet{allred2005}
in the upper and
lower panels, respectively.  The large jump in conductive losses at $\log T=$6.4 is due to the
large inferred electron densities, and hence electron pressures, in the corona.  See \S 6.3 for
discussion.
{\it (right)} Estimation of volumetric heating rate at each temperature in the atmosphere, by combining
the conductive and radiative loss rates under the assumption of energy balance.  The enhancement in 
the low temperature corona is due to large radiative loss rates, while the sharp spike in the high temperature
corona is due to a precipitous jump in conductive loss rates.
\label{fig:losses}}
\end{center}
\end{figure}

At temperatures below 10$^{5.2}$K where our density
diagnostics are insensitive, a constant pressure equal to that derived at $\log T=$5.2 and
5.4 was
used.
We do not calculate the radiative and conductive losses above
10$^{7}$K, as our pressure estimates end there.
The left panels of 
Figure~\ref{fig:losses} displays the run of radiative and conductive fluxes
with temperature from the upper chromosphere through the corona for the DEMs determined using the coronal
and hybrid abundance patterns.  
At temperatures $\log T\leq$6.7 the energy loss rates due to radiation exceed
those from conduction.  However, at higher temperatures the opposite occurs.
This result depends ultimately on the high values of $n_{e}$ at these high temperatures;
reducing $n_{e}$ by an order of magnitude (equivalent to the \ion{Mg}{11}
diagnostic being in the low density limit)
 has a hundred-fold effect on decreasing the conductive loss rates.
%The conductive losses scale as $P_{e}^{4}$, which explains the large jump 
%above $\log T=$6.4 --- the large observed electron densities, and hence pressures,
%in the corona drive this increase. 
%The magnitude of the conductive fluxes in the corona can be
%reduced by 
%changing the $P_{e}(T_{e})$ profile to make it a constant or slowly increasing function of temperature.  
%This might indicate some
%systematics affecting the electron density diagnostics, which push them towards 
%higher densities (and hence pressures).
%If the observed high densities are real, this implies an enormous amount of energy
%being lost in the high temperature coronal material, and 
%requires either a substantial amount of heat input at the highest
%temperatures ($\log T>$6.7) to balance the conductive loss rates
%or demonstrates that energy balance arguments in the corona must be replaced by
%dynamical models.

Figure~\ref{fig:losses} compares the analytic derivation of radiative and conductive losses
with those of preflare numerical calculations of \citet{allred2005}.
The numerical calculations were derived by using a modified form of the Spitzer conduction
formula, with a saturation limit imposed to avoid unphysically large fluxes from the
steep temperature gradient.
The radiative losses show almost no agreement between analytic and numerical calculations.
This has to do with the difference between the observed DEM and the amount of material at
high temperatures which went into the numerical calculations.
The analytic evaluation of conductive fluxes likewise show a peak at $\log T=$5, whereas the 
numerical results show conductive losses peaking at lower temperatures.
The difference may similarly relate to the DEM, whereas above $\log T=$6.4 it most certainly
is related to the jump in pressure inferred from the observations.

The energy heating rate at each temperature can be estimated under the assumption of energy balance
by adding the energy loss rates due to conduction and radiation calculated above.  This is shown
in Figure~\ref{fig:losses} for the two abundance scenarios. 
As pointed out by \citet{cram1982}, the large X-ray fluxes of dMe stars coupled with the assumption of isotropic
assumption means that there is a large X-ray flux directed downward into the atmosphere, which may 
result in an appreciable amount of heating of the upper chromosphere/transition region.
However, recent calculations of the X-ray and EUV backwarming taking place during
stellar flares \citep{allred2006} show that this contributes a negligible amount of heating.

In  the low-temperature
corona, there is a local enhancement in the heating rate which peaks at $\log T=$6.4.
This is due to the increase in radiative loss rate from the peak in the DEM at these temperatures.
The sharp jump in energy heating rate at higher temperatures is a consequence of the large
conductive loss rates at the higher temperatures, which is itself driven by the dramatic rise in
pressure deduced from density-sensitive line ratios.
We can use the emitting volume estimates presented in \S 5.2.1 and Figure~\ref{fig:nete} 
and the volumetric heating rates to determine the power input.  
In the transition region at $\log T=$5.4, the emitting volume is $\sim$10$^{26}$ cm$^{3}$ ($\Delta \log T=$0.1),
with an energy heating rate of $\approx$9$\times$10$^{4}$ erg cm$^{-3}$ s$^{-1}$, for a total
power input of $\approx$9$\times$10$^{30}$ erg s$^{-1}$.
At $\log T=$6.4, the
emitting
volume of 10$^{30}$ cm$^{3}$ coupled with the heating rate of 7$\times$10$^{6}$ erg cm$^{-3}$ s$^{-1}$,
implies 7$\times$10$^{36}$ erg s$^{-1}$ of power input.  At $\log  T=$6.9, there is
an emitting volume of 10$^{25}$ cm$^{-3}$ and heating rate of 6$\times$10$^{10}$ erg cm$^{-3}$ s$^{-1}$,
for a needed power input of 6$\times$10$^{35}$ erg s$^{-1}$.  It is remarkable that despite the
large differences in loss rates, these are within an order of magnitude of each other.
The $\sim$ 7 orders of magnitude difference between the observed
X-ray luminosity of $\sim$10$^{28}$ erg s$^{-1}$ and the total power required to heat the corona
underscores how inefficient radiation alone is in diagnosing the required energy inputs.  
A possible next step would be the investigation of loop models to explain the above results
in terms of atmospheric structure.  
The calculations in this section imply that a large amount of
energy must be deposited at high temperatures, which if realistic, would be hard to
envision under a static energy balance scenario.  In particular the large conductive loss rates
and the steep temperature gradients in the corona imply that a dynamic situation leading to
mass flows is inevitable.
Thus we look to alternative
explanations in the following section.

\subsection{Flare Heating}
This treatment is fundamentally different from the energy balance arguments applied above,
since a flare-heated atmosphere has
mass flows, shocks, particle beams, etc.  A correct treatment requires
a comprehensive flare theory which could account for all the processes known to occur in both
collisional plasmas (e.g. chromospheric evaporation and coronal heating, coronal plasmoid ejections) and 
collisionless plasmas (e.g. particle acceleration, particle beaming), and then combines them
in an ensemble average over a flare distribution to truly describe a flare-heated atmosphere.
Since this is still not possible even for single solar flares, we resort to statistical arguments.

\subsubsection{DEM Considerations}
%In this section we describe an alternative approach to explaining the high temperature
%part of the DEM. 
\citet{gudel2003} studied the 
statistical distribution of coronal flares and their effect on the distribution of plasma with 
temperature in the corona, under the assumption that the coronal heating could be
accounted for entirely by a superposition of flares proceeding to smaller and smaller 
energies (the so-called ``nanoflare heating hypothesis'').  
%Using this methodology,
%assumptions about the distribution of flares with energy and how flares proceed
%(i.e. contribution of any one flare to a given temperature/emission measure bin,
%relationship between peak flare temperature and peak flare DEM, dependence of flare
%decay on flare energy, amount of sustained heating during the flare decay) 
%feed into the shape of the resulting DEM.  
Following the 
arguments outlined in their section 6.4, the shape of the DEM can be used as a diagnostic
for flare-heated coronae:  
%the slope of the DEM below the temperature at which the 
%DEM has its maximum is related to the parameter $\zeta$, which prescribes the amount of
%sustained heating taking place during the flare decay (where temperature decay is
%roughly T$\propto$n$^{\zeta}$; Reale, Serio, \& Peres 1993).  
at temperatures above the
maximum DEM, the decreasing slope can be used to determine $\alpha$, the power-law index
of the flare frequency-energy relation.  
%Through the power-law dependence of the DEM on the quantities described above, \\
%\begin{eqnarray}
%DEM \propto T^{2/\zeta}   \\
%DEM \propto T^{(-b-\phi)(\alpha-2\beta)/(1-\beta)+2b-\phi}
%\end{eqnarray}
%where b is slope of the power-law relationship between coronal temperature and
%emission measure during flares, $\phi$ the power-law index describing radiative losses
%at coronal temperatures, and $\beta$ is the power-law relationship between
%flare decay and energy.
The DEM (Figure~\ref{fig:coronaldem})
appears to have a maximum at $\log T=$6.4, and
we measure a slope from %$\log T=$5.0 to 6.4 (making use of the STIS and FUSE spectral information)
$\log T=$6.4--7.3 of $\sim-$0.6.
Using the parameter values discussed in \citet{gudel2003}
and our measured high temperature DEM slope,
this implies $\alpha \approx$2.2. For values of $\alpha >2$, flares are numerous enough
to provide the observed X-ray luminosity.  The quiescent DEM and this model therefore imply that 
EV~Lac's quiescent corona can be described by a multitude of small-scale flaring events.
%if $\beta$ is 0, i.e. no dependence
%of flare decay timescale on energy, b=5, and $\phi$=0.3 \citep[as used in][]{gudel2003}
This is not a self-consistent treatment of
flare-heated coronae, only a demonstration that the corona can be
interpreted in light of mechanical heating due to flares.  

\subsubsection{Comparison of Flare/Non-Flare Quantities}
\begin{figure}
\begin{center}
\includegraphics[scale=0.4,angle=90]{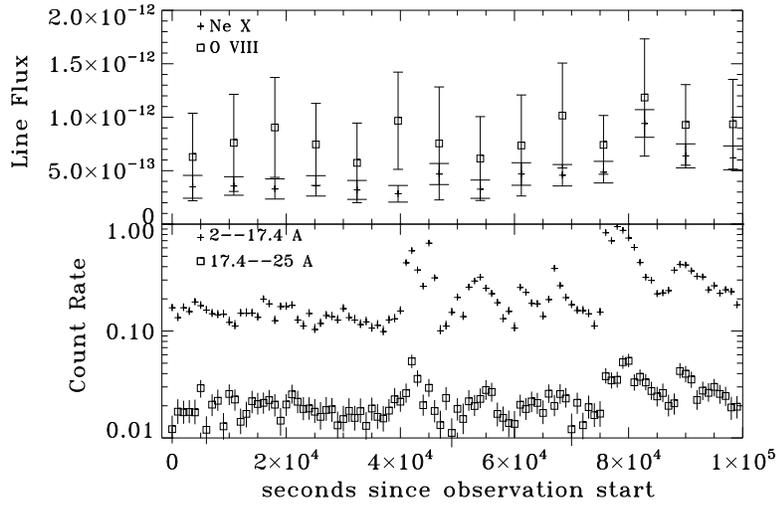}
\caption{{\it (top)} Time variation of two strong lines during the Chandra observation 
--- \ion{Ne}{10} is formed at $\log T_{\rm eff}=$6.7 and \ion{O}{8} is formed at $\log T_{\rm eff}$=
6.4.  Spectra were extracted in two hour time bins.
Despite the large flares seen in the integrated light curve, there is no evidence for 
the high temperature line being more variable than the low temperature line.
{\it (bottom)} Light curve of soft (17.4$<\lambda<$25 \AA) and hard (2$<\lambda<$17.4 \AA)
photons in 1000 s bins during the Chandra observation.  Enhancements
are more noticeable in the hard light curve during large-scale flares, but there is no 
evidence for more variability in the hard light curve outside of large-scale flares.
\label{fig:vary}
}
\end{center}
\end{figure}

We investigated the temporal variability in the X-ray observation to examine whether
there was any difference in behavior of high temperature and low temperature coronal plasmas,
as might be observed if the high temperature component were due to a superposition of flaring
activity.  The two strongest lines in the X-ray spectrum are \ion{Ne}{10} $\lambda$12.14
and \ion{O}{8} $\lambda$18.97, corresponding to $\log T_{eff}=$6.7 and 6.4, respectively.
We extracted spectra in two hour time bins, and performed Gaussian line fits to the spectra,
recording the width, wavelength of line center, and line flux (see Figure~\ref{fig:vary}).  
There is no evidence for
the high temperature line being more variable than the low temperature line, although the
large error bars on particularly the \ion{O}{8} flux hinder a more conclusive result.
An inspection of the spectrum (Figure~\ref{fig:chandraspec}) reveals that longward
of 17.4 \AA, most strong transitions are characterized by $\log T_{\rm eff} <$6.5,
and shortward of 17.4 \AA, the opposite case holds.  We made light curves
of soft (17.4$<\lambda<$25 \AA) and hard (2$<\lambda<$17.4 \AA)
photons in 1000 s bins.  Enhancements
are more noticeable in the hard light curve during large-scale flares, but there is no
evidence for more variability in the hard light curve outside of large-scale flares.
We confirmed this statistically by comparing the means and variances of the soft
and hard light curves outside of times of large-scale flares.

\citet{mk2005} examined UV--X-ray correlations during flares from a sample of dMe stars
observed with XMM-Newton, and found that a comparison of X-ray to UV energy loss rate ratio
during flares was similar to that obtained outside of flaring times, suggesting a common
origin.  
%Previous papers on multi-wavelength studies of individual flares on dMe flare stars have also 
%shown relationships between
%radio \& X-ray \citep{gudel1996}, and
%UV \&X-ray \citep{osten2005} luminosities.  
%Scaling laws also exist between these parameters for main-sequence
%stars and active stars \citep[][for example]{benzgudel1994, ayresetal1995,mj1988,rutten1991}. 
Since the entire outer atmosphere, from photosphere through corona,
is involved in the flare, 
the hypothesis of flare-heating providing the energy requirements
of the corona necessitates a difference in scaling law relationships for stars where
flare-heating is viable.  
However, the variations induced by large-scale flares and our inability to explain those multi-wavelength
correlations in individual cases
hampers an attempt to say anything conclusive about a relationship between time-averaged
quantities and quantities deduced from individual flares.  
A statistical study of flux ratios
in active stars compared to the subset for which flare-heating is a viable hypothesis must be done
to tease out any differences.

%We examined the luminosity ratios 
%during quiescence and during the
%individual large flares on EV~Lac studied in \citet{osten2005}, using
%their figures 11 and 12 to deduce the X-ray to radio luminosities for an X-ray flare and a radio flare, and for two small UV flares, as well as a time-averaged value.   
%There is a large range in these ratios of
%greater than an order of magnitude,
%which may reflect the lack of temporal agreement between flares seen in these wavelength ranges.
%The variations induced by large-scale flares and our inability to explain those multi-wavelength
%correlations hampers an attempt to say anything conclusive about a relationship between time-averaged
%quantities and quantities deduced from individual flares.  
%A statistical study of flux ratios
%in active stars compared to the subset for which flare-heating is a viable hypothesis must be done
%to tease out any differences.

If the quiescent coronal material is heated through a flare-related process which dredges material
from lower in the atmosphere through an evaporation process, then we should
expect to see a correspondence between coronal abundances and those in the chromosphere
where evaporation is presumed to occur.  
%Independent determination of stellar chromospheric
%abundances has not been done.
%Our results cannot discriminate between
%abundance scenarios where the coronal abundance pattern exists to lower in the atmosphere and another
%one in which there is an abundance gradient between the photosphere/lower atmosphere and the corona.
%Further research into abundance determinations at different atmospheric layers
%is needed before any conclusions can be made.
Further, if the physical processes which occur during flares scale with flare parameters
(e.g. flare energy), then large-scale
flares should be able to diagnose processes occuring during the smaller scale flares
which are hypothesized to provide the quiescent luminosity.
%Indeed, the relationship between radio and soft X-ray luminosities has been 
%interpreted as evidence for a correspondence between particle acceleration
%and plasma heating, similar to what is often observed in individual large-scale
%stellar flares.
The large X-ray flares in this Chandra observation, discussed in \citet{osten2005},
show no evidence for abundance enhancements. 
However, some large stellar flares studied in detail show evidence of abundance enhancements
\citep{osten2003}, implying some degree of complexity to both the flare process and
the relationship between flares and flare-heated quiescent conditions.

\section{Radio Emission at High Frequencies }
The negative spectral indices at frequencies higher than 5 GHz points to partially
optically thin emission.  In this section we consider sources of the emission and the implications
for coronal structure.
The high frequency radio spectra of some active M stars has been inferred to provide evidence
of gyroresonance emission \citep{gb1989}, and we address the possible emission mechanisms contributing at 
our highest frequency, 15 GHz.  The three most plausible are: (1) nonthermal gyrosynchrotron emission
from accelerated electrons; (2) gyroresonance emission, and (3) free-free emission.  The last will
be present at all frequencies, and represents the limiting value of stellar radio emission fluxes.
The optical depths of these mechanisms scale as:\\
\begin{eqnarray}
\tau_{NT} \sim n_{\rm tot} B^{2.3+0.98\delta} \\
\tau_{gr} \sim T^{s-1}n_{e} \\
\tau_{ff} \sim \frac{CEM}{T^{3/2}} \\
\end{eqnarray}
where $n_{\rm tot}$ is the total number density of accelerated electrons, $B$ is the magnetic field
in the source, $\delta$ is the index of the nonthermal electron density distribution, $s$ is the harmonic number, and $CEM$ is the emission measure distribution \citep{dulk1985}.
These estimates assume a homogeneous source. 

\begin{figure}
\begin{center}
\includegraphics[scale=0.39]{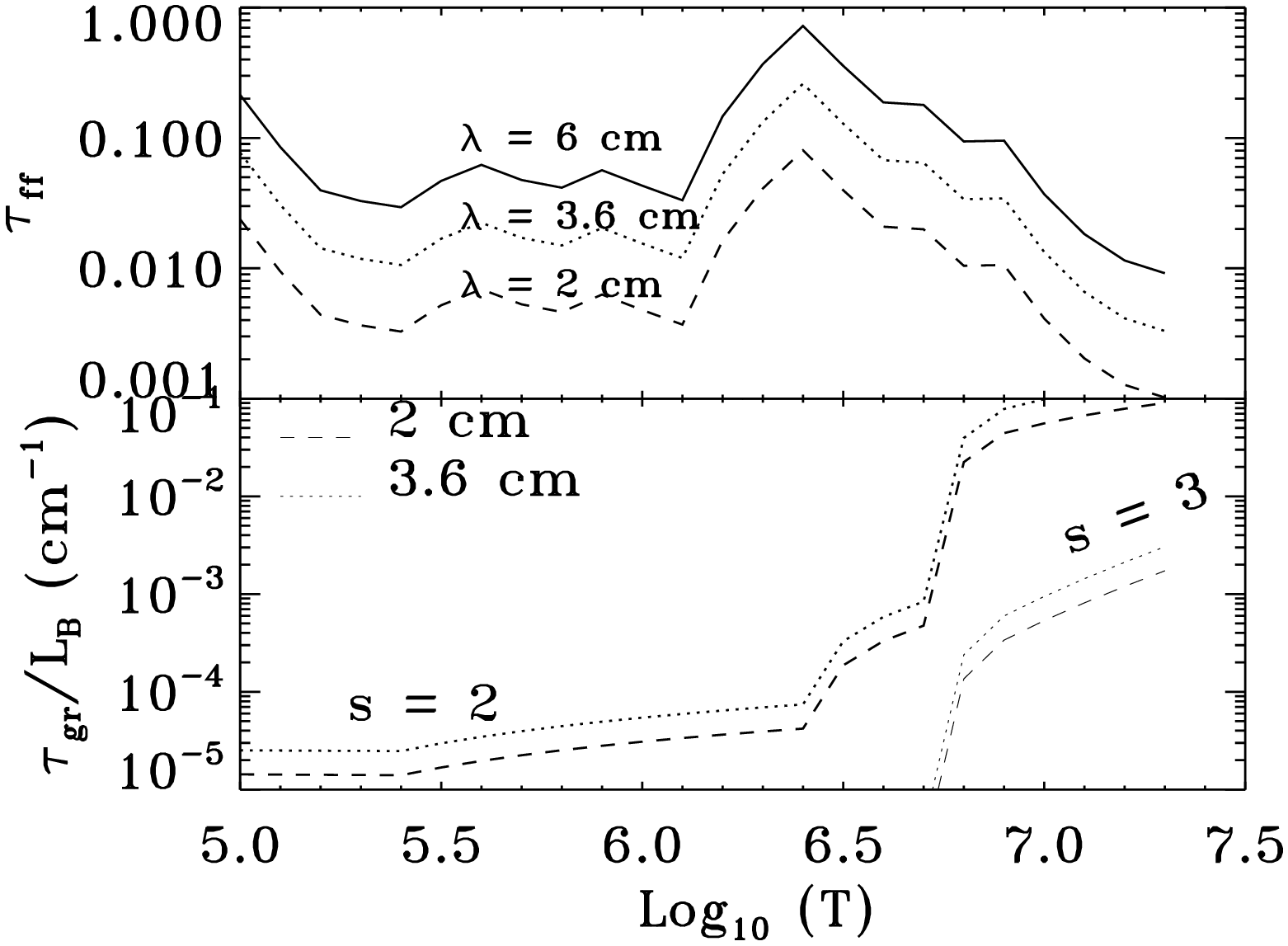}
\includegraphics[scale=0.45]{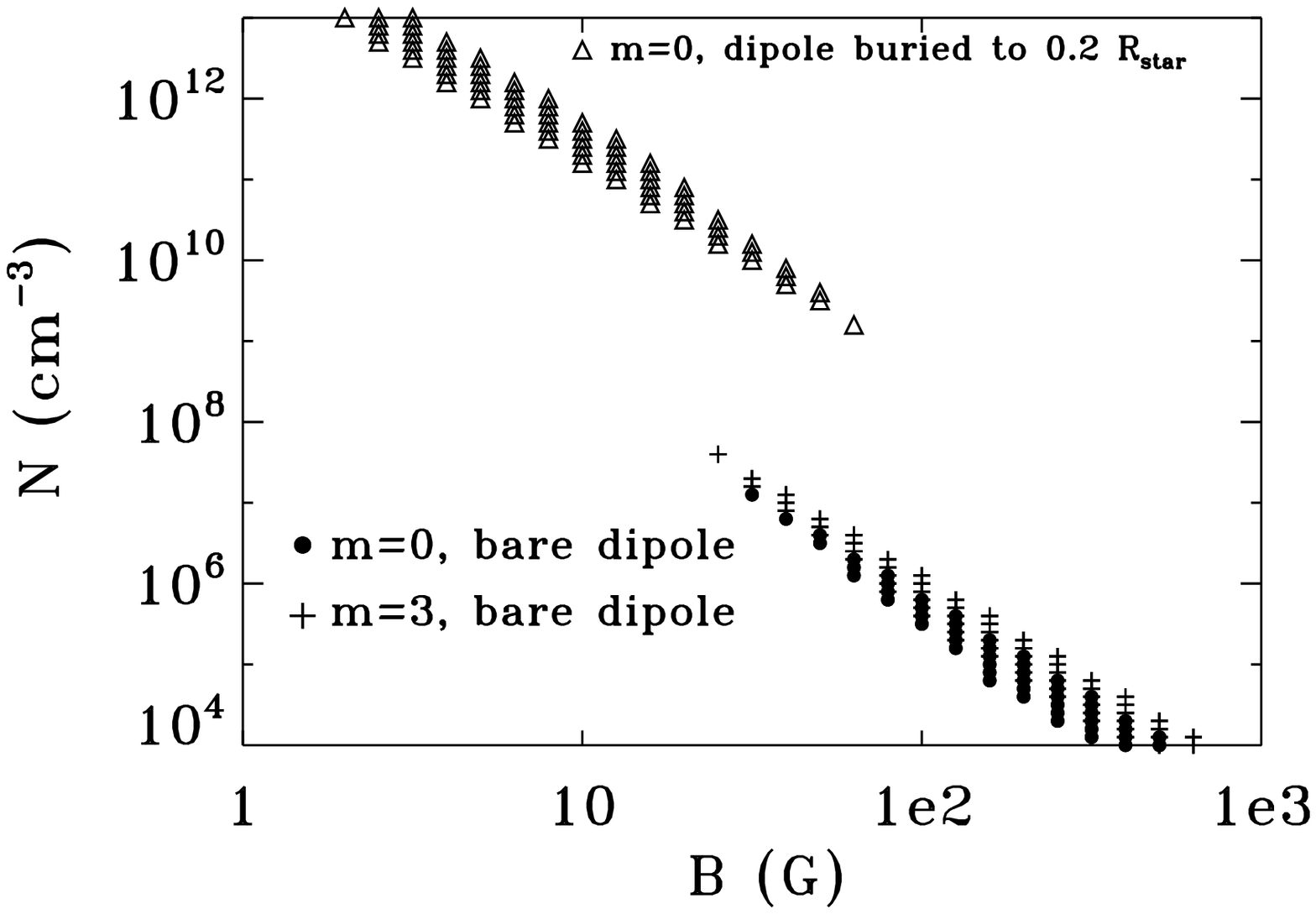}
\caption{{\it (left)} Estimate of optical depths due to free-free (top) and gyroresonance (bottom) emission in the upper atmosphere
of EV~Lac. Gyroresonant optical depths are computed at two harmonics for each of the two highest frequencies detailed
in our observations, relative to the unknown magnetic length scale. 
{\it (right)} Range of total nonthermal electron density and base magnetic field strength
allowed in simple dipole models for the magnetic field geometry to reproduce the observed
15 GHz flux density (to within a factor of two) and spectral peak between 2 and 8 GHz.
See \S 7 for more details. \label{radiotau}}
\end{center}
\end{figure}

For gyroresonance and free-free emission, we have computed
the expected optical depths assuming that the quantities derived from our spectral observations
in \S 4 and \S 5 apply to the source regions of radio emission.  Figure~\ref{radiotau} displays the 
optical depths, calculated using equations (1) and (3) in \citet{leto2000}.  Based on the 
emission measure distribution, at 6 cm, the free-free emission
should approach optical depth unity in the low temperature corona. 
At temperatures below 10$^{5}$K the expected free-free emission optical depth rises above unity,
as expected.  The expected contribution of optically thin free-free emission from transition region
and coronal plasma
($\log T\geq$5) to the observed radio emission is small, however; only $\sim$40 $\mu$Jy.
The magnetic scale length $L_{B}$ is the unknown quantity by
which gyroresonant optical depths are scaled in Figure~\ref{radiotau} and indicates
that the two shortest wavelengths should also suffer from optical depth effects, at either
the second or third harmonic of the gyrofrequency.  Equating the observed frequencies
with the harmonics of the gyrofrequency ($\nu=s\nu_{B}=2.8\times10^{-3}Bs$ GHz) for $s=2,3$, we derive magnetic field strengths in the radio-emitting
source of 2.7, 1.8 kG for $s=2,3$ at 15 GHz (2 cm), and 1.5, 1 kG for $s=2,3$ at 8.4 GHz (3.6 cm).  
These values are consistent with the equipartition magnetic field strengths required to
magnetically confine the hot coronal plasma at temperatures $>$8 MK discussed in \S6.1 and thus
lend some credence to the existence of such strong magnetic fields in the corona.
We emphasize, though, that these
estimates are made for a homogeneous source.  We estimate the optically thick emission due to
gyroresonance emission following equation (5) in \citet{leto2000} as \\
\begin{equation}
S = 810 \frac{T}{2\times 10^{7}} \left( \frac{\nu}{15 GHz} \right)^{2} \left( \frac{R_{s}}{0.3R_{\odot}} \right) \;\;\; \mu Jy
\end{equation}
for a heliocentric distance of 5 pc.
Assuming an optically thick source size equal to 1.1 times the stellar radius of $\approx$ 0.3 R$_{\odot}$, 
we expect a flux density of $\sim$890 $\mu$Jy at 15 GHz and $\sim$280 $\mu$Jy at 8.4 GHz
for emission at $T=2\times10^{7}$K, the highest
temperature revealed by the emission measure analysis (\S5).  This is higher than the observed
flux density at 15 GHz and so is not consistent with the soft x-ray emitting source being
spatially coincident with the radio-emitting source. Thus the high frequency
radio emission is not consistent with
being explained completely by  gyroresonance emission, but spatial inhomogeneities
in radio and/or soft X-ray (SXR) emitting regions
may be account for some of the discrepancies.

For gyrosynchrotron emission from a homogeneous population of mildly relativistic electrons described
by a power-law distribution in energy, the optically thin spectral index $\alpha$ can be related to 
the index of the power law, $\delta$, by $\alpha=1.2-0.895\delta$ \citep{dulk1985}.  For a relatively
hard distribution with $\delta=2$, $\alpha=-$0.59, which is consistent with the
3.6--2 cm spectral index.  However, the total integrated energy does not
converge for such a hard electron distribution.  
If the radio emission is not completely optically thin at one or both
of these frequencies (or includes contributions from gyroresonance emission, see paragraph above), 
this could lead to our inferred $\delta$ being smaller than the actual
$\delta$; thus we explore the effect of $\delta=3$ in the following simple models of the magnetic geometry.

Constraints on coronal properties can be obtained by using
simple analytic models, and a radial decrease in the number of
nonthermal particles.  We use the analytic formulae of \citet{whiteetal1989}, who
parameterized spatial inhomogeneities in the magnetic field and nonthermal number density
distribution. 
%providing formulae for optically thick and thin fluxes in the cases of a bare and
%buried dipole.  
\citet{whiteetal1989} considered a global dipole, which they termed a ``bare'' dipole,
and a ``buried dipole'', in which the depth below the surface at which the dipole is buried
is much less than a stellar radius.
The parameter $m$ describes the power-law index of the radial dependence of the number density.
The $m=0$ case corresponds to a nonthermal electron distribution independent of radius,
appropriate for an isotropic pitch angle distribution.  The other extreme, $m=n=3$, corresponds
to the radial dependence of the electrons being the same as that of the dipolar magnetic field;
this is the situation that would be obtained by conserving both particle and magnetic flux in
the case of open field lines.

By constraining the index of the power-law distribution to be $\delta=3$,
we explored possible scenarios, including a global and buried dipole, the radial
dependence of the nonthermal electron energy (either constant with radius, or having the same radial
dependence as the magnetic field), the base magnetic field strength, and the total number
density of accelerated electrons above a cutoff value.  

For $\delta=3$ and an observed flux density S$_{\nu}$, the magnetic field strength
at the base of the dipole and the total number density of nonthermal electrons
are related as $S_{\nu}\propto n_{\rm tot} B^{2.5}$.  The frequency at which the
spectrum turns over is also a strong function of the magnetic field strength
in the source.  M dwarf radio spectra are consistent with a rising spectrum between
1.4 GHz and 5 GHz, and flat or declining flux density spectra at higher
frequencies \citep{whiteetal1989}.
The right panel of Figure~\ref{radiotau} displays allowed regions of $n_{\rm tot}$ and $B$
which produce the flux density at 15 GHz to within a factor of two, and have a spectral
turnover between 2 and 8 GHz, 
%We use the flux density at 15 GHz and a spectral peak between 2 and 8 GHz
to constrain some of the radio coronal properties of EV Lac.  
There is very little difference between the $m=0$ and $m=3$ models for a bare dipole.
These models imply a relatively large base magnetic field strength and low value of
total nonthermal electron density.
We note that 
 the magnetic field that \citet{jkv1996} observed on EV~Lac is almost certainly not
a global dipole field:  Polarization from optical observations 
is discussed explicitly in \citet{vjkp1998}
where the authors argue that AD~Leo's field is not a global dipole.
Thus we considered dipoles buried at smaller depths, which would be characteristic of
individual active regions.
For a dipole buried to 0.2R$_{\star}$ the base magnetic field strength can be relatively
small ($\leq$ 100 G), but with a larger total nonthermal electron density ($\geq$ 10$^{9}$ cm$^{-3}$).

\section{Constraints on Coronal Structure}
The multi-wavelength data described here pose a conundrum:  the soft X-ray (SXR) observations
require large magnetic field strengths in the hot corona to confine the plasma, but the radio
observations do not support such large fields in the corona providing a significant amount
of gyroresonant opacity.  These two observations can be reconciled if the radio and
SXR emission regions are not coincident.
This conclusion was reached by previous authors who
determined \citep{whiteetal1994} that the hot component of the SXR-emitting plasma was not
coincident with the strong magnetic fields in the low corona, and \citet{leto2000}
who estimated that the hot coronal material must be co-spatial with coronal magnetic fields
less than 0.5--1 kG.  
The results described in this paper represent the first attempt at a
realistic investigation of the problem, using high spectral resolution observations to 
determine the shape of the continuous emission measure distribution and
place constraints on the electron density at several temperatures in the outer atmosphere of
an active M dwarf.  
The coronal electron density determinations place an additional constraint:  
Any radio emission from the high temperature, high density soft X-ray-emitting material is 
below the local plasma frequency ($\nu_{p} \approx 9 \sqrt{(n_{e}/10^{12} cm^{-3})}$ GHz)
and therefore cannot propagate.  For n$_{e}=$10$^{13}$ cm$^{-3}$ $\nu_{p}=$29 GHz so
the electromagnetic wave is damped.  This is an additional argument that
 the high temperature X-ray emitting plasma 
cannot be spatially coincident with the centimeter-wavelength radio emission.

The volumes derived from the emission measure analysis imply remarkably compact coronal structures:
for spherically symmetric coronal emission, V $<$10$^{30}$ cm$^{-3}$ and EV~Lac's radius,
the coronal extent is less than 0.1 R$_{\star}$.
We estimate the filling factors for SXR and radio emission using our observations.  For soft X-ray emission, we compare the observed $V_{X}$ (deduced from VEM and $n_{e}$ in \S 5.2) with the volume available for
a spherically symmetric volume $V\sim 4\pi R_{s}^{2} h$, where $h$ is the height of the emission above the
stellar photosphere and deduce a filling factor $f_{X}$, \\
\begin{equation}
f_{X} = \frac{V_{\rm X}}{4\pi R_{s}^{2}h} \;\;\;.
\end{equation}
and we relate $h$ to the loop length $L$ by $h\sim L/2$.  The volumes shown in Figure~\ref{fig:nete}
display a large discrepancy between the plasma-emitting volume in the low temperature 
corona, $V_{\rm X} \sim$10$^{30}$ cm$^{3}$, and that in the high temperature corona, 
$V_{\rm X}\sim$10$^{25}$ cm$^{3}$. This is consistent with the other significantly different plasma
characteristics of the low- and high-temperature corona discussed above.  We thus derive two filling factors,
for the $\sim$2.5 MK plasma (LTC) and $\sim$ 8MK (HTC) -- \\
\begin{equation}
  f_{X,LTC} = \frac{4\times10^{8}}{L_{LTC}} \;\;\;\; f_{X,HTC}= \frac{4\times10^{3}}{L_{HTC}}
\end{equation}
where $L_{LTC}$ and $L_{HTC}$ are the loop lengths for the low- and high-temperature coronal plasma,
respectively.  Filling factors of order unity are returned if loop lengths $L_{LTC}\sim$few $\times$10$^{8}$ 
cm, whereas dramatically smaller length scales are needed to obtain near unity filling factor
for the high temperature coronal plasma.  Conversely, this seems to indicate a much smaller than unity 
filling factor for the high temperature coronal plasma.
An equivalent argument can also be made using the temperature gradient determined in \S 6.3 (equation 17)
 and 
determining a length scale $l$,\\
\begin{equation}
l=T/(dT/ds) \propto \frac{CEM}{n_{e}^{2}} = \frac{VEM}{n_{e}^{2} 4\pi R_{s}^{2}}
\end{equation}
in which case for filling factor of unity, the loop length must be $\approx 1/l$.

We also estimate the filling factor for radio emission by using the results of resolved radio sizes
from VLBI studies \citep{pesta2000}, and compare this with the volume deduced from optically thin 
emission, where $V_{R} = L_{R}/\eta$, $L_{R}$ is the radio luminosity, and $\eta$ is the 
emissivity for gyrosynchrotron emission, taken from \citet{dulk1985}.  We assume $\delta=3$
and $\theta=$45$^{\circ}$.  The VLBI observations cited were conducted at 3.6 cm, and determined
a source size $h_{R}$ of $\sim$2$\times$10$^{10}$ cm for YZ CMi (dM4.5e).  We hereby assume that this is also a 
characteristic radio emission length scale for EV~Lac.  
The total radio-emitting volume is then $V_{\rm tot}$=$4\pi R_{s}^{2}h_{R}$, with $h_{R}$ the resolved height of radio emission above the stellar radius.  
The  observed radio-
emitting volume is estimated from the radio luminosity and emissivity at 8.5 GHz (3.6 cm).  The 
filling factor for gyrosynchrotron emission at 3.6 cm is thus: \\
\begin{equation}
f_{R}  = \frac{V_{R}}{V_{\rm tot}} = \frac{3.4\times10^{11}}{n_{tot}B^{2.48}}
\end{equation}
where $B$ is the magnetic field strength in the radio-emitting source, and $n_{tot}$ is total
number density of nonthermal particles.  Based on the radio modeling in \S 7, a global dipolar
model for the radio emission can accommodate $B\sim$20G, n$_{tot}\sim$10$^{6}$ cm$^{-3}$, while a dipole buried to
a smaller depth requires elevated number densities, n$_{tot}\sim$10$^{10}$ cm$^{-3}$ at similar magnetic
field strengths.  Using these numbers, we estimate $f_{R}$ to be $\sim$200 for the global
dipole and $\sim$2$\times$10$^{-2}$ for the buried dipole.  
A filling factor larger than unity is unphysical, and thus appears to be in agreement with the optical 
polarization limits
ruling out a global dipolar component.

The results of this section and \S 6
imply different length scales for the cool and hot SXR-emitting
plasma -- the low temperatures should be more extended and therefore undergo more
rotational modulation.  
The radio source is also inhomogeneous, with scale lengths
larger than
the expanded, low
temperature coronal material,
which would place the radio spectral results in agreement
with the soft X-ray constraints. 
The radio source is also likely not coincident with the large
field strengths inferred from optical observations, as the equatorial surface fields estimated from
the simple dipole field geometry for gyrosynchrotron emission are in the range $\sim$30--500 G,
and the filling factor times field strength derived from \citet{saar1994} and \citet{jkv1996} are $\sim$4 kG.
The high temperature corona, from DEM arguments in \S 6.4.1, is likely related to a flare heating
process.  The high pressures derived for this plasma also imply that it is the most compact, with the
smallest timescales for energy loss and the largest energy loss rates.  
The lifetime of accelerated particles to
losses by collisions is 1s or less in the HTC, and only of order 100 times longer in the
LTC.  
In order to maintain a stable, long-lived population of accelerated electrons which is observed
as cm-wavelength radio emission,
continual replenishment is necessary.
If the flare process
which heats plasma to high temperatures also involves the acceleration of particles,
then there should be a link between radio emission from accelerated electrons and high temperature coronal
 plasma --- although from the arguments above the two populations cannot be spatially coincident.
%The discussion in \citet{gb1993}
%does not discriminate between the origin of the X-ray luminosity in describing the correlation
%with radio luminosity for a sample of active stars.  

\subsection{Comparison with Solar Coronal Structures}
The high angular resolution and temperature discrimination of narrow-band filters
available on the Transition Region and Coronal Explorer (TRACE)
 have been instrumental
in making advances in understanding of solar coronal loop structures, heating and dynamics.
EUV-emitting loops have temperatures generally from 0.8--1.6 MK, and thus are cooler than
the temperatures discussed in this paper, but nevertheless the detail with which they have
been investigated motivates this as a starting point for comparison.
The lack of observed temperature variation along loop lengths \citep{lenz1999} in the solar
corona as seen by TRACE has been interpreted as the signature of a highly filamentary
corona \citep{rp2000}.
\citet{asch2000} found nearly isothermal loop threads whose temperature gradients
where much smaller than predicted by static, steady-state models for uniform loop heating.
The loop base pressure was also higher than that predicted for static, steady-state
models invoking uniform heating, and
independent of loop length, implying a 
departure from such models.
\citet{asch2000} showed that coronal heating 
was localized to altitudes less than 20 Mm, which is much smaller than the 
longest loop lengths ($\sim$3$\times$10$^{10}$ cm).  This result suggests that short
loops experience uniform heating, while long loops do not.
%This implies that in EUV-emitting loops with
%lengths up to $\sim$3$\times$10$^{10}$ cm,
%the heating is nonuniform and located near loop footpoints.  
%Shorter loops whose lengths
%are comparable to the heating scale height, are consistent with
%uniform heating.  
The pressure scale height was found to exceed the hydrostatic scale height for all but the 
shortest of loops, supporting
a conclusion that these loops are not in hydrostatic equilibrium.
\citet{neupert1998} showed that radiative losses exceed conductive losses by factors of 100. 
Comparison of results for EUV-emitting loops with 
studies of soft X-ray emitting loops suggests a physical distinction between the two.
\citet{priest2000} determined that a larger temperature gradient exists in soft X-ray-emitting
loops, and they ruled out footpoint-dominated heating, preferring instead uniform heating.

Our conclusion calling into doubt the applicability of hydrostatic, steady-state loops
appears to resonate with the results obtained from TRACE for solar coronal loops, albeit for 
markedly different temperatures.  However, other aspects of coronal structures on the Sun
and EV Lac appear vastly different.  The temperature gradients inferred from an analytical treatment
of the spatially integrated emission (assumed spherically symmetric)
in EV~Lac suggest that at low temperatures in the corona,
T$\sim$2.5MK, the gradient will be small, $\sim$10$^{-2}$ K/cm.  But the temperature gradient
appears to increase by a factor of $\sim$10$^{6}$K over a range of 3 in temperature.
This would presumably place severe requirements on the types of structures and heat inputs
which can sustain such a large increase.
At coronal temperatures below 5 MK radiative losses dominate conductive losses, as seen also in
solar coronal observations, however the magnitude by which radiative losses exceed conductive losses
is a maximum of 10$^{6}$ at 2.5 MK.
At temperatures above 5 MK however conductive losses dominate, and by a large amount.

\section{Conclusions}
Through high quality spectroscopic observations and extensive wavelength coverage,
we are able to determine a number of properties of the quiescent atmosphere of the dMe
flare star EV~Lac. 
%which confirm previous conjectures based on less well-defined observational
%constraints.

Nonthermal broadening in transition region line profiles appears to peak at temperatures $\leq$
10$^{5}$K, in contrast with the Sun.  This behavior is also manifested in another active dMe star, AD~Leo.
Further investigation is needed to determine if this is a characteristic of most active stars;
the enhancement due to turbulent pressure support in the transition region has implications for 
structure and dynamics.

We are able to constrain electron densities across two orders of magnitude in the transition region
and corona, detailing a nearly constant pressure in the transition region and low temperature corona.
The high temperature coronal material appears to be characterized by a large jump in pressure.
Based on modeling the observed $P(T_{e})$ from the emission measure distribution, we find no evidence
to support a corona in hydrostatic equilibrium.  Large magnetic field strengths are required to
confine the high temperature coronal plasma.
Coronal spectral analysis reveals a continuous distribution of temperatures and evidence for sub-Solar
abundances. 
We extend the differential emission measure to transition region and upper chromospheric temperatures
with the use of STIS and FUSE spectra; there is a deep emission measure minimum in the transition region.
We cannot at this point determine whether an abundance gradient exists in EV~Lac's outer atmosphere; we
explored two possibilities, a coronal abundance pattern in the corona and
solar photospheric abundances elsewhere, in addition to a coronal abundance pattern
throughout EV~Lac's upper atmosphere, and find little difference in the atmospheric properties.

Using  low density ionization balance calculations reveals a discrepancy in Li- and Na-like
ions in comparing flux predictions from the differential emission measure with observed fluxes.  
This may arise from ignoring the density dependence of dielectronic recombination rates in
calculating ionization balance, or dynamical effects, or both.  There is an urgent need to resolve
this discrepancy, particularly for stellar studies, as these brightest transitions are used heavily
in estimating the amount of material in transition region plasmas.  The discrepancy can only be
revealed when a suitably exposed spectrum contains diagnostic information at the same temperatures as
these transitions, a situation which occurs only for a limited number of stars.  Thus there is the
potential for misinterpretation of the data if this effect is not resolved, and the pessimistic
prospect for future UV spectroscopic missions means astronomers will be relying heavily on 
already obtained archival data.

Based on observed electron densities and emission measure distributions, the timescale for upper
atmospheric material to lose energy is short, requiring significant continued heating.  
Calculations of analytic expressions for radiative and conductive loss rates show that a large
amount of energy is being lost at high temperatures, which argues for an enhanced amount of energy
input at the highest coronal temperatures.  
These implications rest to a large degree on the high electron pressures found in the corona
from electron-density sensitive line ratios.
The DEM shape is consistent with flare heating
arguments, although statistical comparisons of flare and non-flare quantities requires a better
defined knowledge of stellar flare physics than is revealed by current flare observations.

EV~Lac is detected at a radio frequency of 15 GHz (2 cm) for the first time.  This places
constraints on the amount of gyroresonance emission, and rules out the high temperature
thermal coronal plasma being spatially coincident with the radio emission source.  By
assuming that the bulk of cm-wavelength radio emission arises from accelerated electrons
in a dipolar field configuration, we can constrain the field strengths and total number density
of accelerated electrons for both a global dipole and one buried to a smaller depth. 
Generally equatorial surface field strengths $\leq$ a few hundred Gauss are required to explain the
radio spectrum.

The constraints on electron pressure, timescales for loss, volumes and filling factors 
suggests a two-phase thermal corona:  a low temperature corona occupied by a large volume of
plasma, and a high temperature corona with small length scales.  The radio-emitting
corona also appears to be inhomogeneous, but have a large length scale.  The relationship between
high temperature coronal plasma and an origin in a flare heating process suggests a link with
the radio emission source if accelerating electrons and heating coronal plasmas are a fundamental
consequence of flares.  However, the high frequency radio observations raise constraints
on the spatial relationship between the high temperature coronal plasma and accelerated
particles.
More detailed models of coronal structure are needed to explain this
hypothesis within the observational constraints discussed here.

This represents the results of VLA project AO160.
\appendix
\section{Appendix: Derivation of Analytic Expressions for Divergences of Conductive Losses}
Here we derive the analytic expressions for the conductive flux.
Using equation (6-9) for the radiative flux, the radiative loss rate is \\
\begin{equation}
\nabla \cdot F_{r}(T_{e}) =\frac{\phi(T_{e}) \psi(T_{e}) \Delta \log T}{4\pi R^{2}} \;
\end{equation}
%Using equation (6-10) for the classical conductive losses, \\
%\begin{equation}
%\nabla \cdot F_{c}(T_{e}) = -\kappa \frac{5}{2}T^{3/2} \left( \frac{dT}{ds} \right)^{2}
%-\kappa T^{5/2} \frac{d^{2}T}{ds^{2}}
%\end{equation}
%where we estimate the second derive of T(s) using equation (6-9), \\
%\begin{eqnarray}
%\frac{d^{2}T}{ds^{2}} = \left( \frac{dT}{ds} \right)^2 \left[ \frac{2}{P} \frac{dP}{dT}
%-\frac{1}{T} - \frac{dCEM}{dT}\frac{1}{CEM} \right]
%\end{eqnarray}
%and this simplifies to : \\
Using equation (6-13) for the conductive flux, we 
calculate the divergence analytically as \\
\begin{eqnarray}
\nabla \cdot F_{c} = \frac{(1+\frac{F_{c,cl}}{F_{c,sat}})\frac{dF_{c,cl}}{ds} - F_{c,cl}\frac{d}{ds}(\frac{F_{c,cl}}{F_{c,sat}}) }{(1+\frac{F_{c,cl}}{F_{c,sat}})^2}
\end{eqnarray}
where 
$dF_{c,cl}/ds$ is
(using equation (6-11)) \\
%\begin{eqnarray}
%\nabla \cdot F_{c,cl}(T_{e}) = -\kappa T^{3/2} \left( \frac{dT}{ds} \right)^{2}
%\left[ \frac{3}{2} +\frac{2T}{P} \frac{dP}{dT} + \frac{T_{e}}{CEM}
%\frac{dCEM}{dT} \right ] ,
%\end{eqnarray}
%after simplifying,
\begin{eqnarray}
\frac{dF_{c,cl}}{ds} = -\kappa T^{3/2} \left( \frac{dT}{ds} \right)^{2}
\left[ \frac{3}{2} +\frac{d\ln P}{d\ln T}+\frac{d\ln CEM}{d\ln T} \right]
\end{eqnarray}
and \\
\begin{equation}
\frac{d}{ds} \left(\frac{F_{c,cl}}{F_{c,sat}} \right) = \frac{4\sqrt{m_{e}}\kappa}{k_{B}^{3/2}} \frac{T}{n_{e}}
\frac{dT}{ds}
\end{equation}

%\clearpage

%\bibliographystyle{natbib}
%\bibliography{/export/users/rosten/papers/evlac_quiet/evlac}

\clearpage
\begin{deluxetable}{llllll}
\tablenum{1}
\tablewidth{0pt}
\tablecolumns{6}
\tablecaption{VLA Multifrequency Observations \label{tbl:vla}}
\tablehead{\colhead{$\nu$} & \colhead{tos} & \colhead{S$_{\nu}$} & \colhead{Beam Params} & \colhead{V Flux} & \colhead{Percent Pol} \\
\colhead{(MHz)} & \colhead{(s)} & \colhead{(mJy)} & \colhead{(maj. x min., PA)\tablenotemark{a}} & \colhead{(mJy)} & \colhead{(\%)}  }
\startdata
4860 & 10570 & 0.51$\pm$0.04 & 5.2x3.5, 87 & $-0.11\pm$0.02 & $-22.5\pm$4 \\
8460 &10450 & 0.47$\pm$0.14 & 3.1x2.1, 89 & $-0.07\pm$0.01 & $-16\pm$2 \\
14939 & 11940 & 0.34$\pm$0.04 & 7.7x5.5, 81 & $<$0.11 & $\ldots$ \\
\enddata
\tablenotetext{a}{Beam major axis, beam minor axis in arcsec; PA=position angle in degrees}
\end{deluxetable}

\begin{deluxetable}{lcccclll}
%\rotate
\tablewidth{0pt}
\tablenum{2}
\tablecolumns{8}
\tablecaption{Detected Lines in HST/STIS Spectrum \label{tbl:stislines}}
\tablehead{ 
\colhead{Transition} &\colhead{$\lambda_{\rm lab}$}  & \colhead{$\lambda_{\rm fit}$} &
\colhead{v} & \colhead{FHWM} & \colhead{flux} & \colhead{$\chi^{2}_{\nu}$} &
\colhead{$\log T_{\rm eff}$} \\
\colhead{} & \colhead{(\AA)} & \colhead{(\AA)} & \colhead{(km s$^{-1}$)} & \colhead{(km s$^{-1}$)} 
& \colhead{(erg cm$^{-2}$ s$^{-1}$)} & \colhead{}  & \colhead{(K)} }
\startdata
\multicolumn{8}{c}{Single Gaussian Fits} \\
\hline
\ion{C}{3} & 1174.935 & 1174.935\tablenotemark{1} & 0  & 37.5$\pm$6.9 & 5.33$E-$15$\pm$1.04$E-$15 & 0.51 & 4.5 \\
\ion{C}{3} & 1175.265 & 1175.265\tablenotemark{1} & 0 & 32.9$\pm$6.6 & 4.53$E-$15$\pm$9.65$E-$16 & '' & '' \\
\ion{C}{3} & 1175.592 & 1175.592\tablenotemark{1} & 0 & 43.4$\pm$9.2 & 6.54$E-$15$\pm$1.47$E-$15 & '' & ''\\
\ion{C}{3} & 1175.713 & 1175.713\tablenotemark{1} & 0 & 31.9$\pm$3.8 & 1.07$E-$14$\pm$1.37$E-$15 & '' & ''\\
\ion{C}{3} & 1175.989 & 1175.989\tablenotemark{1} & 0 & 33.6$\pm$7.1 & 4.58$E-$15$\pm$1.01$E-$15 & '' & ''\\
\ion{C}{3} & 1176.372 & 1176.372\tablenotemark{1} & 0 & 47.9$\pm$11.2 & 4.79$E-$15$\pm$1.19$E-$15 & '' & '' \\
\ion{O}{5}\tablenotemark{2} & 1218.390 & 1218.339 & $-$11.0$\pm$1.5 & 30.0$\pm$3.2 & 4.81$E-$15$\pm$5.21$E-$16 & 0.54 & 5.4 \\
\ion{N}{5} & 1242.804 & 1242.793 & $-$1.1$\pm$0.7 & 33.3$\pm$1.7 & 8.77$E-$15$\pm$4.56$E$-16 & 0.59 & 5.3 \\
\ion{Si}{2} & 1264.737 & 1264.723 & $-$1.8$\pm$2.1 & 28.9$\pm$4.3 & 1.32$E-$15$\pm$2.09$E-$16 & 0.43 & 4.4 \\
\ion{Si}{2} & 1265.001 & 1264.990 & $-$1.1$\pm$3.6 & 30.6$\pm$7.4 & 7.54$E-$16$\pm$1.88$E-$16 & '' & 4.4 \\
\ion{Si}{3} & 1294.543 & 1294.534 & $-$0.6$\pm$3.0 & 35.2 $\pm$4.6 & 1.28$E-$15$\pm$2.34$E-$16 & 0.82 & 4.8 \\
\ion{Si}{3} & 1296.726 & 1296.755 & 8.2$\pm$3.0 & 22.0$\pm$6.2 & 5.20$E-$16$\pm$1.54$E-$16 & 0.46 & 4.8 \\
\ion{Si}{3}\tablenotemark{2,3} & 1298.944 & 1298.925 & $-$2.9$\pm$2.1 & 30.2$\pm$4.4 & 1.32$E-$15$\pm$2.03$E-$16 & 0.70 & 4.8 \\
\ion{Si}{3} & 1303.323 & 1303.328 & 2.7$\pm$2.8 & 24.6$\pm$5.5 & 5.75$E-$16$\pm$1.38$E-$16 & 0.60 & 4.8 \\
\ion{Si}{2} & 1309.274 & 1309.275 & 1.7$\pm$2.7 & 28.2$\pm$6.0 & 8.20$E-$16$\pm$1.78$E-$16 & 1.04 & 4.4 \\
\ion{Fe}{21} & 1354.080 & 1354.072 & $-$0.3$\pm$6.4 & 108.5$\pm$13.5 & 3.27$E-$15$\pm$4.27$E-$16 & 0.71 & 7.0 \\
\ion{O}{5} & 1371.292 & 1371.296 & 2.4$\pm$1.7 & 23.6$\pm$3.5 & 1.25$E-$15$\pm$1.91$E-$16 & 0.66 & 5.4 \\
\ion{O}{4}\tablenotemark{2} & 1401.171 & 1401.168 & 0.9$\pm$3.4 & 26.1$\pm$6.8 & 5.89$E-$16$\pm$1.63$E-$16 & 0.48 & 5.1 \\
\ion{O}{4} & 1399.779 & 1399.759 & $-$2.8$\pm$6.0 & 26.1\tablenotemark{4} & 3.19$E-$16$\pm$1.62$E-$16 & 0.51 & 5.1 \\
\ion{Si}{2} & 1526.706 & 1526.687 & $-$2.4$\pm$2.8 & 26.7$\pm$5.7 & 1.36$E-$15$\pm$3.07$E-$16 & 0.49 & 4.4 \\
\ion{Si}{2} & 1533.430 & 1533.410 & $-$2.4$\pm$3.7 & 24.2$\pm$7.6 & 1.70$E-$15$\pm$5.6$E-$16 & 0.44 & 4.4 \\
\hline
\multicolumn{8}{c}{Double  Gaussian Fits} \\
\hline
\ion{Si}{3} & 1206.499 & 1206.508 & 3.7$\pm$1.0 & 42.2$\pm$2.0 & 1.71$E-$14$\pm$8.9$E-$16 & 0.54 & 4.8 \\
	   &		& 1206.427 & $-$16.4$\pm$10.0 & 130.2$\pm$21 & 6.86$E-$15$\pm$1.14$E-$15 & '' & '' \\
\ion{N}{5} & 1238.821 & 1238.805 & $-$2.4$\pm$0.5 & 29.8$\pm$1.0 & 1.27$E-$14$\pm$4.86$E-$16 & 0.82 & 5.3 \\
           &		& 1238.816 & 0.3$\pm$2.2 & 77.7$\pm$4.8 & 8.49$E-$15$\pm$5.47$E-$16 & '' & ''\\
\ion{Si}{4} & 1393.755 & 1393.754 & 1.3$\pm$0.6 & 25.6$\pm$1.3 & 9.18$E-$15$\pm$4.77$E-$16 & 0.76 & 4.9 \\
           &		& 1393.796 & 10.3$\pm$3.2 & 88.4$\pm$6.7 & 7.52$E-$15$\pm$6.0$E-$16 & '' & ''\\
\ion{Si}{4} & 1402.770 & 1402.766 & 0.6$\pm$0.6 & 23.9$\pm$1.2 & 5.27$E-$15$\pm$2.89$E-$16 & 0.97 & 4.9 \\
	   &		& 1402.761 & $-$0.4$\pm$3.8 & 63.7$\pm$8.0 & 3.17$E-$15$\pm$4.14$E-$16 & '' & ''\\
\ion{C}{4} & 1548.201 & 1548.200 & 1.3$\pm$0.5 & 31.8$\pm$1.0 & 4.30$E-$14$\pm$1.38$E-$15 & 0.89 & 5.0 \\
           &		& 1548.217 & 4.6$\pm$4.3 & 135.9$\pm$9.0 & 2.08$E-$14$\pm$1.4$E-$15 & '' & ''\\
\ion{C}{4} & 1550.772 & 1550.758 & $-$1.2$\pm$0.7 & 28.2$\pm$1.4 & 2.11$E-$14$\pm$1.1$E-$15 & 0.56 & 5.0 \\
	   &		& 1550.824 & 11.5$\pm$5.9 & 100.3$\pm$12.2 & 1.10$E-$14$\pm$1.4$E-$15 & '' & ''\\
%\ion{C}{2} & 1335.709 & 1335.688 & $-$4.7$\pm$0.4 & 26.9$\pm$0.9 & 1.86$E-$14$\pm$6.4$E-$16 & 0.86 & 4.4 \\
   	   %&		& 1335.738 &6.5$\pm$2.4 & 73.8$\pm$5.1 & 9.35$E-$15$\pm$6.71$E-$16 & '' & ''\\
%\ion{C}{2} & 1334.524 & 1334.489 & $-$7.9$\pm$0.4 & 12.4$\pm$1.1 & 3.29$E-$15$\pm$2.98$E-$16 & 0.93 & 4.5 \\
	  %&		& 1334.559 & 7.9$\pm$2.0 & 77.3$\pm$4.5 & 7.89$E-$15$\pm$4.74$E-$16 & '' & ''\\
\enddata
\tablenotetext{1}{Wavelengths of transitions fixed; width and peak flux allowed to vary.}
\tablenotetext{2}{transition is sensitive to electron density}
\tablenotetext{3}{feature is blended with another transition}
\tablenotetext{4}{Width fixed to that of \ion{O}{4} $\lambda$1401.}
\tablenotetext{f}{forbidden transition; not used in DEM analysis}
\tablenotetext{b}{feature is blended with another transition of the same ion}
\end{deluxetable}
\clearpage

\begin{deluxetable}{llll}
\tablewidth{0pt}
\tablenum{3}
\tablecolumns{4}
\tablecaption{Expected and Observed Effectively Thin Line Ratios in STIS Spectrum \label{tbl:linerat}}
\tablehead{ \colhead{ion} & \colhead{Wavelengths} & \colhead{theo.} & \colhead{Obs.} }
\startdata
\ion{Si}{2} & 1533.430/1526.706 & 2 & 1.25$\pm$0.5 \\
\ion{C}{3} & 1175.265/1176.372 & 0.8 & 0.9$\pm$0.3 \\
\ion{Si}{3} & 1296.726/1303.323 & 0.8 & 0.9$\pm$0.3 \\
\ion{C}{3} & 1175.713/1174.935 & 3 & 2.0$\pm$0.5 \\
\ion{Si}{4} & 1393.755/1402.770 & 2 & 1.7$\pm$0.1\tablenotemark{N},2.4$\pm$0.4\tablenotemark{B} \\
\ion{C}{4} & 1548.201/1550.772 & 2 & 2$\pm$0.1\tablenotemark{N},1.9$\pm$0.3\tablenotemark{B} \\
\ion{N}{5} & 1238.821/1242.804 & 2 & 1.4$\pm$0.1\tablenotemark{N1},2.4\tablenotemark{N2} \\
\enddata
\tablenotetext{N}{Ratio value refers to flux ratio of the two narrow components.}
\tablenotetext{B}{Ratio value refers to flux ratio of the two broad components.}
\tablenotetext{N1}{Ratio value refers to flux ratio of narrow component of 1238 \AA\ line
to total flux of 1242 \AA\ line.}
\tablenotetext{N2}{Ratio value refers to flux ratio of narrow and broad components of
1238 \AA\ line to total flux of 1242 \AA\ line.}
\end{deluxetable}

\begin{deluxetable}{llcccc}
\tablewidth{0pt}
\tablenum{4}
\tablecolumns{5}
\tablecaption{Detected Lines in FUSE Spectra \label{tbl:fuselines}}
\tablehead{ \colhead{$\lambda_{\rm lab}$ } &
\colhead{Transition} & \colhead{$\lambda_{\rm fit}$} &\colhead{flux} & \colhead{$\log T_{\rm eff}$}
   \\
\colhead{(\AA)} & \colhead{} & \colhead{(\AA)} & \colhead{10$^{-14}$ erg cm$^{-2}$ s$^{-1}$} 
&\colhead{} &   }
\startdata
977.022 & \ion{C}{3} & \ldots & 7.9$\pm$1.1 & 4.9\\
1037.615 & \ion{O}{6} & \ldots & 11.1$\pm$0.7 &5.5 \\
1031.914 & \ion{O}{6} & \ldots & 22.9$\pm$1.1 &5.5 \\
974.860 & \ion{Fe}{18} & \ldots & 0.53$\pm$0.31 &6.8 \\
\enddata
\end{deluxetable}

\begin{deluxetable}{llllll}
\tablewidth{0pt}
\tablenum{5}
\tablecolumns{6}
\tablecaption{Detected Lines in EUVE Spectra \label{tbl:euvelines}}
\tablehead{ \colhead{$\lambda_{\rm lab}$ } &
\colhead{Transition} & \colhead{$\lambda_{\rm fit}$} &\colhead{flux\tablenotemark{a}}  
&\colhead{$\log T_{\rm eff}$}   \\
\colhead{(\AA)} & \colhead{} & \colhead{(\AA)} & \colhead{(erg cm$^{-2}$s$^{-1}$)} &\colhead{} 
 }
\startdata
93.923 & \ion{Fe}{18} & 93.923$\pm$0.038 & 4.52E-14$\pm$9.90E-15&6.8 \\
128.73 & \ion{Fe}{21} & 128.730$\pm$0.045 & 3.67E-14$\pm$9.40E-15 &7.0 \\
132.85 & \ion{Fe}{23},\ion{Fe}{20} & 132.85$\pm$0.032 & 6.90E-14$\pm$1.27E-14& \\
284.16 & \ion{Fe}{15} & 284.16$\pm$0.11 &1.54E-13$\pm$3.13E-14&6.4  \\
335.41 & \ion{Fe}{16} & 335.41$\pm$0.10 & 1.57E-13$\pm$2.91E-14&6.4  \\
360.761 & \ion{Fe}{16} & 360.76$\pm$0.18 & 7.47E-14$\pm$2.65E-14&6.4  \\
\enddata
\tablenotetext{a}{Fluxes have been corrected for interstellar absorption, using N$_{H}$=4$\times$10$^{18}$ cm$^{-2}$.}
\end{deluxetable}

\begin{deluxetable}{lllcccc}
\tablewidth{0pt}
\tablenum{6}
\tablecolumns{7}
\rotate
\tablecaption{Detected Lines in Chandra Spectra \label{tbl:xraylines}}
\tablehead{ \colhead{$\lambda_{\rm lab}$ } &
\colhead{Transition} & \colhead{$\lambda_{\rm HEG}$} &\colhead{$f_{\rm HEG}$\tablenotemark{a}} 
& \colhead{$\lambda_{\rm MEG}$} & \colhead{$f_{\rm MEG}$\tablenotemark{a}} & 
\colhead{$\log T_{\rm eff}$}  \\
\colhead{(\AA)} & \colhead{} & \colhead{(\AA)} & \colhead{(erg cm$^{-2}$ s$^{-1}$)} & \colhead{(\AA)} & 
\colhead{(erg cm$^{-2}$ s$^{-1}$)} & \colhead{}   }
\startdata
%5.0387  &\ion{S}{15} &  5.0383$\pm$0.0017 &  6.47E$-$14$\pm$1.97E$-$14  &   5.0358$\pm$0.0036 &  3.06E$-$14$\pm$1.05E$-$14 && \\
6.1804   &\ion{Si}{14} & 6.1821$\pm$0.0004 &  8.28E$-$14$\pm$5.99E$-$15    &    6.1808$\pm$0.0009 &  6.86E$-$14$\pm$5.81E$-$15   &7.3 \\
6.6479 &\ion{Si}{13}&   6.6487$\pm$0.0005 &  8.84E$-$14$\pm$7.72E$-$15    &    6.6480$\pm$0.0007 &  8.87E$-$14$\pm$6.02E$-$15   &6.9 \\
6.6882 & \ion{Si}{13}&   $\ldots$& $\ldots$ &      6.6854$\pm$0.0018 &  2.49E$-$14$\pm$4.35E$-$15  & '' \\
6.7403 &\ion{Si}{13} &   6.7411$\pm$0.0004&   6.38E$-$14$\pm$4.65E$-$15    &    6.7380$\pm$0.0010 &  5.23E$-$14$\pm$4.77E$-$15 &''\\
8.4246 &\ion{Mg}{12} &    8.4204$\pm$0.0011 &  3.00E$-$14$\pm$5.91E$-$15    &    8.4203$\pm$0.0014 &  3.17E$-$14$\pm$4.18E$-$15 &6.9 \\
9.1687 &\ion{Mg}{11} &   9.1702$\pm$0.0007 &  3.15E$-$14$\pm$3.98E$-$15    &    9.1675$\pm$0.0012  & 3.59E$-$14$\pm$3.99E$-$15 &6.9 \\
9.2312&\ion{Mg}{11} &    9.2306$\pm$0.0034 &  7.51E$-$15$\pm$4.70E$-$15    &    9.2305$\pm$ 0.0034 &  9.62E$-$15$\pm$3.13E$-$15 & ''\\
9.3143 &\ion{Mg}{11} &   9.3156$\pm$0.0023 &  1.39E$-$14$\pm$5.75E$-$15    &    9.3131$\pm$0.0021  & 1.66E$-$14$\pm$3.29E$-$15  & ''\\
9.4797 & \ion{Fe}{21} &   9.4782$\pm$0.0010 &  2.41E$-$14$\pm$4.21E$-$15    &    9.4776$\pm$0.0024 &  1.39E$-$14$\pm$3.13E$-$15  &7.0\\
9.7085 &\ion{Ne}{10} &   9.7097$\pm$0.0013 &  1.76E$-$14$\pm$4.08E-15    &    9.7096$\pm$0.0014 &  2.85E$-$14$\pm$3.72E$-$15 &6.9\\
10.2385 &\ion{Ne}{10} &  10.2399$\pm$0.0004 &  6.45E$-$14$\pm$4.82E$-$15    &   10.2382$\pm$0.0009 &  5.80E$-$14$\pm$4.78E$-$15&6.9   \\
%10.6190 & \ion{Fe}{24}&  10.6205$\pm$0.0014 &  2.49E$-$14$\pm$6.34E$-$15    &   10.6153$\pm$0.0019 &  2.15E$-$14$\pm$3.81E$-$15 &&  \\
%10.8160 & \ion{Fe}{19} &   $\ldots$ &  $\ldots$   &   10.8188$\pm$0.0034&   1.12E$-$14$\pm$3.57E$-$15  && \\
%10.9810 & \ion{Fe}{23} &  $\ldots$ &   $\ldots$   &   10.9797$\pm$0.0033 &  1.38E$-$14$\pm$4.25E$-$15 && \\
11.0010 &\ion{Ne}{9} & $\ldots$ &  $\ldots$   &   10.9981$\pm$0.0027 &  1.82E$-$14$\pm$4.70E$-$15 &6.7  \\
11.1760 & \ion{Fe}{24} &  $\ldots$ &  $\ldots$  &   11.1777$\pm$0.0026\tablenotemark{B} &  1.71E$-$14$\pm$4.24E$-$15&  7.3 \\
11.7360 & \ion{Fe}{23} &  11.7375$\pm$0.0016 &  2.86E$-$14$\pm$8.31E$-$15 &      11.7363$\pm$0.0022 &  2.49E$-$14$\pm$5.12E$-$15  & 7.2\\
11.7700  & \ion{Fe}{22} & 11.7685$\pm$0.0010 &  3.83E$-$14$\pm$7.27E$-$15 &      $\ldots$ &  $\ldots$  & 7.1\\
12.1375 & \ion{Ne}{10} &  12.1350$\pm$0.0004 &  3.82E$-$13$\pm$2.65E$-$14  &   12.1314$\pm$0.0004 &  4.13E$-$13$\pm$1.39E$-$14  & 6.7\\
12.1610 & \ion{Fe}{23} &  12.1614$\pm$0.0009 &  3.93E$-$14$\pm$6.67E$-$15    &   $\ldots$ &  $\ldots$ &7.1\\
12.2660 & \ion{Fe}{17} &  $\ldots$  &  $\ldots$    & 12.2623$\pm$0.0020 &   3.33E$-$14$\pm$6.33E$-$15 &6.7\\
12.2840 & \ion{Fe}{21} &  12.2855$\pm$0.0012 &  3.42E$-$14$\pm$7.56E$-$15    &   12.2835$\pm$0.0017 &  4.21E$-$14$\pm$6.67E$-$15 &7.0  \\
13.4473 & \ion{Ne}{9} &  13.4480$\pm$0.0003 &  2.47E$-$13$\pm$1.34E$-$14    &   13.4459$\pm$0.0005 &  2.67E$-$13$\pm$1.37E$-$14  &6.5\\
13.5180 & \ion{Fe}{19} &  13.5185$\pm$0.0014 &  6.06E$-$14$\pm$1.52E$-$14    &   $\ldots$ &  $\ldots$  & 6.9 \\
13.5531 & \ion{Ne}{9} &  13.5556$\pm$0.0022 &  3.44E$-$14$\pm$1.39E$-$14    &   13.5566$\pm$0.0027 &  3.05E$-$14$\pm$7.65E$-$15  &6.5\\
13.6990 & \ion{Ne}{9} &  13.7005$\pm$0.0005 &  1.10E$-$13$\pm$1.07E$-$14    &   13.6971$\pm$0.0008  & 1.37E-13$\pm$1.04E-14  & ''\\
14.2080& \ion{Fe}{18} &   14.2066$\pm$0.0008 &  9.60E$-$14$\pm$1.47E$-$14     &  14.2043$\pm$0.0013  & 8.71E$-$14$\pm$1.03E$-$14&6.9  \\
14.3730& \ion{Fe}{18}&  $\ldots$&  $\ldots$ &  14.3719$\pm$0.0031  & 2.75E$-$14$\pm$8.00E$-$15 &''  \\
14.5340&\ion{Fe}{18}&   $\ldots$ &  $\ldots$ &  14.5378$\pm$0.0033  & 2.35E$-$14$\pm$7.38E$-$15 &''  \\
15.0140&  \ion{Fe}{17} & 15.0149$\pm$0.0003 &  2.51E$-$13$\pm$1.15E$-$14    &  15.0119$\pm$0.0007  & 2.83E$-$13$\pm$1.86E$-$14  &6.7 \\
%15.0790&  \ion{Fe}{19} & $\ldots$ &  $\ldots$ &  15.0821$\pm$0.0029  & 3.14E$-$14$\pm$8.54E$-$15 &&  \\
15.1980& \ion{Fe}{19} &  $\ldots$ &  $\ldots$ &  15.1966$\pm$0.0033  & 3.26E$-$14$\pm$1.02E$-$14 &6.9  \\
15.2610&\ion{Fe}{17} &   15.2627$\pm$0.0017\tablenotemark{B} &  6.69E$-$14$\pm$2.05E$-$14     &  15.2607$\pm$0.0010\tablenotemark{B}  & 1.06E$-$13$\pm$1.03E$-$14 &6.7 \\
%15.2797& \ion{Fe}{17} &  $\ldots$ &  $\ldots$ &  15.2786$\pm$0.0024  & 4.03E$-$14$\pm$9.06E$-$15  &&  \\
15.4530&  \ion{Fe}{17} & $\ldots$ &  $\ldots$ &  15.4470$\pm$0.0028  & 2.31E$-$14$\pm$6.22E$-$15  &6.7  \\
15.6250& \ion{Fe}{18} &  $\ldots$ &  $\ldots$ &  15.6244$\pm$0.0022  & 3.75E$-$14$\pm$7.89E$-$15  &6.9  \\
16.0710&\ion{Fe}{18} &    $\ldots$ &  $\ldots$ &  16.0739$\pm$0.0015  & 6.84E$-$14$\pm$9.90E$-$15 &''   \\
16.7800&  \ion{Fe}{17} &  $\ldots$ &  $\ldots$ &  16.7763$\pm$0.0009  & 1.54E$-$13$\pm$1.35E$-$14  &6.7  \\
17.0510&\ion{Fe}{17}&    $\ldots$ &  $\ldots$ &  17.0502$\pm$0.0008  & 2.37E$-$13$\pm$1.78E$-$14   &'' \\
17.0960& \ion{Fe}{17} &   $\ldots$ &  $\ldots$ &  17.0955$\pm$0.0009  & 2.01E$-$13$\pm$1.73E$-$14  &''  \\
%17.6230& \ion{Fe}{18} &   $\ldots$ &  $\ldots$ &  17.6250$\pm$0.0035  & 2.69E$-$14$\pm$8.87E$-$15  &&  \\
18.9671 &  \ion{O}{8} & $\ldots$  & $\ldots$  & 18.9686$\pm$0.0005   &9.29E$-$13$\pm$3.98E$-$14  &6.4  \\
21.6015 &\ion{O}{7} &   $\ldots$  & $\ldots$  & 21.6031$\pm$0.0012   &3.13E$-$13$\pm$3.58E$-$14  &''  \\
21.8036 & \ion{O}{7} & $\ldots$  & $\ldots$  & 21.8065$\pm$0.0026   &1.01E$-$13$\pm$2.47E$-$14 & ''\\
22.0977 & \ion{O}{7} &  $\ldots$  & $\ldots$  & 22.0977$\pm$0.0018   &1.71E$-$13$\pm$2.84E$-$14 & ''  \\
24.7792 & \ion{N}{7} &  $\ldots$  & $\ldots$  & 24.7786$\pm$0.0029   &7.49E$-$14$\pm$2.09E$-$14  &''  \\
\enddata
\tablenotetext{a}{Fluxes have been corrected for interstellar absorption using N$_{H}$=4$\times$10$^{18}$ cm$^{-2}$.}
\tablenotetext{B}{Blend}
\end{deluxetable}

\begin{deluxetable}{lcccc}
\tablewidth{0pt}
\tablenum{7}
\tablecolumns{4}
\tablecaption{Coronal Abundance Determinations\label{tbl:abund}}
\tablehead{ \colhead{FIP } &
\colhead{Element} & \colhead{Ions\tablenotemark{a}} &\colhead{Value}  \\
\colhead{(eV)} & \colhead{} & \colhead{} &  \colhead{} }
\startdata
%11.26 & C & II,III,IV &    \\
14.534 & N & VII &  N/Fe=0.95$\pm$0.32 \\
13.618 & O & VII, VIII & O/Fe=0.80$\pm$0.04 \\
21.564 & Ne & IX,X & Ne/Fe=1.72$\pm$0.14\\
7.646 & Mg & XI,XII & Mg/Fe=0.49$\pm$0.05 \\
8.151 & Si & XIII,XIV & Si/Fe=1.48$\pm$0.09 \\
7.87 & Fe & XV--XXIV & Fe/H=0.53$\pm$0.16 \\
\enddata
\tablenotetext{a}{Ionization stages used in coronal abundance determination}
\end{deluxetable}

\begin{deluxetable}{llll}
\tablewidth{0pt}
\tablenum{8}
\tablecolumns{4}
\tablecaption{Variability check: \ion{Fe}{21} and \ion{Fe}{18} ratios \label{tbl:var}}
\tablehead{ \colhead{ratio} & \colhead{Observed\tablenotemark{1}} & \colhead{Predicted\tablenotemark{1}  } & \colhead{Log$_{10}$ T$_{\rm max}\pm0.15$ } }
\startdata
\cline{1-4}
\multicolumn{4}{c}{--- Fe XXI ---} \\
\cline{1-4}
$\lambda$1354\tablenotemark{S}/$\lambda$9.48\tablenotemark{C} & 0.22$\pm$0.04 & 0.5--1.2 & 6.85--7.15 \\
$\lambda$1354\tablenotemark{S}/$\lambda$12.284\tablenotemark{C} & 0.11$\pm$0.02 & 0.08--0.15 & 6.85--7.15 \\
$\lambda$128.73\tablenotemark{E}/$\lambda$9.48\tablenotemark{C} & 1.93$\pm$0.56 & 5.--13. & 6.85--7.15 \\
$\lambda$128.73\tablenotemark{E}/$\lambda$12.284\tablenotemark{C} & 0.96$\pm$0.28 & 0.8--1.6 & 6.85--7.15 \\
$\lambda$1354\tablenotemark{S}/$\lambda$128.73\tablenotemark{E} & 0.12$\pm$0.03 & 0.09--0.10 & 6.85--7.15 \\
\cline{1-4}
\multicolumn{4}{c}{--- Fe XVIII ---} \\
\cline{1-4}
$\lambda$974\tablenotemark{F}/$\lambda$14.208\tablenotemark{C} & 0.06$\pm$0.04     & 0.04--0.12 & 6.65--7.05 \\
$\lambda$974\tablenotemark{F}/$\lambda$16.07\tablenotemark{C} &  0.08$\pm$0.05    & 0.14--0.30 & 6.65--7.05 \\
$\lambda$93.92\tablenotemark{E}/$\lambda$14.208\tablenotemark{C} & 0.49$\pm$0.12 & 0.32--1.05 & 6.65--7.05 \\
$\lambda$93.92\tablenotemark{E}/$\lambda$16.07\tablenotemark{C} & 0.66$\pm$0.17 & 1.2--2.8 & 6.65--7.05 \\
$\lambda$974\tablenotemark{F}/$\lambda$93.92\tablenotemark{E} &   0.12$\pm$0.07     & 0.109-0.118 & 6.65--7.05 \\
\enddata
\tablenotetext{1}{Energy flux ratios}
\tablenotetext{S}{HST/STIS observation 2001 September}
\tablenotetext{F}{FUSE observation 2002 July}
\tablenotetext{E}{EUVE observation 1993 September}
\tablenotetext{C}{Chandra observation 2001 September}
\end{deluxetable}

\begin{deluxetable}{lcccc}
\tablewidth{0pt}
\tablenum{9}
\tablecolumns{5}
\tablecaption{Densities Derived from UV and X-ray Line Ratios\label{tbl:dens}}
\tablehead{ \colhead{Ion} & \colhead{ratio } &
\colhead{R=f$_{1}$/f$_{2}$\tablenotemark{1}} & \colhead{$\log_{10}$T$_{0}$} &\colhead{n$_{e}$} \\
\colhead{} & \colhead{} & \colhead{} & \colhead{} & \colhead{(cm$^{-3}$)}  }
\startdata
\ion{O}{4} & $\lambda$1401/$\lambda$1399 & 2.2$\pm$0.9 &   5.2 (T$_{max}$)   &$<$1.6$\times$10$^{11}$ \\
\ion{O}{5} & $\lambda$1218/$\lambda$1371 & 4.6$\pm$0.7 &  5.4 (T$_{max}$) &1.0$\times$10$^{11}$ (7.4$\times$10$^{10}$--1.6$\times$10$^{11}$ \\
\ion{O}{7} & $\lambda$22.10/$\lambda$21.80 & 2.0$\pm$1.0 &6.4   & 3.0$\times$10$^{10}$(8.7$\times$10$^{9}$--9.5$\times$10$^{10}$ \\
\ion{Ne}{9} & $\lambda$13.70/$\lambda$13.55 & 3.6$\pm$0.9 &6.7    & $<$1.7$\times$10$^{11}$\\
\ion{Mg}{11} & $\lambda$9.31/$\lambda$9.23 & 1.4$\pm$0.6 & 6.9     &1.0$\times$10$^{13}$ (4.3$\times$10$^{12}$--2.6$\times$10$^{13}$) \\
\ion{Si}{13} & $\lambda$6.74/$\lambda$6.69 & 2.1$\pm$0.4 &6.9    &$<$1.8$\times$10$^{13}$  \\
\ion{Fe}{17} & $\lambda$17.051/$\lambda$17.096 & 1.18$\pm$0.13 & 7.0 & $<$ 10$^{13}$ \\
\enddata
\tablenotetext{1}{Energy flux ratio}
\end{deluxetable}
\clearpage
\begin{deluxetable}{ccc}
\tablenum{10}
\tablecolumns{3}
\tablecaption{Using the narrow line profile component in DEM analyses \label{tbl:ncbc}}
\tablehead{\colhead{Transition} & \colhead{$f_{obs}/f_{pred}$\tablenotemark{a}} & \colhead{$f_{obs}/f_{pred}$\tablenotemark{b}} }
\startdata
\cline{1-3}
\multicolumn{3}{c}{Coronal Abundance Pattern} \\
\cline{1-3}
\ion{N}{5} $\lambda$1238 & 3.2 & 5.3 \\
\ion{Si}{3} $\lambda$1206 & 0.3 & 0.4 \\
\ion{Si}{4} $\lambda$1393 & 0.7 & 1.3 \\
\ion{Si}{4} $\lambda$1402 & 0.8 & 1.3 \\
\ion{C}{4} $\lambda$1548 & 2.1 & 3.1 \\
\ion{C}{4} $\lambda$1550 & 2.1 & 3.1 \\
\cline{1-3} \\
\multicolumn{3}{c}{Hybrid Abundance Pattern} \\
\cline{1-3} \\
\ion{N}{5} $\lambda$1238 & 2.2 & 3.6 \\
\ion{Si}{3} $\lambda$1206 & 0.4 & 0.5 \\
\ion{Si}{4} $\lambda$1393 & 1.5 & 2.8 \\
\ion{Si}{4} $\lambda$1402 & 1.7 & 2.8 \\
\ion{C}{4} $\lambda$1548 & 1.5 & 2.3 \\
\ion{C}{4} $\lambda$1550 & 1.5 & 2.3 \\
\enddata
\tablenotetext{a}{Agreement between observed and predicted flux using integrated line flux of
narrow Gaussian component}\\
\tablenotetext{b}{Agreement between observed and predicted flux using integrated line flux of
both Gaussian components}
\end{deluxetable}
%\begin{deluxetable}{llll}
%\tablenum{11}
%\tablecolumns{4}
%\tablecaption{Comparison of Collisional Length Scales and Pressure Scale Height in
%the Outer Atmosphere of EV~Lac \label{tbl:lengths}}
%\tablehead{ \colhead{Log T} & \colhead{H$_{n}$} & \colhead{$\lambda_{\rm mfp}$\tablenotemark{1}} & \colhead{H$_{n}$/
%$\lambda_{\rm mfp}$}   \\
%\colhead{} & \colhead{ (cm)} & \colhead{(cm)} & \colhead{} }  
%\startdata
%5.2 & 1.0$\times$10$^{4}$ & 8.2$\times$10$^{2}$ & 10 \\
%5.4 & 1.6$\times$10$^{4}$ & 3.3$\times$10$^{3}$ & 5 \\
%6.4 & 1.6$\times$10$^{5}$ & 1.1$\times$10$^{6}$ & 0.1 \\
%6.7 & 3.2$\times$10$^{5}$ & 7.7$\times$10$^{5}$ & 0.4 \\
%6.9 & 5.1$\times$10$^{5}$ & 1.8$\times$10$^{4}$ & 30 \\
%7.0 & 6.4$\times$10$^{5}$ & 5.2$\times$10$^{4}$ & 10 \\
%\enddata
%\tablenotetext{1}{Coulomb mean free path calculated using densities from Table~\ref{tbl:dens}}
%\end{deluxetable}
%\begin{figure}
%\begin{center}
%\includegraphics[scale=0.5]{chandralc.ps}
%\caption{Light curve of Chandra observation of EV~Lac.  The 100 ks observation was marked by numerous
%flaring events; we selected the initial $\sim$ 40 ks plus another interval composed of small
%enhancements for detailed spectral analysis. \label{fig:chandralc}}
%\end{center}
%\end{figure}
\clearpage

%\begin{figure}
%\begin{center}
%\includegraphics[scale=0.7]{evlac_hybrid_cpress_fluxagree.ps}
%\caption{{\it (top)} DEM reconstructed from STIS, FUSE, EUVE, and Chandra spectra
%using a hybrid abundance:  coronal abundance pattern at T$>$1MK, and 
%solar photospheric for T$<$1MK. Symbols and legend are as in Figure~\ref{fig:coronaldem}.
%{\it (bottom)} Same as Figure~\ref{fig:coronaldem}bottom, for the hybrid abundance case.
  %\label{fig:hybriddem}}
%\end{center}
%\end{figure}

\clearpage

\end{document}